\documentclass{article} \usepackage[onecolumn]{emulateapj}
\usepackage{epsf}

\newcommand{\wepsilon}{\hbox{$\widehat\epsilon$}}

\begin{document}

\submitted{Accepted for publication in The Astronomical Journal}

\title{THE EFFICIENCY OF GLOBULAR CLUSTER FORMATION}
\author{Dean E. McLaughlin\altaffilmark{1}}
\affil{Department of Astronomy, 601 Campbell Hall\\
University of California, Berkeley, CA 94720-3411\\
dean@crabneb.berkeley.edu}

\altaffiltext{1}{Hubble Fellow}

\lefthead{{\sc McLAUGHLIN}}
\righthead{THE EFFICIENCY OF CLUSTER FORMATION}

\begin{abstract}

The specific frequencies of globular cluster systems, $S_N\propto {\cal N}_
{\rm tot}/L_{V,{\rm gal}}\propto M_{\rm gcs}/M_{\rm stars}$, are discussed in
terms of their connection to the efficiency of globular cluster formation in
galaxy halos, which is claimed to reflect a generic aspect of the star
formation process as it operates even at the current epoch. It is demonstrated
that the total {\it masses} of GCSs are little affected by the dynamical
destruction of low-mass clusters at small galactocentric radii. This permits
direct, empirical estimates of the cluster formation efficiency by mass,
$\epsilon_{\rm cl}\equiv M_{\rm gcs}^{\rm init}/M_{\rm gas}^{\rm init}$, even
after $10^{10}$ years of GCS evolution. However, the standard practice of
using only the stellar luminosities of galaxies as indicators of their initial
total gas masses (and thus relating $S_N$ to $\epsilon_{\rm cl}$ in one step)
leads to serious conceptual problems, that are reviewed here. The {\it first
specific frequency problem}, which is the well known tendency for many
brightest cluster galaxies to have higher than average $S_N$, is a global one;
the {\it second specific frequency problem} is a local one, in which the more
extended spatial distribution of GCSs relative to halo stars in some (not all)
bright ellipticals leads to $S_N$ values that increase with radius inside the
galaxies.  Extending similar suggestions in the recent literature, it is
argued that these trends in $S_N$ do not reflect any such behavior in the
underlying $\epsilon_{\rm cl}$; rather, {\it both} of these problems stem from
neglecting the hot, X-ray emitting gas in and around many large ellipticals,
and both may be alleviated by including this component in estimates of
$M_{\rm gas}^{\rm init}$.

This claim is checked and confirmed in each of M87, M49, and NGC 1399, all of
which have been thought to suffer from one or the other of these $S_N$
problems. Existing data are combined to construct GCS surface density profiles
that extend over nearly the whole extents of these three galaxies, and a
non-parametric, geometrical deprojection algorithm is developed to afford a
direct comparison between the {\it volume} density profiles of their GCSs,
stars, and gas. It is found, in each case, that $\rho_{\rm cl}\propto(\rho_
{\rm gas}+\rho_{\rm stars})$ at radii beyond roughly a stellar effective
radius, inside of which dynamical evolution may have depeleted the initial
GCSs. The constant of proportionality is the same in all three galaxies:
$\epsilon_{\rm cl}=0.0026\pm0.0005$. Taken together, these results suggest that
GCSs generally should be more spatially extended than stellar halos only in
gas-rich galaxies that also have a high global specific frequency.

The implication that $\epsilon_{\rm cl}$ might have had a universal value is
supported by global GCS data for a sample of 97 giant ellipticals,
brightest cluster galaxies, and faint dwarfs. The total globular cluster
populations in all of these early-type systems are in excellent
agreement with the predictions of a constant $\epsilon_{\rm cl}$ at the level
observed directly in M87, M49, and NGC 1399; all {\it systematic} variations
in GCS specific frequency between galaxies are shown to result entirely from
different relations, in different magnitude ranges, between $M_{\rm gas}^{\rm
init}$ and the present-day $L_{V,{\rm gal}}$. An identical $\epsilon_{\rm cl}$
is also calculated for the Pop.~II spheroid of the Milky Way, and is indicated
(although less conclusively) for the ongoing formation of open clusters. The
inferred universal cluster formation efficiency, of $\simeq$0.25\% by mass,
should serve as a strong constraint on general theories of star and cluster
formation. The associated inference of a non-universal formation efficiency
for unclustered stars is considered, particularly in terms of the suggestion
that this might result, both in dwarf galaxies and at large galactocentric
radii in the brightest ellipticals, from feedback and galactic winds.
Implications for a merger-formation model of early-type GCSs, and for the
proposed existence of intergalactic globulars in clusters of galaxies, are
briefly discussed.

\end{abstract}


\section{Introduction}

As the study of galaxy formation comes more and more to require a reliable
description of star-gas interactions and feedback on relatively small
spatial scales, it is forced towards questions that overlap with
fundamental issues in star formation theory. Conversely, as evidence
accumulates that most stars today---in environments ranging from the disk of
the Milky Way to circumnuclear starbursts---are born not in isolation but in
groups, the study of star formation in general seems increasingly to demand a
focus on larger scales than might previously have been considered. Although
these two disciplines have not yet converged---the smallest scales currently
of importance for galaxy formation can still be one or two orders of magnitude
greater than the ``large'' scales in star formation---they will necessarily
continue to approach one another as further progress is made. It is therefore
important to identify and understand those areas of truly common interest, on
the largest stellar scales and the smallest galactic ones. In many ways, the
systems of old globular clusters in galaxy halos are positioned just at this
interface; they can potentially be exploited to constrain aspects of both
local and global star-formation processes, in both protogalactic and
present-day settings.

It has, indeed, long been recognized that the global properties of globular
cluster systems (GCSs) might be used to good effect as probes of galaxy
formation and evolution. (This is especially true of bright early-type
galaxies, whose GCSs are more populous and better studied than those in typical
spirals like the Milky Way.) For example, following the suggestion that large
fractions of the GCSs in ellipticals may have formed in major merger events
(e.g., \markcite{sch87}Schweizer 1987; \markcite{ash92}Ashman \& Zepf 1992;
\markcite{zep93}Zepf \& Ashman 1993), and the related discovery of young super
star clusters in systems like the Antennae galaxies (\markcite{whs95}Whitmore
\& Schweizer 1995), much recent discussion has centered on the interpretation
of the distribution of metallicities (or broad-band colors) among E-galaxy
globulars as clues to their hosts' dynamical histories (see
\markcite{gei96}Geisler, Lee, \& Kim 1996; \markcite{for97}Forbes, Brodie \&
Grillmair 1997; \markcite{cot98}C\^ot\'e, Marzke, \& West 1998;
\markcite{kis98}Kissler-Patig, Forbes, \& Minniti 1998). A somewhat older
approach has used just the total population of GCSs, as a function of parent
galaxy luminosity, Hubble type, and local environment, to make arguments
relating to issues such as biasing in a cold dark matter cosmology
(\markcite{wes93}West 1993); the merger history of ellipticals
(\markcite{har81}Harris 1981; \markcite{vdb84}van den Bergh 1984;
\markcite{ash92}Ashman \& Zepf 1992; \markcite{for97}Forbes et al.~1997);
the formation of cD galaxies and their envelopes (\markcite{mcl93}McLaughlin,
Harris, \& Hanes 1993, \markcite{mcl94}1994); star formation in galaxies at the
centers of cluster cooling flows (\markcite{har95}Harris, Pritchet, \& McClure
1995); the influence of early galactic winds on the evolution of giant
ellipticals (\markcite{har98}Harris, Harris, \& McLaughlin 1998); and even
larger-scale aspects of galaxy clusters such as populations of intergalactic
stars (e.g., \markcite{wes95}West et al.~1995; \markcite{har98}Harris et
al.~1998). Yet a third characteristic of GCSs---their radial distributions
(the projected number density of globulars around a galaxy as a function of
galactocentric radius)---has also served as the basis for discussions of
elliptical galaxy formation (e.g., \markcite{har86}Harris 1986;
\markcite{ash92}Ashman \& Zepf 1992; \markcite{kis97}Kissler-Patig 1997;
\markcite{vdb98}van den Bergh 1998).

Rather less developed, however, is a clear sense of the relevance of GCSs
to {\it local} star formation on smaller scales: What relation, if any, do the
fundamental properties of globular cluster systems bear to the way stars form
now, in the dense clumps within Galactic giant molecular clouds? Can the
study of one of these phenomena possibly shed light on the other? Larson
(\markcite{lar88}1988, \markcite{lar93}1993, \markcite{lar96}1996), has
emphasized the possible parallels between the basic pattern of local star
formation and the birth of globular clusters within large protogalactic clouds
in a scenario for the collapse of galaxies from clumpy initial conditions (as
in, e.g., the classic picture of \markcite{sea78}Searle \& Zinn 1978, and
current models of hierarchical galaxy formation). A more quantitative
discussion along these lines, including an exposition of links to several of
the gross characteristics of GCSs, is given by \markcite{har94}Harris \&
Pudritz (1994). Most recently, \markcite{mcl96}McLaughlin \& Pudritz (1996)
and \markcite{elm97}Elmegreen \& Efremov (1997) have focussed in detail on the
globular cluster luminosity function (the number of clusters per unit
magnitude, from which may be derived the mass function $d{\cal N}/dm$ for the
entire system of globulars in a given galaxy) in attempts to model this
property of GCSs using our current understanding of present-day star formation
in molecular clouds and their clumps.

The work presented in this paper further explores the connections between the
formation of stars, globular clusters, and galaxies, by concentrating on two
of the primary attributes of GCSs: their specific frequencies (or total
populations) and radial distributions. The next Section discusses the
significance of these GCS properties as they relate to general questions on the
formation of stellar clusters. In \S3 below, data are compiled from the
literature, and a geometrical deprojection algorithm applied, to construct
{\it volume} density profiles for the stars, the X-ray gas, and the globular
clusters in three galaxies with well studied GCSs: M87 (the cD at the center
of the Virgo Cluster), M49 (giant elliptical in Virgo), and NGC 1399 (cD
galaxy in the Fornax Cluster). An intercomparison of the density profiles of
the three halo components yields a point-by-point measure of the GCS
contribution to the total luminous mass in each system, as a function of {\it
three-dimensional} galactocentric radius, $r_{\rm gc}$. These GCS mass ratios,
which are constant throughout each galaxy (at large $r_{\rm gc}$) and the same
in all three, are interpreted as an indication of the formation efficiency of
globular clusters. In \S4, these results are supplemented with broader
considerations of the GCSs in other early-type galaxies and in the Milky Way
halo, and with a direct estimate for the relative formation rate of open
clusters in the Galactic disk, to argue for the existence of a {\it universal
efficiency, or probability, for the formation of a bound stellar cluster from
a dense cloud of gas}. Section 5 discusses the implications of this for a few
specific issues in galaxy and GCS formation and evolution, and \S6 concludes
with a summary.

\section{The Efficiency of Cluster Formation from Globular Cluster Systems}

\subsection{Cluster Formation Efficiencies}

The possibility that globular cluster systems can be connected not only to
galaxy formation, but to ongoing star formation as well, is suggested by the
fact that this latter process operates largely in a {\it clustered
mode}: by mass, most new stars in the Milky Way appear in groups within the
largest clumps in molecular clouds, and only rarely in true isolation. This is,
by now, an established empirical fact (e.g., \markcite{lad91}Lada et al.~1991;
\markcite{zin93}Zinnecker et al.~1993), and it is easily seen
(\markcite{pat94}Patel \& Pudritz 1994) to be a direct consequence of the
different power-law slopes in the mass function of molecular clumps ($d{\cal
N}/dm\propto m^{-1.6}$, so that the largest clumps, which weigh in at
$10^2$--$10^3\,M_\odot$, contain  most of the star-forming gas mass in any
molecular cloud) and the stellar initial mass function ($d{\cal N}/dm\propto
m^{-2.35}$, putting most of the mass in young stars into $\la1\,M_\odot$
objects). Similarly, young super star clusters, which tend to have sizes and
masses that are generally comparable to the old globulars in the Milky Way,
can account for as much as $\sim$20\% of the UV light in starburst galaxies
(\markcite{meu95}Meurer et al.~1995).

This is, however, {\it not} to say that all, or even most, stars are born
into bona fide clusters that exist as coherent dynamical units for any
length of time; on the contrary, it would appear that very few
multiple-star systems actually form as gravitationally bound clusters. (A
quantification of ``very few'' is one of the goals of this paper.) At some
point during the collapse and fragmentation of a cluster-sized cloud
of gas, the massive stars which it has formed will expel any remaining gas
by the combined action of their stellar winds, photoionization, and supernova
explosions. If the {\it cumulative} star formation efficiency of the cloud,
${\rm SFE}=M_{\rm stars}/ (M_{\rm stars}+M_{\rm gas})$, is below a critical
threshold when the gas is lost, then the blow-out carries away sufficient
energy that the stellar group remaining is unbound, and disperses into the
field. The precise value of this threshold SFE depends on details of the
dynamics and magnetic field in the gas cloud before its self-destruction, and
on the timescale over which the massive stars dispel the gas; but various
estimates place it in the range ${\rm SFE}_{\rm crit}\sim 0.2$--0.5 (e.g.,
\markcite{hil80}Hills 1980; \markcite{mat83}Mathieu 1983;
\markcite{elm85}Elmegreen \& Clemens 1985; \markcite{ver90}Verschueren 1990;
\markcite{lad84}Lada, Margulis, \& Dearborn 1984).

A complete theory of star formation must therefore be able to anticipate the
final cumulative SFE in any single piece of gas with (say) a given mass and
density, and thereby predict whether or not it will form a bound cluster. No
such theory yet exists. More within reach are considerations of a statistical
nature, in which a {\it distribution} of cumulative SFEs (and, thus, the
fraction of clouds that achieve ${\rm SFE}\ge {\rm SFE}_{\rm crit}$ before
self-destruction) is calculated for an ensemble of star-forming clouds, in
either a protogalactic or a present-day context (see, e.g.,
\markcite{elm83}Elmegreen 1983; \markcite{elm85}Elmegreen \& Clemens 1985;
\markcite{elm97}Elmegreen \& Efremov 1997). But even here, the systematic
effects of a range in the initial masses of the gaseous clouds, or of changes
in the assumed star formation law (which is probably the most outstanding
``free'' parameter here) have not been explored fully. It is therefore not
clear that the dominant factor governing the formation of bound stellar
clusters vs.~unbound associations has been identified.

It is nonetheless possible to constrain any forthcoming theories of star and
cluster formation with
empirical evaluations of the probability that a cluster-sized cloud of gas is
able to achieve a cumulative SFE of at least the critical 20\%--50\%. This
probability---which can equally well be interpreted as that fraction of an
ensemble of massive star-forming clouds which manages to produce bound stellar
systems---is referred to here as the efficiency of cluster formation. As was
indicated above, an estimate of this will be taken first from a detailed
investigation of the old globular cluster populations in three giant
elliptical galaxies; general arguments will then be used to show that the
result also applies to globular cluster formation in other galaxies, and to
the formation of disk clusters today.

A first (and very rough) assessment of the formation efficiency of
globular clusters appeals to the most basic and most easily measured observable
of a GCS: namely, its total population, ${\cal N}_{\rm tot}$, summed over all
cluster luminosities and galactocentric positions. This is usually compared to
the integrated $V$-band luminosity of the parent galaxy's halo (as a measure
of its unclustered field star population) through the so-called specific
frequency (\markcite{hvd81}Harris \& van den Bergh 1981):
\begin{equation}
S_N\equiv {\cal N}_{\rm tot}\times10^{0.4(M_V^T+15)} \propto
{\cal N}_{\rm tot}/L_{V,{\rm gal}}\ .
\label{eq:21}
\end{equation}
Even though any observational derivation of $S_N$ necessarily involves
(possibly large) extrapolations of data that directly count only the brightest
clusters in some spatially limited subset of a GCS, these estimates tend
usually to be surprisingly robust. To date, specific frequencies have
been obtained for of order 100 galaxies, some 80\% of which are early-type
systems (see the compilations of \markcite{har91}Harris 1991;
\markcite{dur96}Durrell et al.~1996; \markcite{kis97}Kissler-Patig 1997;
\markcite{bla97}Blakeslee 1997; \markcite{btm97}Blakeslee, Tonry, \& Metzger
1997; \markcite{har98}Harris et al.~1998). Given this bias in the
observational database, some of what follows will make use of results that are
really specific to elliptical galaxies; spirals are brought more explicitly
into the discussion in \S4 below.

Specific frequency is easily related to the total masses of a galaxy's GCS
and of its unclustered halo stars, if both the mean mass of the individual
globulars in the system and the mass-to-light ratio of the overall {\it
stellar} component are known. As is discussed further in \S3.1 below, the
basic distribution of globular cluster masses is quite similar from galaxy to
galaxy, so that the mean cluster mass in the Milky Way GCS---$\langle m\rangle_
{\rm cl}=2.4\times10^5\,M_\odot$---serves as a reliable estimate for other
systems as well. Meanwhile, the stellar mass-to-light ratios of large galaxies
are essentially equal to the dynamical $M/L$ in their {\it cores}, which are
baryon-dominated. As is also discussed below (\S4.1), observations show that
the core $M/L$ of hot galaxies increases with luminosity; but, even though
this turns out to be rather an important consideration in detail, it will
suffice for now to use a single value of $\Upsilon_V\equiv M/L_V=7\,M_\odot\,
L_\odot^{-1}$ as a rough characterization of large ellipticals
(\markcite{vdm91}van der Marel 1991; cf.~\markcite{fab79}Faber \& Gallagher
1979; \markcite{bin87}Binney \& Tremaine 1987). The definition of $S_N$ in
equation (\ref{eq:21}) may then be rewritten as (cf.~\markcite{zep93}Zepf \&
Ashman 1993; \markcite{har94}Harris \& Pudritz 1994; \markcite{har98}Harris et
al.~1998)
\begin{equation}
S_N\simeq2500\,\left({{\langle m\rangle_{\rm cl}}\over{2.4\times10^5\,M_\odot}}
\right)^{-1}\,\left({{\Upsilon_V}\over{7\,M_\odot\,L_\odot^{-1}}}\right)\,
{{M_{\rm gcs}}\over{M_{\rm stars}}}\ .
\label{eq:22}
\end{equation}
To be clear, it is stressed again that the mass-to-light ratio
applied here refers to measurements in the cores of ellipticals, and is not
meant to include any dark matter other than stellar remnants. In any event, an
inversion of this last expression yields
\begin{equation}
{{M_{\rm gcs}}\over{M_{\rm stars}}} = 4.0\times10^{-4}\,S_N\,
\left({{\langle m\rangle_{\rm cl}}\over{2.4\times10^5\,M_\odot}}\right)\,
\left({{\Upsilon_V}\over{7\,M_\odot\,L_\odot^{-1}}}\right)^{-1}\,
\sim\,2\times10^{-3}\,\left({{S_N}\over{5}}\right)\ ,
\label{eq:23}
\end{equation}
where $S_N\approx5$ is suggested by the studies cited above as a
representative value for early-type galaxies ranging in luminosity from
$M_V^T\simeq-15$ to $M_V^T\simeq-22$. Of course, there is some uncertainty in
this ``mean'' $S_N$, and some intrinsic scatter about it. Moreover, the data
suggest that $S_N$ varies systematically with $L_{V,{\rm gal}}$---the simple
scaling ${\cal N}_{\rm tot}\propto L_{V,{\rm gal}}$ that corresponds to a
constant specific frequency is not strictly obeyed---particularly among dwarf
ellipticals and between brightest cluster galaxies. These issues have been
discussed in detail in the literature, and they are revisited here, in \S2.2.
With this caveat in mind, however, $M_{\rm gcs}/M_{\rm stars}\sim 2\times
10^{-3}$ is useful as a crude, order-of-magnitude guide to a ``typical''
global GCS mass ratio in galaxy halos.

The advantage of working in terms of this {\it mass} ratio
is that, whatever its precise value in any one system at the
current epoch, it is expected to be a fairly well conserved quantity that has
not changed drastically over the course of GCS and galaxy evolution in a
Hubble time. {\it This is true even though the total population ${\cal N}_
{\rm tot}$ of any GCS is bound to decrease over time}, as individual globulars
are disrupted by the cumulative effects of external gravitational shocks and
internal two-body relaxation (see, e.g., \markcite{agu88}Aguilar, Hut, \&
Ostriker 1988; \markcite{ost89}Ostriker, Binney, \& Saha 1989;
\markcite{cap93}Capuzzo-Dolcetta 1993; Murali \& Weinberg
\markcite{mwa97}1997a, \markcite{mwb97}1997b, \markcite{mwc97}1997c;
\markcite{gne97}Gnedin \& Ostriker 1997; \markcite{veh97}Vesperini \& Heggie
1997; but cf.~\markcite{por98}Portegies Zwart et al.~1998). These
dynamical processes discriminate preferentially against low-mass and
low-density globular clusters, and it turns out that even if large {\it
numbers} of such objects are lost, no great change is effected in
the total {\it mass} of the remaining GCS. This conclusion rests on the fact
that the mass functions of GCSs---particularly those in early-type
galaxies---are shallow enough that most of the total $M_{\rm gcs}$ is
contained in the most massive members of the system; but these are the least
susceptible to dynamical destruction. The implication is that the total mass
of a GCS may not change much over its lifetime; thus, {\it if} the current
stellar mass of a galaxy were a reliable indication of the total amount of gas
that was initially available to form stars, then the presently observed ratio
$M_{\rm gcs}/M_{\rm stars}\sim2\times10^{-3}$ would serve as a rough, {\it
global} estimate of the efficiency of globular cluster formation. Note that
this reasoning also shows that galactic halos cannot be built up from
disrupted globulars; see, e.g., \markcite{ash92}Ashman \& Zepf (1992) and
\markcite{har94}Harris \& Pudritz (1994). (One of the main conclusions of this
paper will be that, in fact, $M_{\rm stars}$ is {\it not} always an adequate
measure of an initial gas supply, and that $M_{\rm gcs}/M_{\rm stars}$
therefore cannot always be equated directly to a cluster formation efficiency.
However, since most previous discussions of GCS specific frequencies do draw
on this simple assumption at some level, it is of interest to develop some of
its implications.)

A simple numerical example should emphasize the robustness of $M_{\rm gcs}$.
Thus, consider a newly formed system of globulars whose masses are distributed
according to a power-law mass function between some lower and upper limits:
\begin{equation}
\left({{d{\cal N}}\over{dm}}\right)_{\rm init}\,=\,K\,m^{-\gamma_2}\,,
\ \ \ \ \ \ \ \ m_\ell \leq m\leq m_u\ ,
\label{eq:24}
\end{equation}
where $\gamma_2>0$ and reasonable choices for $m_\ell$ and $m_u$ may be
$500\,M_\odot$ and $5\times10^6\,M_\odot$. (The precise values are not
critical here.) Now let this GCS be eroded over $10^{10}$ yr, keeping in mind
that each cluster in the system has a definite, mass-dependent lifetime
against the dynamical mechanisms mentioned above. In a rough approximation to
a combination of complicated processes, there is then a mass scale (say $m_*$)
above which few or no globulars will have disappeared by the current epoch.
Thus, the GCS mass function at $m>m_*$ is effectively unchanged, even after a
Hubble time, from the initial distribution. However, clusters with initial
masses $m<m_*$ {\it can} have been removed from the system, and the GCS mass
function below $m_*$ may differ significantly from its initial form. If this
low-mass end of the evolved $d{\cal N}/dm$ can also be described by a
power-law, then the full, present-day GCS mass function is just
\begin{equation}
\left({{d{\cal N}}\over{dm}}\right)_{\rm obs}\,=\,\left\{
\begin{array}{cl}
K\,m_*^{\gamma_1-\gamma_2}\,m^{-\gamma_1}\,, & \ \ \ \ m_\ell\leq m\leq m_* \\
K\,m^{-\gamma_2}\,, & \ \ \ \ m_*\leq m\leq m_u\ ,
\end{array}
\right.
\label{eq:25}
\end{equation}
where $\gamma_1<\gamma_2$ for consistency.

Observed GCSs do, in fact, have mass functions that conform to roughly this
sort of double power law (\markcite{har94}Harris \& Pudritz 1994;
\markcite{dem94}McLaughlin 1994; \markcite{mcl96}McLaughlin \& Pudritz
1996)---although there is somewhat more structure than this at the highest
cluster masses in $d{\cal N}/dm$, it can be safely ignored for the purposes of
the present argument. Empirically, the mass scale $m_*$ is {\it always} about
$1.6\times10^5\,M_\odot$; at this mass, the logarithmic slope of $d{\cal N}/
dm$ changes rather abruptly from $\gamma_2>1$ to $\gamma_1<1$, with precise
values that can vary slightly from galaxy to galaxy.\footnotemark
\footnotetext{It is this sharp change from an exponent of $>1$ to one of $<1$
at a well defined and nearly universal cluster mass that results in the
characteristic peak of the globular cluster {\it luminosity} function in
galaxies (the GCLF: the number of clusters per unit absolute magnitude, which
is proportional to $d{\cal N}/d\,\ln\,m$; see
\markcite{dem94}McLaughlin 1994). This peak occurs at the luminosity
corresponding to the mass $m_*$, i.e., at $M_V^0\simeq-7.4$, for the
$\Upsilon_V\simeq2\,M_\odot\,L_\odot^{-1}$ typical of globulars; its
universality is what makes the GCLF attractive as a distance indicator (e.g.,
\markcite{jac92}Jacoby et al.~1992; \markcite{whi97}Whitmore 1997).}
There are strong arguments to support the notion that observed GCS mass
functions at $m>m_*$ are indeed accurate reflections of the formation
distributions (namely, the high-mass sides of $d {\cal N}/dm$ in observed GCSs
show no significant variation with galactocentric radius inside a single
system, and are quite similar---though not identical---from galaxy to galaxy;
see \markcite{har94}Harris \& Pudritz 1994; \markcite{mcl96}McLaughlin \&
Pudritz 1996; \markcite{elm97}Elmegreen \& Efremov 1997). In addition, some
theoretical studies (e.g., \markcite{sur79}Surdin 1979; \markcite{oka95}Okazaki
\& Tosa 1995; \markcite{ves97}Vesperini 1997; \markcite{elm97}Elmegreen \&
Efremov 1997) have suggested that the shallower slopes at lower masses might
be caused entirely by dynamical evolution from an initial $d{\cal N}/dm\propto
m^{-\gamma_2}$, roughly as described above. In reality, this view could very
well be too extreme: empirically, the system of young super star clusters in
the Antennae, which is not likely to have been severely affected by dynamical
destruction, may have a mass function of the form (\ref{eq:25}) rather than
(\ref{eq:24}) (\markcite{fva99}Fritze-von Alvensleben 1999); and theoretically,
such an initial shape to $d {\cal N}/dm$ may in fact be {\it preserved} during
a Hubble time of dynamical evolution (\markcite{ves97}Vesperini 1997). Thus,
the intent here is not necessarily to argue for an explanation of
present-day GCS mass spectra in purely evolutionary terms, but to show that
even if such is adopted as an extreme possibility, the implied time dependence
of total GCS masses is quite weak.

Given equations (\ref{eq:24}) and (\ref{eq:25}), then, straightforward
integrations allow for comparisons of the present GCS population to the
initial one:
\begin{equation}
{{{\cal N}_{\rm tot}^{\rm obs}}\over{{\cal N}_{\rm tot}^{\rm init}}} =
\left({{m_u}\over{m_*}}\right)^{\gamma_2-1}\,\left\{
{{1-(m_*/m_u)^{\gamma_2-1}+[(\gamma_2-1)/(1-\gamma_1)][1-(m_\ell/m_*)^
{1-\gamma_1}]}\over{(m_u/m_\ell)^{\gamma_2-1}-1}}\right\}\ ,
\label{eq:26}
\end{equation}
and of the current total mass to its initial value:
\begin{equation}
{{M_{\rm gcs}^{\rm obs}}\over{M_{\rm gcs}^{\rm init}}} =
\left({{m_*}\over{m_u}}\right)^{2-\gamma_2}\,\left\{
{{(m_u/m_*)^{2-\gamma_2}-1+[(2-\gamma_2)/(2-\gamma_1)][1-(m_\ell/m_*)^
{2-\gamma_1}]}\over{1-(m_\ell/m_u)^{2-\gamma_2}}}\right\}\ .
\label{eq:27}
\end{equation}
Values of $\gamma_1=0.4$ and $\gamma_2=1.7$ may be taken as representative of
the typical situation in elliptical galaxies (\markcite{har94}Harris \&
Pudritz 1994; McLaughlin \markcite{dem94}1994, \markcite{mcl95}1995;
\markcite{mcl96}McLaughlin \& Pudritz 1996). With $m_\ell=500\,M_\odot$,
$m_*=1.6\times10^5\,M_\odot$, and $m_u=5\times10^6\,M_\odot$, it is then easily
seen that
\begin{equation}
{{{\cal N}_{\rm tot}^{\rm obs}}\over{{\cal N}_{\rm tot}^{\rm init}}}=0.036
\ \ \ \ \ \ \ \ {\rm and}\ \ \ \ \ \ \ \ \ 
{{M_{\rm gcs}^{\rm obs}}\over{M_{\rm gcs}^{\rm init}}}=0.76
\label{eq:28}
\end{equation}
in this simplistic scenario. Thus, {\it even though dynamical evolution
reduces the total population of this hypothetical GCS by more than an order of
magnitude, it diminishes the total mass by only} $\sim$25\%. (The same line of
argument also shows that the often appreciable uncertainties in the exact form
of $d{\cal N}/dm$ at cluster masses $m<m_*$---or even blind extrapolations from
observations of only brighter clusters---do not seriously bias observational
estimates of $M_{\rm gcs}$ in any galaxy.)

This result is best interpreted as an estimate for the possible depletion of a
GCS {\it globally} (i.e., spatially averaged over the whole of a parent
galaxy), because the values of $\gamma_1$, $\gamma_2$, and $m_*$ used are
taken from observations of $d{\cal N}/dm$ in projection, often over fairly
large areas on the sky. Thus, the normalization $K$ in equations (\ref{eq:24})
and (\ref{eq:25}) can be taken to include an implicit integration over
galactocentric radius, and the exponents $\gamma$ and mass scale $m_*$ to be
appropriate averages. The importance of these considerations is not actually
clear, since radial dependences are not expected theoretically in current
models for the initial GCS mass function (\markcite{mcl96}McLaughlin \&
Pudritz 1996; \markcite{elm97}Elmegreen \& Efremov 1997) and observational
evidence for such trends at the current epoch is slight to non-existent (e.g.,
\markcite{mcl94}McLaughlin et al.~1994 and \markcite{har98}Harris et al.~1998;
though see also \markcite{kav97}Kavelaars \& Hanes 1997 and
\markcite{mwb97}Murali \& Weinberg 1997b). Nevertheless, it should be noted
that dynamical friction, gravitational shocks, and even evaporation couple to
galactic tidal fields in such a way that if a GCS's present-day mass function
{\it were} sculpted to any great extent by these processes, then the {\it
local} ratios ${\cal N}_{\rm tot}^{\rm obs}/{\cal N}_{\rm tot}^{\rm init}$ and
$M_{\rm gcs}^{\rm obs}/M_{\rm gcs}^{\rm init}$ would be expected to increase
outwards from the center of the system, from values less than equation
(\ref{eq:28}) at small galactocentric radii to near unity at large distances
(e.g., beyond a stellar effective radius; see especially
\markcite{mwb97}Murali \& Weinberg 1997b).

Returning now to the main issue, the global efficiency of globular cluster
formation in a galaxy may be defined as
\begin{equation}
\epsilon_{\rm cl}\,\equiv\,
M_{\rm gcs}^{\rm init}/M_{\rm gas}^{\rm init}\ ,
\label{eq:29}
\end{equation}
where $M_{\rm gas}^{\rm init}$ refers to the total gas supply that was
available to the protogalaxy, whether in a rapid monolithic collapse or, more
likely, in a slower assembly (at high redshift) of many distinct, subgalactic
clouds. From the preceding discussion, it is clear that if a newly formed GCS
were able to evolve ``passively''---that is, if its mass were potentially
vulnerable only to reductions caused by forces internal to an isolated parent
galaxy---then a present-day observation of $M_{\rm gcs}$ would suffice, to at
least a $\sim$25\% level of accuracy globally, as a measure of $M_{\rm gcs}^
{\rm init}$. Passive evolution also implies mass conservation for the galaxy
as a whole, so that a reliable value for $M_{\rm gas}^{\rm init}$ could be
obtained, at any epoch, from an inventory of the total mass in gas plus
unclustered stars and stellar remnants. Thus, a good {\it observational
estimate} of the global cluster formation efficiency in this simple situation
is provided by
\begin{equation}
\wepsilon_{\rm cl}\,\equiv\,{{M_{\rm gcs}}\over{M_{\rm gas}+M_{\rm stars}}}\ ,
\label{eq:210}
\end{equation}
which is expected to be nearly a time-independent quantity.\footnotemark
\footnotetext{To be fully consistent, of course, the observational estimate
of cluster formation efficiency should read $\wepsilon_{\rm cl}=M_{\rm gcs}/
(M_{\rm gas}+M_{\rm stars}+M_{\rm gcs})$; but, as \S\S3 and 4 below will show,
$M_{\rm gcs}$ turns out {\it always} to be so small, relative to ($M_{\rm gas}
+ M_{\rm stars}$), that its contribution to the denominator of $\wepsilon_{\rm
cl}$ is utterly negligible.}

Realistically, of course, galaxies are not such pristine entities; interactions
between systems may effect significant changes in both $M_{\rm gcs}$ and
$(M_{\rm gas}+M_{\rm stars})$ over time, making it difficult to evaluate (or
even clearly define) initial GCS and gas masses. For example, the formation
history of an elliptical may include major mergers with gas-rich spiral
systems, and it has been proposed that these might involve the formation of
new globular clusters and substantial increases in $M_{\rm gcs}$ (e.g.,
\markcite{ash92}Ashman \& Zepf 1992; \markcite{kbf93}Kumai, Basu, \& Fujimoto
1993a). Alternatively, additional globulars might be captured by a large
galaxy that accretes small, gas-poor systems (e.g., \markcite{cot98}C\^ot\'e
et al.~1998) or accumulates tidal debris from other members in a cluster of
galaxies (see \markcite{muz87}Muzzio 1987). However, processes such as these
also increase the total masses of gas and/or stars in the final system.
In {\it large} galaxies (i.e., in those that do not suffer significant gas
loss from strong galactic winds; cf.~\S4.2), any changes in the ratio $M_{\rm
gcs}/(M_{\rm gas}+M_{\rm stars})$ are bound to be smaller than changes in
either of these quantities individually. The presently observed $\wepsilon_
{\rm cl}$ in any large galaxy with a complex dynamical history is then a
mass-weighted average of the true $\epsilon_{\rm cl}$ in some number of
discrete star-formation events and individual accreted systems; and in the
limit that the underlying efficiency of cluster formation is independent of
environment and epoch (as the results of \S\S3 and 4 will suggest is nearly
correct), it is still true that $\wepsilon_{\rm cl}\simeq\epsilon_{\rm
cl}$---just as in the case of passive evolution, and subject only to the small
error associated there with the effects of GCS erosion by cluster evaporation
and tidal disruption.

In gas-poor elliptical galaxies and the halos of spirals, equation
(\ref{eq:210}) reduces to $\wepsilon_{\rm cl}\simeq M_{\rm gcs}/M_{\rm stars}$,
so that---as was mentioned above---the cluster formation efficiency
can be obtained directly from obervations of the global GCS specific frequency
(cf.~eq.~[\ref{eq:23}]). More generally, however, ellipticals can contain
large amounts of hot, X-ray emitting gas (e.g., \markcite{for85}Forman, Jones,
\& Tucker 1985), and when it happens that $M_{\rm gas}$ is non-negligible
relative to $M_{\rm stars}$, measurements of $S_N$ are only low-order
approximations to $\wepsilon_{\rm cl}$ (which is itself only an estimate of
the real efficiency $\epsilon_{\rm cl}$). Keeping this in mind, the observed
systematics of both global {\it and local} GCS specific frequencies provide a
useful starting point for an examination of the extent to which cluster
formation efficiencies might vary from galaxy to galaxy, or from place to
place within a single system.

\subsection{The Specific Frequency ``Problems''}

As was intimated in the discussion around equation (\ref{eq:23}), not
all galaxies have the same globular cluster specific frequency. There are
several aspects to this: (1) Spiral galaxies tend to have lower $S_N$---at
least when their GCS populations are normalized to their {\it total} (bulge
plus disk) light---than do average ellipticals (\markcite{har91}Harris 1991).
(2) Although a constant $S_N\approx5$ is a fair {\it approximation} for many
normal giant ellipticals, fits to the data suggest that, in fact, $S_N\propto
L_{V,{\rm gal}}^{0.2-0.3}$ (\markcite{kis97}Kissler-Patig 1997;
cf.~\markcite{san93}Santiago \& Djorgovski 1993). At the extremes of the
early-type galaxy sequence, moreover, (3) the $S_N$ values for centrally
dominant or first-ranked members of galaxy clusters show a much steeper
dependence on galaxy luminosity---$S_N\propto L_{V,{\rm gal}}^{0.8}$ or so is
implied by the correlations of \markcite{bla97}Blakeslee (1997) and
\markcite{har98}Harris et al.~(1998)---and (4) specific frequencies actually
appear to increase towards {\it lower} luminosities in dwarf ellipticals:
$S_N\propto L_{V,{\rm gal}}^{-0.4}$ (\markcite{mil98}Miller et al.~1998; see
also \markcite{dur96}Durrell et al.~1996). Thus, some brightest cluster
galaxies, and some of the faintest dwarfs, have global specific
frequencies that exceed the ``average'' by factors of 3 or 4. All in all,
observed $S_N$ values span a range from $\sim$1 to $\sim$20.\footnotemark
\footnotetext{It has also been claimed (\markcite{har91}Harris 1991;
\markcite{wes93}West 1993; \markcite{khf93}Kumai, Hashi, \& Fujimoto 1993b)
that early-type galaxies in sparse groups have systematically lower $S_N$ than
those in high-density cluster environments. However, this correlation is
uncertain, and apparently rather weak; if it does prove to be important,
it will most likely be as a higher-order correction to the more basic trends
discussed here.}

As will be shown in \S4 below, the first two of these points can be understood
fairly easily. Here, it suffices to note that (1) when only the {\it halo}
stars and globular clusters in the Milky Way are considered, and allowances
made for the lower Pop.~II mass-to-light ratio in our Galaxy vs.~large
ellipticals, $\wepsilon_{\rm cl}$ here is no different from that in ``normal''
early-type systems. It is reasonable to expect that this result---which is
contrary to the claim made by \markcite{zep93}Zepf \& Ashman (1993)---should
hold in other spirals as well. Also, (2) the observed dependence of $S_N$ on
$L_{V,{\rm gal}}$ among non-brightest cluster ellipticals is fully consistent
with that expected from equation (\ref{eq:22}) for gas-poor systems with a
uniform efficiency of cluster formation ($\wepsilon_{\rm cl}\simeq M_{\rm gcs}/
M_{\rm stars}$) and a mass-to-light ratio ($\Upsilon_V$) that scales with
luminosity in the way observed for galaxies in the fundamental plane. However,
if taken at face value, items (3) and (4) just above appear to imply
that GCS formation efficiencies can vary by as much as an order of magnitude
from galaxy to galaxy, in a non-monotonic fashion and for reasons that have
never been understood. These issues are not so readily dismissed. The high
$S_N$ ``phenomenon'' in many first-ranked cluster ellipticals, especially, has
been discussed at length in the literature (e.g., \markcite{mcl94}McLaughlin
et al.~1994; \markcite{har95}Harris et al.~1995; \markcite{wes95}West et
al.~1995; \markcite{for97}Forbes et al.~1997; \markcite{bla97}Blakeslee 1997;
\markcite{btm97}Blakeslee et al.~1997; \markcite{har98}Harris et al.~1998), and
is referred to here as the {\it first specific frequency problem}. Equally
non-trivial, though only more recently recognized, is the {\it inverse}
variation of $S_N$ with $L_{V,{\rm gal}}$ among dwarf ellipticals.

Local specific frequencies, which are obtained for a single GCS by comparing
its projected radial distribution to its galaxy's surface brightness profile
(normalized as in eq.~[\ref{eq:21}]), are also of interest here; they
are related in the obvious way to the efficiency of cluster formation as a
function of (projected) galactocentric radius, $R_{\rm gc}$. Of particular
importance is the fact that $S_N$ is seen to be an {\it increasing} function
of $R_{\rm gc}$ in many galaxies. This is equivalent to the well known fact
that GCSs are often less centrally concentrated than the stellar halos of their
parent galaxies. This statement has two contexts, which must be separated.
First, it seems that the GCS of most any galaxy---spiral or elliptical,
small or large---has an apparent core radius that is significantly larger
than that of the unclustered field stars (e.g., \markcite{djo94}Djorgovski
\& Meylan 1994; \markcite{lau86}Lauer \& Kormendy 1986;
\markcite{mcl95}McLaughlin 1995; \markcite{for96}Forbes et al.~1996;
\markcite{min96}Minniti, Meylan, \& Kissler-Patig 1996;
\markcite{cot98}C\^ot\'e et al.~1998); that is, the {\it central regions} of
GCSs show projected radial distributions with slopes that are significantly
shallower than those of the light profiles on kpc scales, so that the density
ratio of the two halo components is an increasing function of radius. Second,
the GCSs of some large galaxies are spatially more extended than the stellar
halos on {\it larger scales}, $R_{\rm gc}\sim10$--100 kpc
(\markcite{har79}Harris \& Racine 1979; \markcite{har86}Harris 1986).
However, this is a less general result than that relating to the large core
radii of globular cluster systems. It is true of many dominant cluster
galaxies like M87 in Virgo (see the discussion in \markcite{msh95}McLaughlin
et al.~1995), and also of the M31 halo (\markcite{pri94}Pritchet \& van den
Bergh 1994); but many elliptical systems (e.g., \markcite{dur96}Durrell et
al.~1996; \markcite{kpa97}Kissler-Patig et al.~1997), and the Milky Way itself
(\markcite{har76}Harris 1976; \markcite{zin85}Zinn 1985), show GCS and
halo-star distributions that are in quite close agreement (and, thus, a
constant local $S_N$) outside of any core regions. The physical factors that
govern whether the GCS density falls off at the same rate as or more slowly
than the unclustered halo density in a given system have not yet been
identified. (No case has ever been found in which a GCS radial distribution is
{\it steeper} than a galaxy's light profile.)

The first of these points is not easily related to the radial dependence of
cluster formation efficiencies because, as was suggested in \S2.1, the
dynamical mechanisms that act to destroy globulars work most efficiently at
small $R_{\rm gc}$, where the stellar densities are highest. Although the
cores of GCSs could quite plausibly have always been somewhat larger than
those of halos or bulges generally (e.g., \markcite{har94}Harris \& Pudritz
1994; \markcite{cot98}C\^ot\'e et al.~1998), it seems inevitable that any
initial discrepancies must have been augmented by an inside-out erosion of the
cluster system from tidal shocking, dynamical friction, and evaporation
(\markcite{ost89}Ostriker et al.~1989; \markcite{cap93}Capuzzo-Dolcetta 1993;
\markcite{mwb97}Murali \& Weinberg 1997b). Detailed modelling is therefore
required to disentangle the effects of formation and evolutionary processes in
this instance. At galactocentric radii greater than a few kpc, however, the
timescales on which the evolutionary mechanisms operate can exceed a Hubble
time, and the studies cited here show that dynamics will not significantly
affect GCS radial distributions beyond about one effective radius in the
stellar halo (see also \markcite{agu88}Aguilar et al.~1988). Thus, to the
extent that the assumption $\wepsilon_{\rm cl}\simeq M_{\rm gcs}/M_{\rm stars}$
is valid, the distension of GCSs relative to stellar halos and the concomitant
systematic increase in local $S_N$ in the outer parts of {\it some} galaxies
would seem to say that bound clusters were sometimes (but not always, and for
unknown reasons) {\it more} likely to form at larger $R_{\rm gc}$, in gas that
was presumably at lower ambient densities and pressures. This situation will
be referred to as the {\it second specific frequency problem}.

It is clearly of interest, from the points of view of both galaxy and star
formation, to verify whether these global and local $S_N$ variations really do
reflect differences in $\epsilon_{\rm cl}$---and, if so, to understand their
origins. Accordingly, much attention has been paid to the first $S_N$ problem 
in particular, usually in efforts to empirically correlate the $S_N$
measurements for brightest cluster galaxies (BCGs) with global properties of
the galaxies themselves or of the clusters they inhabit. One such correlation
does appear to be of fundamental importance, as will be discussed momentarily,
but it suggests that $S_N$ varies in the face of {\it constant} $\epsilon_{\rm
cl}$. Indeed, no proposed solution of the first specific frequency problem
has identified a physical mechanism that could possibly lead to the
superefficient production of bound star clusters in some galaxies
(see the references given above). In fact, one suggestion---that high global
specific frequencies (meaning $S_N\ga5$) in centrally located BCGs result from
the acquisition of globulars and stars which are tidally stripped from the
outer, high-$S_N$ regions of other galaxies in a cluster---actually appears
to rely on the existence of the {\it second} specific frequency problem (see
\markcite{for97}Forbes et al.~1997; \markcite{cot98}C\^ot\'e et al.~1998); but
this is itself essentially unexplained.\footnotemark
\footnotetext{It should be noted that the second $S_N$ problem, and local
specific frequencies generally, are not typically discussed in terms of cluster
formation efficiencies, even though it is well understood that global $S_N$
are of interest in this regard. The reason for this may be that mismatches
between GCS and stellar radial distributions in E galaxies are usually
attributed either to the globulars having formed ``slightly before'' the bulk
of the unclustered halo stars (in a dissipational collapse scenario;
\markcite{har86}Harris 1986), or to the expectation that a cluster system will
be ``puffed up'' relative to the final light profile of an elliptical that
forms in one or more major mergers of gas-rich systems with pre-existing
GCSs (\markcite{ash92}Ashman \& Zepf 1992; \markcite{zep93}Zepf \& Ashman
1993). In either case, no particular radial variation in $\epsilon_{\rm cl}$
would necessarily be required. However, the first of these arguments has not
been sufficiently well developed to account quantitatively for the observed
magnitude of the second $S_N$ problem; and the second is actually inconsistent
with the observation that systems with higher global $S_N$ have spatially more
extended GCSs (see the discussion of \markcite{for97}Forbes et al.~1997). Nor
is it apparent, from either these suggestions, why there are also many
galaxies whose GCSs and stellar halos have the {\it same} density profiles.}
Nor, to date, has any explanation (or prediction) been given for the high $S_N$
that have now been observed in many faint dwarf galaxies.

The view is taken here that {\it the second specific frequency problem is
inseparable from the first}, in the specific sense that both relate to the
possible influence of environment on the probability of bound cluster
formation. Any real insight into either one of these issues must therefore
include a better understanding of the other. This view is in keeping with
recent evidence (\markcite{for97}Forbes et al.~1997;
\markcite{kis97}Kissler-Patig 1997; \markcite{vdb98}van den Bergh 1998) for at
least a rough correlation between global $S_N$ values and GCS density-profile
slopes in elliptical galaxies: those GCSs with the highest specific
frequencies tend also to have the shallowest radial distributions. Such a
correspondence hints strongly at (although it does not directly show) an
intimate connection between an increasing local $S_N$ and a high global $S_N$.

A significant development in this field has come with the recent recognition
of a correlation between the global specific frequencies of BCGs and the X-ray
luminosity of the hot gas in their parent clusters (\markcite{wes95}West et
al.~1995; \markcite{bla97}Blakeslee 1997; \markcite{btm97}Blakeslee et
al.~1997). A relatively large and homogeneous sample of BCGs shows $S_N$ values
that increase from about 4, in clusters with X-ray luminosities (on 500-kpc
scales) $L_X\sim10^{42}$ erg s$^{-1}$ at 0.5--4.5 keV, to $S_N\simeq12$ for
$L_X\sim10^{45}$ erg s$^{-1}$. Three different interpretations have already
been attached to this result:

\markcite{wes95}West et al.~(1995) propose that it points to the existence of
intergalactic globulars which collect naturally at the centers of galaxy
clusters and are superimposed in projection on any galaxy there. This would
certainly boost the apparent total population ${\cal N}_{\rm tot}$ of a
central BCG, so that globulars need not have formed in situ with abnormally
high specific frequency; but they would still have had to do so elsewhere in
the cluster (see also \markcite{cot98}C\^ot\'e et al.~1998;
\markcite{har98}Harris et al.~1998).

In a different approach, \markcite{bla97}Blakeslee (1997) and
\markcite{btm97}Blakeslee et al.~(1997) consider $L_X$ to be mainly a
reflection of the temperature of intracluster gas, and thus a measure of a
cluster's mass and the depth of its potential. They then claim that the
observed $S_N$--$L_X$ correlation is consistent with the globulars in BCGs
having formed in numbers directly proportional to the {\it total} mass,
including dark matter, of their {\it entire} galaxy clusters
(see \markcite{kav98}Kavelaars 1998 for a similar suggestion on the scale of
individual galaxies). In this scenario, cluster formation would obey a
universal efficiency of sorts---although one which is quite different from
that discussed here---and the first $S_N$ problem would result from some
process that caused the BCGs in more massive clusters to be systematically
underluminous (i.e., underefficient in the production of unclustered field
stars). Here, however, the role that the dark matter content of a galaxy
cluster might play in the formation of globular clusters is not clear.
Moreover, any rule such as $M_{\rm gcs}^{\rm init}/(M_{\rm gas}^{\rm init}+
M_{\rm dark})=constant$ clearly can apply over only a limited range of
conditions.

Thus, \markcite{har98}Harris et al.~(1998) instead pursue the hypothesis that
GCSs formed with a constant $\epsilon_{\rm cl}$ defined essentially as in
equation (\ref{eq:29}) above. This is adopted as a working {\it assumption}
by \markcite{har98}Harris et al., who construct plausibility
arguments for a universal value of $\epsilon_{\rm cl}\sim10^{-3}$ (as could be
expected) and suggest that the higher global $S_N$ in more X-ray luminous
BCGs might have resulted from stronger (partial) galactic winds which heated 
larger fractions of their initial gas supplies to keV temperatures after an
early bout of star and globular cluster formation. Such hot gas would be
largely removed from further star formation---but still confined to the
vicinity of these central cluster galaxies---and the final BCGs would be
optically underluminous as a result.

It is this basic idea that is closest to the discussion in this paper (although
many aspects of the quantitative analysis here differ significantly
from those in \markcite{har98}Harris et al.~1998). That is, the
fundamental point of the global $S_N$--$L_X$ correlation for BCGs is the
associated implication that the first specific frequency problem may stem, not
from intrinsic variations in the global $\epsilon_{\rm cl}$ of these systems,
but from the fact that they can contain large amounts of gas. High gas mass
ratios invalidate the assumption that $\wepsilon_{\rm cl}\simeq M_{\rm gcs}/
M_{\rm stars}$, and therefore cause $S_N$ to overestimate the true globular
cluster formation efficiency. If it is correct to infer from this that the
``excess'' globulars in systems with high global $S_N$ are associated with hot
gas, then the fact that the distribution of gas is generally more extended
than that of the stars in ellipticals (e.g., \markcite{tri86}Trinchieri,
Fabbiano, \& Canizares 1986; \markcite{fab89}Fabbiano 1989) suggests a possible
explanation for the second specific frequency problem as well as the first. 
These two issues would then be closely related---possibly even just different
aspects of a single phenomenon---as was suggested above; and both might be
settled simply by measuring the full estimator $\wepsilon_{\rm cl}=M_{\rm gcs}
/(M_{\rm gas}+M_{\rm stars})$ both locally and globally in individual systems.

In the next Section, this universal-$\epsilon_{\rm cl}$ hypothesis is {\it
tested}, for the first time, by evaluating the observable
$\wepsilon_{\rm cl}$ as a function of galactocentric radius in each of the
galaxies M87, M49, and NGC 1399. It turns out that the inclusion of the
X-ray gas in M87 is indeed key to the alleviation of both the first and the
second $S_N$ problems there: once this is done, $\wepsilon_{\rm cl}$
is constant locally, and essentially the same globally, in all three of these
systems. The discussion is expanded to include other giant ellipticals and
BCGs, as well as dwarf ellipticals, in \S4. It is shown directly that all
available GCS data are consistent with a single efficiency of cluster
formation. Notably, the increase of $S_N$ towards low $L_{V,{\rm gal}}$ in the
dwarfs (\markcite{mil98}Miller et al.~1998; \markcite{dur96}Durrell et
al.~1996) can be explained in terms of this universal $\epsilon_{\rm cl}$ once
allowances are made for the effects of the violent gas outflows (a more extreme
version of the feedback envisioned by \markcite{har98}Harris et al.~1998 for
BCGs) that were driven by early bursts of star formation in these low-mass
systems.

\section{Case Studies of Three Galaxies}

M87, M49, and NGC 1399 are ideally suited to a closer examination of these
issues, for three reasons: (1) They are relatively nearby, making it possible
to observe internal trends within the systems in the first place. M87 and M49
both are assumed here to lie at a distance of $D=15$ Mpc, appropriate for the
core of the Virgo Cluster (\markcite{pie94}Pierce et al.~1994;
\markcite{fre94}Freedman et al.~1994). A relative distance modulus of $\Delta
(m-M)[{\rm Fornax}-{\rm Virgo}]=0.2$ (e.g., \markcite{mcm93}McMillan,
Ciardullo, \& Jacoby 1993; \markcite{ton97}Tonry et al.~1997) then puts NGC
1399 at $D\simeq16.5$ Mpc. (2) Each is a giant elliptical (M87 and NGC 1399 are
actually type cD), and is therefore accompanied by a large GCS: ${\cal N}_{\rm
tot}\simeq13\,500$ for M87; $6\,900$ for M49; and $4\,700$ in NGC 1399 (e.g.,
\markcite{har91}Harris 1991; see also \S5 below). And, most importantly, (3)
even this small sample of three galaxies suffices to bring out all aspects of
the two specific frequency problems summarized in \S2.2: M87, as is well
known, suffers from both; M49 does not show the first, but has long been
thought to exhibit the second; and NGC 1399 has been claimed as an example of
the first problem, but not the second.

Because of their proximity, these systems have been well studied in all
respects of importance here. $B$-band surface photometry extends in M87 to a
projected galactocentric radius of about $22\arcmin\simeq95$ kpc (de
Vaucouleurs \& Nieto \markcite{dvn78}1978, \markcite{dvn79}1979); in M49, to a
similar distance (\markcite{kin78}King 1978; \markcite{cao94}Caon, Capaccioli,
\& D'Onofrio 1994); and in NGC 1399, to $R_{\rm gc}\simeq14\farcm5=70$ kpc
(\markcite{kil88}Killeen \& Bicknell 1988; \markcite{cao94}Caon et al.~1994).
These measurements are easily converted to projected stellar mass density
profiles $\Sigma_{\rm stars}(R_{\rm gc})$, as mass-to-light ratios have been
obtained from a spectroscopic analysis applied uniformly to all three systems
(\markcite{vdm91}van der Marel 1991). X-ray observations, which have been made
with several different satellites, yield gas densities over still larger areas
around each galaxy. Here, use will be made of the {\it ROSAT} data for M87
(taken from \markcite{nul95}Nulsen \& B\"ohringer 1995); of {\it ROSAT} data
for M49 (\markcite{irw96}Irwin \& Sarazin 1996); and of {\it Einstein}
observations in NGC 1399 (\markcite{kil88}Killeen \& Bicknell 1988). These
studies all give the {\it volume} electron density $n_e(r_{\rm gc})$, as a
function of {\it three-dimensional} galactocentric radius, derived either from
fits of a parametric function to the projected X-ray brightness profile or
from a direct deprojection of the same. Finally, integrated optical magnitudes
have been measured for globular clusters in these galaxies at radii $R_{\rm
gc}\sim1-100$ kpc---generally, over most of the areas covered by the surface
photometry of unresolved halo stars. These are used to construct projected GCS
number densities, $N_{\rm cl}(R_{\rm gc})$, which are trivially converted to
mass profiles, $\Sigma_{\rm cl}$, once a mean globular cluster mass is
specified. For M87, relevant investigations are those of
 \markcite{hsm76}Harris \& Smith (1976), \markcite{har86}Harris (1986),
\markcite{gri86}Grillmair, Pritchet, \& van den Bergh (1986),
\markcite{lau86}Lauer \& Kormendy (1986), \markcite{coh88}Cohen (1988),
McLaughlin et al.~(\markcite{mcl93}1993, \markcite{mcl94}1994),
\markcite{mcl95}McLaughlin (1995), \markcite{whi95}Whitmore et al.~(1995), and
\markcite{har98}Harris et al.~(1998). Data on M49 can be found in
\markcite{har78}Harris \& Petrie (1978), \markcite{hvd81}Harris \& van den
Bergh (1981), \markcite{har86}Harris (1986), \markcite{hap91}Harris et
al.~(1991), and \markcite{lee98}Lee, Kim, \& Geisler (1998). The GCS of NGC
1399 has been observed by \markcite{han86}Hanes \& Harris (1986),
\markcite{bri91}Bridges, Hanes, \& Harris (1991), \markcite{wag91}Wagner et
al.~(1991), and \markcite{kpa97}Kissler-Patig et al.~(1997).

While surface brightness profiles and X-ray gas densities often can be obtained
for most of an entire galaxy with a single set of observations---as is the
case in the studies cited above---this is generally {\it not} so for GCS
radial distributions. Older, photographic surveys, which did have large
($R\sim100$ kpc) fields of view, were hampered by rather bright limiting
magnitudes that became systematically more so towards smaller galactocentric
radii. Newer, CCD photometry achieves deeper limiting magnitudes, and digital
data reduction techniques can quantify and correct for any spatial variations;
but until very recently, CCD fields of view were quite small, and best suited
to analyses of the central regions ($R_{\rm gc}\la10$ kpc) of GCSs. Thus, the
first step in comparing the stellar, gaseous, and globular cluster contents of
any large galaxy is to combine existing photographic and CCD GCS data into a
single, composite radial distribution that gives the projected density $N_{\rm
cl}(R_{\rm gc})$, to a uniform limiting magnitude, over as large a range of
galactocentric radius as possible. This is done for each of M87, M49, and
NGC 1399 in \S3.1. But then, the surface densities of stars and GCSs are
related to their volume densities by an integral of the form $\Sigma\propto
\int\rho\,dz$, while the observed X-ray brightness along some line of sight is
proportional to $\int\rho_{\rm gas}^2\,dz$. A direct intercomparison of the
{\it raw} GCS, optical, and X-ray imaging data in a galaxy therefore makes
little sense. Thus, to discuss these three components simultaneously in M87,
M49, and NGC 1399, either (1) their GCS and stellar surface densities can be
deprojected and compared to their derived gas volume densities, or (2) the gas
surface densities, $\Sigma_{\rm gas}(R_{\rm gc})\propto\int\rho_{\rm gas}
(r_{\rm gc})\,dz$, can be obtained from the published volume density profiles
and compared to the directly observed $\Sigma_{\rm stars}(R_{\rm gc})$ and
$\Sigma_{\rm cl}(R_{\rm gc})$. The results of these two separate exercises are
presented in \S\S3.2 and 3.3.

\subsection{GCS Surface Densities}

When joining GCS radial distributions from two or more separate studies of the
same galaxy, two important factors must be taken into account. First, the GCS
of a distant galaxy (essentially, any with $D\ga5$ Mpc) is usually identified
as a centrally concentrated excess of point-like sources, above a more uniform
distribution of unresolved background galaxies and foreground stars. The
surface density of this ``background,'' $N_b$, is subtracted from the total
density of point sources at any $R_{\rm gc}$ to form the surface density
profile $N_{\rm cl}$ of the GCS alone; but $N_b$ is a function of seeing
conditions, detector characteristics, and limiting magnitude. Background
corrections must therefore be applied separately to each of the datasets to be
combined. Second, differences in limiting magnitude (call this $V_{\rm lim}$)
are of further importance because a deep survey will clearly see more
globulars, at a given $R_{\rm gc}$, than one with a bright limit. Before they
can be simply matched onto one another, then, distinct background-corrected
profiles $N_{\rm cl}(R_{\rm gc})$ must be normalized to the same effective
$V_{\rm lim}$.

The published GCS observations that will be used here for M87, M49, and NGC
1399 are summarized in Table \ref{tab1}. These papers were selected
from the more numerous references cited above, simply because they tabulate
all of the relevant data in full. The third column of Table \ref{tab1} gives
the minimum and maximum $R_{\rm gc}$ contained in each study's field of view
(note that $1\arcmin=4.363$ kpc for a distance of $D=15$ Mpc to M87 and M49,
while $D=16.5$ Mpc for NGC 1399 implies $1\arcmin=4.800$ kpc); Column 4 gives
the background surface density to the limiting magnitude in Column 5; and
Column 6 lists the scalings applied to each dataset to bring them all to a
common $V_{\rm lim}$.

{\small
\begin{deluxetable}{llcccl}
\tablecaption{Published GCS Radial Distributions\label{tab1}}
\tablehead{Galaxy and & Source & \colhead{Radial Coverage} &
\colhead{Background Density} & \colhead{$V_{\rm lim}$} & \colhead{Scale}
\nl
GCLF Parameters & & \colhead{(arcmin)} & \colhead{(arcmin$^{-2}$)} &
\colhead{(mag)} & Factor }
\startdata
M87 & McLaughlin (1995)\tablenotemark{1} & $0.153-1.965$ & $3.5\pm0.8$
& 23.9~ & 1.000 \nl
$V^0=23.70\ \ \sigma_V=1.39$ & McLaughlin et al.~(1993) & $1.210-9.090$
& $6.3\pm0.4$ & 24.0~ & 0.954 \nl
 & Harris (1986) & $1.000-22.52$ & $5.8\pm0.3$ & 23.4~ & 1.35~ \nl
 & & & & & \nl
 & & & & & \nl
M49 & Harris \& Petrie (1978)\tablenotemark{2} & $2.000-19.69$ & $4.3\pm0.3$
& 23.3~ & 1.000 \nl
$V^0=23.75\ \ \sigma_V=1.30$ & Harris \& van den Bergh (1981)\tablenotemark{2}
& $0.732-12.15$ & $3.3\pm0.5$ & $\ldots$ & 0.86\tablenotemark{3}~ \nl
 & Harris et al.~(1991) & $0.600-2.304$ & $2.88\pm0.45$ & 23.8~ & 0.707 \nl
 & Lee et al.~(1998) & $0.900-7.167$ & $2.95\pm0.30$ & 23.45 & 0.892 \nl
 & & & & & \nl
 & & & & & \nl
NGC 1399 & Kissler-Patig et al.~(1997) & $0.761-9.463$ & $6.1\pm0.3$ & 24.0~
& 1.000 \nl
$V^0=23.90\ \ \sigma_V=1.20$ & Hanes \& Harris (1986) & $0.734-13.97$
& $5.40\pm0.45$ & 23.4~ & 1.575 \nl
\enddata
\tablenotetext{1}{See also Harris et al.~(1998).}
\tablenotetext{2}{As given by Harris (1986).}
\tablenotetext{3}{See the discussion of Harris (1986).}
\end{deluxetable}
}

Generally, the background levels $N_b$ have been determined and applied by the
original authors, either from observations of a blank field near the observed
GCS, or from ``on-frame'' extrapolations of counts at the edges of the GCS
field itself (see \markcite{har86}Harris 1986); these values are reproduced in
Table \ref{tab1} only for completeness. The single exception to this is the
M49 study of \markcite{lee98}Lee et al.~(1998), for which $N_b$ has been
derived here, using the method of \markcite{har86}Harris (1986). (It is
possible, particularly with on-frame estimates, to overestimate $N_b$ and
produce spurious curvature at large radii in a subtracted $N_{\rm cl}$
profile. As is discussed below, however, this is not a major concern in the
composite GCS radial distributions obtained here.) The limiting magnitudes
$V_{\rm lim}$ in Table \ref{tab1} are also those quoted in the original
papers.\footnotemark
\footnotetext{Not all of the photometry in these studies is in the $V$ band,
however. A mean globular cluster color of $V-R=0.5$ has been adopted to
convert the $R$-band data  of \markcite{mcl95}McLaughlin (1995) to $V$
magnitudes; a mean $V-T_1=0.45$ brings the Washington photometry of
\markcite{lee98}Lee et al.~(1998) onto the same system; and approximate
transformations between photographic and $V$-band magnitudes are discussed in
Harris' earlier papers on M87 and M49.}
These have been used to renormalize the individual $N_{\rm cl}$ profiles,
by adopting---for computational convenience---a standard Gaussian
approximation to the globular cluster luminosity function (GCLF, $d{\cal N}/
d\,\ln\,m$; cf.~\S2.1). That is, with the number of globulars at a given
apparent magnitude given roughly by $d{\cal N}/dV\propto\exp [-(V-V^0)^2/2
\sigma_V^2]$ (see \markcite{har91}Harris 1991), then the density of clusters
to some limiting magnitude, say $N_{\rm cl}(V\le V_{{\rm lim},2})$, is
referred to the density up to some other limit, i.e., $N_{\rm cl}(V\le
V_{{\rm lim},1})$, through
\begin{equation}
{{N_{\rm cl}(V\le V_{{\rm lim},1})}\over{N_{\rm cl}(V\le V_{{\rm lim},2})}}
\,=\,
{{1+{\rm erf}\left[(V_{{\rm lim},1}-V^0)/\sqrt{2}\,\sigma_V\right]}\over
{1+{\rm erf}\left[(V_{{\rm lim},2}-V^0)/\sqrt{2}\,\sigma_V\right]}}\ .
\label{eq:31}
\end{equation}
The right-hand side of this expression is just the scaling factor given in
Column 6 of Table \ref{tab1}, where $V_{{\rm lim},1}$ for each GCS has been
taken as the observed limiting magnitude in the first of each set of
references. Thus, the published radial distributions for M87 have all been
normalized to an effective $V_{\rm lim}=23.9$; for M49, to $V_{\rm lim}=23.3$;
and to $V_{\rm lim}=24.0$ in NGC 1399. The Gaussian peaks $V^0$ and dispersions
$\sigma_V$ used in equation (\ref{eq:31}) are given in the first column of
Table \ref{tab1}. These parameters are taken from the fits of
\markcite{whi95}Whitmore et al.~(1995) and \markcite{har98}Harris et al.~(1998)
to the M87 GCLF, from \markcite{lee98}Lee et al.~(1998) for M49, and from
\markcite{koh96}Kohle et al.~(1996) for NGC 1399. Note that all of the scale
factors derived here are rather modest, and thus do not introduce significant
uncertainty into the final, combined $N_{\rm cl}$ profiles.

\begin{figure*}[b]
\centering \leavevmode
\epsfysize=4.0truein
\epsfbox{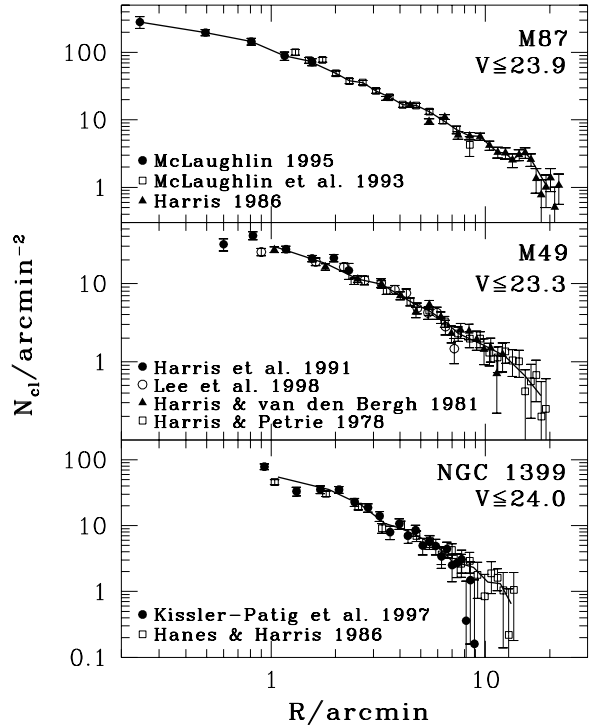}
\caption{Published surface density profiles $N_{\rm cl}
(R_{\rm gc})$ for the GCSs of M87, M49, and NGC 1399. Points refer to
individual studies, normalized as in Table \ref{tab1} to the effective
limiting magnitudes indicated; bold lines describe the composite distributions
given in Tables \ref{tab2}, \ref{tab3}, and \ref{tab4}.
\label{fig1}}
\end{figure*}

A few practical notes should be made regarding this scaling procedure. First,
it is applied to the GCS densities at every observed galactocentric radius, and
it therefore assumes that there is no intrinsic radial variation in the GCLF
or the underlying GCS mass function. As was also mentioned in \S2.1 above, this
is a reasonable assumption that is directly supported by the observations
themselves. Second, the limiting magnitudes for the photographic studies of
\markcite{har78}Harris \& Petrie (1978) and \markcite{har86}Harris (1986) are
rather poorly determined; Harris quotes only approximate ranges for $V_{\rm
lim}$. The values adopted here, which are fully consistent with these ranges,
have been chosen to give scaled $N_{\rm cl}(R_{\rm gc})$ that agree well with
more recent CCD studies with accurately known $V_{\rm lim}$. And third, the
scaling applied to the M49 GCS counts of \markcite{hvd81}Harris \& van den
Bergh (1981) has not been calculated in this way at all; rather, the factor
of 0.86 is derived, by \markcite{har86}Harris (1986), by requiring a good
overall agreement with the $N_{\rm cl}$ profile of \markcite{har78}Harris \&
Petrie (1978).

The results of this exercise are illustrated in Fig.~\ref{fig1}. Clearly, the
different projected density profiles in each GCS corroborate each other. It
remains only to
extract a single $N_{\rm cl}$ at each $R_{\rm gc}$, in each system, from the
overlapping data that have been collected. It is not generally possible to
simply average different densities at a given $R_{\rm gc}$, since profiles
such as these are constructed by counting clusters in a series of wide,
circular annuli that need not coincide from study to study. Instead,
Fig.~\ref{fig1} shows that a perfectly acceptable approach is to simply adopt
one of the (scaled) profiles over some range in $R_{\rm gc}$, and then switch
to any other that might extend beyond that range. At this point, it must be
kept in mind that ultimately the data will be deprojected to yield volume GCS
densities, $n_{\rm cl}(r_{\rm gc})$. The method developed in the Appendix to
do this requires that the inner and outer radii be known for the annuli
used to define the composite $N_{\rm cl}$. This further restricts the useful
published datasets to those which contain this information on the radial bins
as well (that is, {\it in addition} to just a density and a mean radius for
each annulus).

In M87, then, a composite GCS profile is derived by using the data of
\markcite{mcl95}McLaughlin (1995) from $R_{\rm gc}=0\farcm153$ to $R_{\rm gc}=
1\farcm870$; those of \markcite{mcl93}McLaughlin et al.~(1993) from
$1\farcm870$ to $7\farcm870$; and those of \markcite{har86}Harris (1986) from
$7\farcm870$ to $22\farcm52$. (In order to make annuli that do not overlap in
the final $N_{\rm cl}$, some interpolation is applied to the datasets around
the matching points at $1\farcm870$ and $7\farcm870$.) For M49, the scaled
profile of \markcite{hvd81}Harris \& van den Bergh (1981) is used over
$0\farcm732\le R_{\rm gc}\le 10\farcm96$, and that of \markcite{har78}Harris \&
Petrie (1978) for $10\farcm96\le R_{\rm gc}\le 19\farcm69$. And in NGC 1399,
the densities of \markcite{kis97}Kissler-Patig et al.~(1997) are adopted
for $0\farcm761\le R_{\rm gc}\le 7\farcm571$, with the data of
\markcite{han86}Hanes and Harris (1986) continuing on out to $R_{\rm gc}=13
\farcm97$. In all cases, CCD data have usually been chosen over photographic
data in any regions of overlap. However, superseding this preference is a
concern over possible errors in the background estimation $N_b$; where this
seems too high---for example, when the background-subtracted $N_{\rm cl}$ at
the largest radii in a smaller-scale CCD study drop significantly below the
densities indicated by a more extensive photographic survey (as in the
outermost data points of \markcite{kis97}Kissler-Patig et al.~1997 for NGC
1399)---the photographic profiles are given precedence. Enforcing consistency
in this way between independent GCS datasets ensures that the question of
potentially overestimated $N_b$ values remains an issue only at the largest
radii (beyond the regime of any overlap) in the widest-field observations of
each galaxy.

The composite radial distributions thus obtained are listed in the third
columns of Tables \ref{tab2} through \ref{tab4}, as average surface densities
in series of concentric circular annuli (with inner, outer, and average
projected radii given in the first and second columns of the Tables). They are
also drawn as the bold lines in the three panels of Fig.~\ref{fig1}. There are
two details of particular note. First, the densities for the innermost two
annuli in M49 have been scaled up from the (already renormalized) GCS profile
of \markcite{hvd81}Harris \& van den Bergh (1981), by a further factor of 1.1,
to put the composite $N_{\rm cl}$ at small radii into better agreement with
the CCD data of \markcite{hap91}Harris et al.~(1991) and \markcite{lee98}Lee et
al.~(1998). This ad hoc correction most likely just reflects
some residual incompleteness in the photographic GCS counts at small $R_{\rm
gc}$, and in any event does not significantly affect the results which follow.
Second, all of the $N_{\rm cl}$ profiles have been smoothed somewhat by
rebinning, in order to allow for a more stable deprojection in \S3.2 below:
the outermost six annuli of \markcite{har86}Harris (1986) are combined into
one for the M87 profile; the outermost 11 rings of \markcite{har78}Harris \&
Petrie (1978) are blended into 4 for M49; and the last 8 radial bins of
\markcite{han86}Hanes \& Harris (1986) become 4 in the NGC 1399 GCS. (As an
added benefit of this smoothing, any possible errors in the subtracted $N_b$
are expected to affect only the outermost {\it one or two points} in the final
$N_{\rm cl}$ profiles.) Similarly, counts in the ``Northeast'' and
``Northwest'' sectors of NGC 1399 in the analysis of
\markcite{kis97}Kissler-Patig et al.~(1997; see their Table 8) have been added
together in every annulus they define; and these annuli have been combined in
pairs to give the coarser profile in Table \ref{tab4} here.

\begin{deluxetable}{ccccc}
\tablecaption{Density Profile of the M87 GCS\tablenotemark{1} \label{tab2}}
\tablewidth{0pt}
\tablehead{
\colhead{Radius} & $R_{\rm gc}$ & \colhead{$N_{\rm cl}$} & $r_{\rm gc}$ &
\colhead{$n_{\rm cl}$} \nl
\colhead{(arcmin)} & (ave.) & \colhead{(arcmin$^{-2}$)} & (ave.) &
\colhead{(arcmin$^{-3}$)}
}
\startdata
0.153 -- 0.328 & 0.244 & 282.6$\pm$55.8 & 0.245 & 243.7$\pm$155.6 \nl
0.328 -- 0.655 & 0.494 & 197.2$\pm$19.9 & 0.505 & 89.34$\pm$34.01 \nl
0.655 -- 0.983 & 0.808 & 145.3$\pm$16.5 & 0.827 & 68.58$\pm$20.55 \nl
0.983 -- 1.310 & 1.157 & 89.10$\pm$12.40& 1.152 & 18.16$\pm$12.94 \nl
1.310 -- 1.870 & 1.500 & 74.46$\pm$8.78 & 1.602 & 19.33$\pm$5.23~ \nl
1.870 -- 2.160 & 2.010 & 48.92$\pm$4.25 & 2.018 & 11.36$\pm$3.56~ \nl
2.160 -- 2.490 & 2.320 & 37.81$\pm$3.26 & 2.328 & 4.015$\pm$2.391 \nl
2.490 -- 2.880 & 2.680 & 35.75$\pm$2.80 & 2.689 & 6.404$\pm$1.722 \nl
2.880 -- 3.320 & 3.090 & 26.93$\pm$2.05 & 3.104 & 3.423$\pm$1.183 \nl
3.320 -- 3.830 & 3.570 & 21.74$\pm$1.65 & 3.580 & 2.875$\pm$0.818 \nl
3.830 -- 4.430 & 4.120 & 16.78$\pm$1.34 & 4.135 & 0.926$\pm$0.587 \nl
4.430 -- 5.110 & 4.760 & 16.38$\pm$1.23 & 4.776 & 1.447$\pm$0.469 \nl
5.110 -- 5.900 & 5.490 & 13.26$\pm$1.13 & 5.512 & 1.202$\pm$0.360 \nl
5.900 -- 6.820 & 6.350 & ~9.79$\pm$0.93 & 6.368 & 0.768$\pm$0.271 \nl
6.820 -- 7.870 & 7.330 & ~7.16$\pm$0.96 & 7.354 & 0.442$\pm$0.219 \nl
7.870 -- 8.973 & 8.403 & ~5.82$\pm$0.73 & 8.431 & 0.184$\pm$0.166 \nl
8.973 -- 9.962 & 9.454 & ~5.68$\pm$0.69 & 9.474 & 0.346$\pm$0.154 \nl
9.962 -- 10.95 & 10.44 & ~4.20$\pm$0.65 & 10.46 & 0.222$\pm$0.135 \nl
10.95 -- 11.93 & 11.43 & ~3.32$\pm$0.61 & 11.45 & 0.079$\pm$0.124 \nl
11.93 -- 12.92 & 12.42 & ~3.23$\pm$0.61 & 12.43 & 0.156$\pm$0.120 \nl
12.92 -- 13.90 & 13.40 & ~2.55$\pm$0.58 & 13.41 & 0.008$\pm$0.113 \nl
13.90 -- 14.87 & 14.38 & ~3.05$\pm$0.57 & 14.39 & 0.067$\pm$0.104 \nl
14.87 -- 15.84 & 15.35 & ~3.27$\pm$0.57 & 15.36 & 0.143$\pm$0.103 \nl
15.84 -- 16.81 & 16.32 & ~2.59$\pm$0.55 & 16.33 & 0.212$\pm$0.093 \nl
16.81 -- 22.52 & 19.46 & ~1.01$\pm$0.43 & 19.77 & 0.024$\pm$0.010 \nl
\enddata
\tablenotetext{1}{Limiting magnitude $V_{\rm lim}=23.9$; divide by 0.5572 to
obtain the density over all magnitudes.}
\end{deluxetable}

\begin{deluxetable}{ccccc}
\tablecaption{Density Profile of the M49 GCS\tablenotemark{1} \label{tab3}}
\tablewidth{0pt}
\tablehead{
\colhead{Radius} & $R_{\rm gc}$ & \colhead{$N_{\rm cl}$} & $r_{\rm gc}$ &
\colhead{$n_{\rm cl}$}
\nl
\colhead{(arcmin)} & (ave.) & \colhead{(arcmin$^{-2}$)} & (ave.) &
\colhead{(arcmin$^{-3}$)}}
\startdata
0.732 -- 1.464 & 1.035 & 29.83$\pm$2.53 & 1.129 & 8.041$\pm$1.684 \nl
1.464 -- 2.196 & 1.793 & 18.00$\pm$1.61 & 1.848 & 3.818$\pm$0.874 \nl
2.196 -- 2.928 & 2.536 & 11.40$\pm$1.11 & 2.575 & 1.163$\pm$0.543 \nl
2.928 -- 3.660 & 3.274 & ~9.82$\pm$1.06 & 3.304 & 1.307$\pm$0.426 \nl
3.660 -- 4.392 & 4.009 & ~7.02$\pm$0.80 & 4.034 & 0.759$\pm$0.292 \nl
4.392 -- 4.996 & 4.684 & ~5.48$\pm$0.63 & 4.699 & 0.464$\pm$0.279 \nl
4.996 -- 5.992 & 5.471 & ~4.51$\pm$0.59 & 5.505 & 0.384$\pm$0.169 \nl
5.992 -- 6.988 & 6.471 & ~3.29$\pm$0.52 & 6.500 & 0.276$\pm$0.138 \nl
6.988 -- 7.981 & 7.467 & ~2.30$\pm$0.48 & 7.493 & 0.120$\pm$0.120 \nl
7.981 -- 8.973 & 8.462 & ~1.97$\pm$0.45 & 8.484 & 0.092$\pm$0.097 \nl
8.973 -- 10.95 & 9.912 & ~1.62$\pm$0.31 & 9.985 & 0.070$\pm$0.044 \nl
10.95 -- 13.90 & 12.33 & ~1.18$\pm$0.23 & 12.47 & 0.054$\pm$0.026 \nl
13.90 -- 16.81 & 15.28 & ~0.66$\pm$0.22 & 15.39 & 0.028$\pm$0.023 \nl
16.81 -- 19.69 & 18.19 & ~0.37$\pm$0.21 & 18.28 & 0.010$\pm$0.005 \nl
\enddata
\tablenotetext{1}{Limiting magnitude $V_{\rm lim}=23.3$; divide by 0.3646 to
obtain the density over all magnitudes.}
\end{deluxetable}

\begin{deluxetable}{ccccc}
\tablecaption{Density Profile of the NGC 1399 GCS\tablenotemark{1}
\label{tab4}}
\tablewidth{0pt}
\tablehead{
\colhead{Radius} & $R_{\rm gc}$ & \colhead{$N_{\rm cl}$} & $r_{\rm gc}$ &
\colhead{$n_{\rm cl}$}
\nl
\colhead{(arcmin)} & (ave.) & \colhead{(arcmin$^{-2}$)} & (ave.) &
\colhead{(arcmin$^{-3}$)}}
\startdata
0.761 -- 1.518 & 1.075 & 54.88$\pm$4.60 & 1.171 & 13.74$\pm$0.31~ \nl
1.518 -- 2.274 & 1.858 & 35.30$\pm$3.10 & 1.915 & 7.719$\pm$0.163 \nl
2.274 -- 3.031 & 2.625 & 20.64$\pm$2.11 & 2.666 & 4.288$\pm$0.949 \nl
3.031 -- 3.788 & 3.388 & 10.81$\pm$1.48 & 3.420 & 1.141$\pm$0.607 \nl
3.788 -- 4.544 & 4.149 & ~8.77$\pm$1.27 & 4.175 & 0.957$\pm$0.472 \nl
4.544 -- 5.301 & 4.908 & ~6.72$\pm$1.10 & 4.930 & 0.618$\pm$0.393 \nl
5.301 -- 6.058 & 5.667 & ~5.31$\pm$0.97 & 5.686 & 0.496$\pm$0.314 \nl
6.058 -- 6.814 & 6.425 & ~3.93$\pm$0.86 & 6.442 & 0.402$\pm$0.267 \nl
6.814 -- 7.571 & 7.183 & ~2.61$\pm$0.79 & 7.197 & 0.108$\pm$0.261 \nl
7.571 -- 8.104 & 7.833 & ~2.60$\pm$0.99 & 7.840 & 0.093$\pm$0.332 \nl
8.104 -- 9.573 & 8.808 & ~2.32$\pm$0.71 & 8.854 & 0.175$\pm$0.134 \nl
9.573 -- 11.04 & 10.21 & ~1.39$\pm$0.68 & 10.32 & 0.045$\pm$0.111 \nl
11.04 -- 12.50 & 11.75 & ~1.32$\pm$0.55 & 11.78 & 0.095$\pm$0.097 \nl
12.50 -- 13.97 & 13.21 & ~0.65$\pm$0.62 & 13.24 & 0.024$\pm$0.023 \nl
\enddata
\tablenotetext{1}{Limiting magnitude $V_{\rm lim}=24.0$; divide by 0.5332 to
obtain the density over all magnitudes.}
\end{deluxetable}

Given these radial distributions, it is a simple matter to derive the total
{\it mass} surface density profiles of the GCSs. This requires, first, another
multiplicative scaling of $N_{\rm cl}(R_{\rm gc})$ in each case, so that the
number density of {\it all} globulars is properly reflected. (Note that the
observations used here reach limiting magnitudes which are just at, or brighter
than, the peak of the symmetric Gaussian that best fits the GCLF. This implies
that most of the faint globulars are lost in the noise in these studies, and 
that half or fewer have been directly observed at any $R_{\rm gc}$.) These
final correction factors are obtained from equation (\ref{eq:31}) above, by
setting $V_{{\rm lim},1}=\infty$ and putting $V_{{\rm lim},2}$ at the
effective limit for each composite $N_{\rm cl}$. In this way, it is found that
the M87 distribution in Table \ref{tab2} should be multiplied by 1.795; the
M49 densities in Table \ref{tab3}, by 2.743; and the NGC 1399 numbers, by
1.875. Once this is done, the total number densities are multiplied by a mean
globular cluster mass to obtain the mass densities $\Sigma_{\rm cl}(R_{\rm
gc})$.

The mean cluster mass adopted here is $\langle m\rangle_{\rm cl}=2.4\times
10^5\,M_\odot$, which (as was mentioned in \S2.1 above) is the value for the
Milky Way GCS. In both its shape {\it and} its mass scale, the Galactic GCLF,
or mass function, bears enough similarity to those of the cluster systems
being discussed here (and, indeed, to most others: \markcite{har91}Harris
1991) that the same $\langle m\rangle_{\rm cl}$ may safely be applied to them
as well. The only issue is whether or not there is any systematic radial trend
in mean globular cluster masses. It was also pointed out in \S2.1 that there
are no strong radial dependences observed in the GCLF of M87
(\markcite{mcl94}McLaughlin et al.~1994; \markcite{har98}Harris et al.~1998),
suggesting that $\langle m\rangle_{\rm cl}$ is constant with $r_{\rm gc}$ in
this galaxy, and thus likely also in similar systems like M49 and NGC 1399.
The question can be addressed in more detail only for the Milky Way,
which---although clearly an imperfect substitute for these giant
ellipticals---is the only galaxy where accurate luminosities and
three-dimensional galactocentric distances both can be assigned to globulars
on an individual basis.

\begin{figure*}[tb]
\centering \leavevmode
\epsfysize=4.0truein
\epsfbox{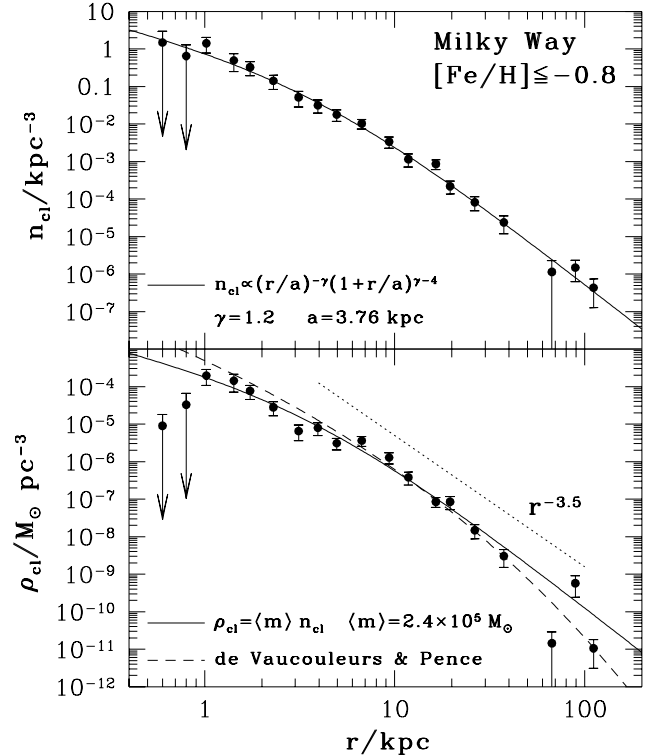}
\caption{Volume number- and mass-density profiles of the
metal-poor (halo) globular clusters in the Milky Way. Individual magnitudes,
metallicities, and galactocentric radii for 112 clusters are taken from the
compilation of Harris (1996). Solid curves represent a parametric fit to the
$n_{\rm cl}$ distribution, which is multiplied by a mean cluster mass of
$\langle m\rangle_{\rm cl}=2.4\times10^5M_\odot$ to obtain the mass densities
$\rho_{\rm cl}$. The good fit to the data in both the top and bottom panels
shows that $\langle m\rangle_{\rm cl}$ is not a function of $r_{\rm gc}$ in the
Milky Way. Broken curves in the bottom panel refer to the density profile
$\rho_{\rm stars}(r_{\rm gc})$ of the Pop.~II spheroid; see \S4.3 of the text.
\label{fig2}}
\end{figure*}

Figure \ref{fig2} shows that there is no dependence on radius in the
observed $\langle m\rangle_{\rm cl}$ of the Milky Way. The top panel plots the
volume number density of globulars in the Galactic halo, derived from the
database of \markcite{har96}Harris (1996); drawn as the solid line is a
least-squares fit of a parametric density distribution in the family discussed
by \markcite{deh93}Dehnen (1993) and \markcite{tre94}Tremaine et al.~(1994).
(Only metal-poor clusters have been counted, as these are the ones that
display true halo kinematics: \markcite{zin85}Zinn 1985;
\markcite{min95}Minniti 1995.) In the bottom panel, the total $V$-band
luminosities of the clusters (taken again from \markcite{har96}Harris 1996)
have been summed separately in each radial bin, and converted to total cluster
masses by applying a mass-to-light ratio of $\Upsilon_V=2$
(\markcite{man91}Mandushev, Spassova, \& Staneva 1991; \markcite{pry93}Pryor
\& Meylan 1993). The mass densities $\rho_{\rm cl}$ that result are shown as
the circular points. The solid line now corresponds to the same $n_{\rm cl}$
fit from the top panel, but multiplied by the average mass calculated for all
globulars with $2\,{\rm kpc}\le r_{\rm gc}\le40\,{\rm kpc}$. This shows that
$\rho_{\rm cl}=\langle m\rangle_{\rm cl}\,n_{\rm cl}$ to a good approximation
at every $r_{\rm gc}$, and therefore that the mean globular cluster mass is
not a strong function of galactocentric radius. The other curves in the bottom
panel of this Figure are discussed in \S4.3. For now, the point is simply to
lend some circumstantial support to the claim that $\langle m \rangle_{\rm cl}
\simeq2.4\times10^5\,M_\odot$ should not vary significantly with
galactocentric radius in M87, M49, and NGC 1399.

\begin{figure*}[tb]
\centering \leavevmode
\epsfysize=4.0truein
\epsfbox{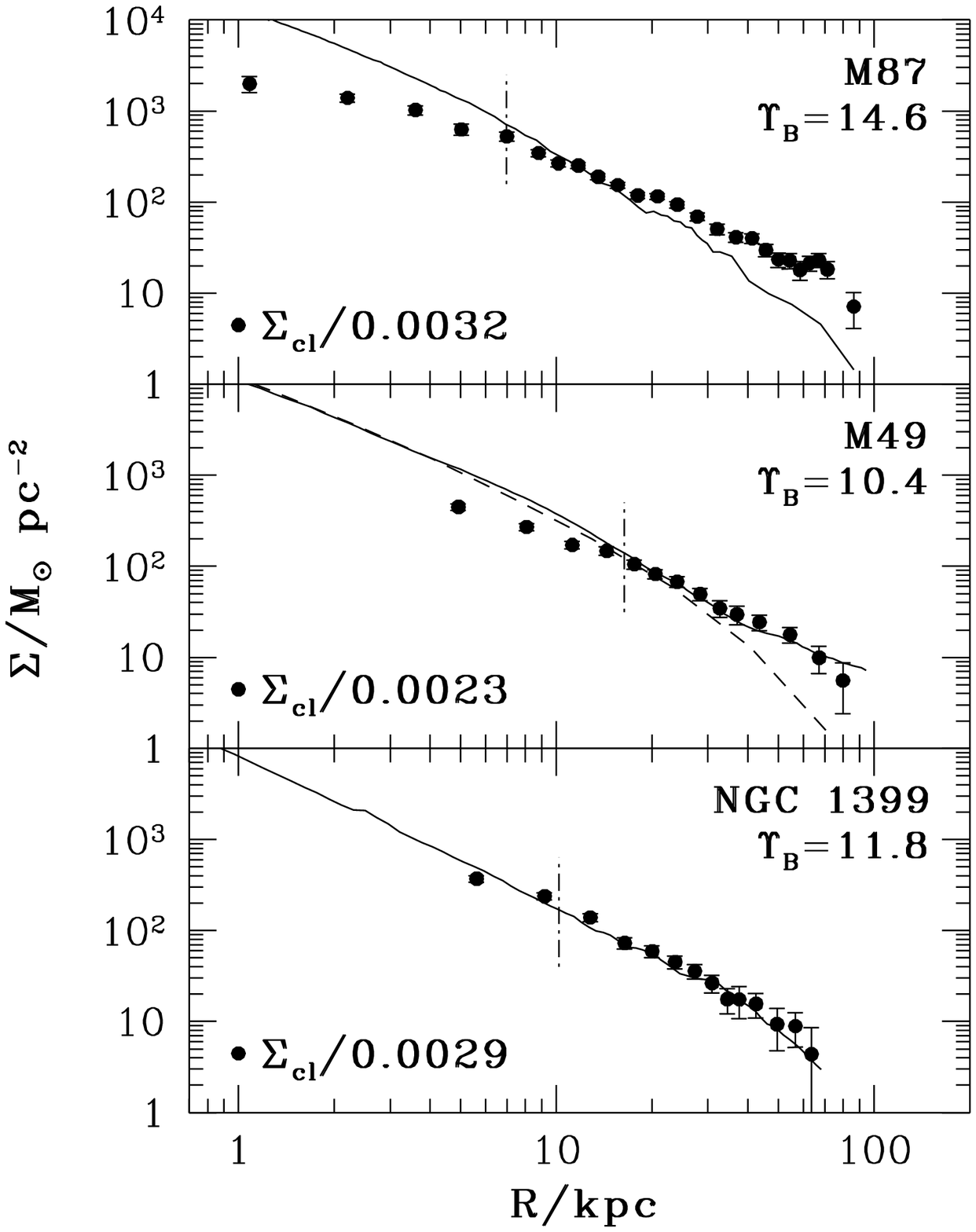}
\caption{Comparison of the projected mass density profiles of
globular clusters (points) with those of the unclustered halo stars (lines)
in M87, M49, and NGC 1399. GCS densities have been corrected for
photometric incompleteness (see text), and scaled up by the inverse of the
cluster formation efficiencies derived in \S3.2 below. Mass-to-light ratios
for the stars are taken in each galaxy from van der Marel (1991). The broken
vertical lines mark the effective radius of each galaxy: $R_{\rm eff}=7.0$
kpc for M87 (de Vaucouleurs \& Nieto 1978), 16.4 kpc for M49 (Caon et
al.~1994), and 10.2 kpc in NGC 1399 (Caon et al.~1994). The disparity between
$\Sigma_{\rm cl}$ and $\Sigma_{\rm stars}$ in M87 (top panel) embodies the
specific frequency ``problems'' there. 
\label{fig3}}
\end{figure*}

The total projected mass densities for these three GCSs are shown as the
filled circles in Fig.~\ref{fig3}, where angular distances and areas have been
converted to linear units using $D({\rm Virgo})=15$ Mpc and $D({\rm Fornax})=
16.5$ Mpc. The final $\Sigma_{\rm cl}$ have also been scaled upwards by
additional factors of $300-400$ for this Figure, to facilitate direct
comparisons with the mass surface densities of the galaxies' stars (drawn as
lines). These latter distributions are derived from published $B$-band surface
brightness profiles (M87: de Vaucouleurs \& Nieto \markcite{dvn78}1978,
\markcite{dvn79}1979; M49: \markcite{kin78}King 1978, \markcite{cao94}Caon et
al.~1994; NGC 1399: \markcite{cao94}Caon et al.~1994), according to the
standard relation
\begin{equation}
\Sigma_{\rm stars}/(M_\odot\,{\rm pc}^{-2})=
\Upsilon_B\times10^{0.4(27.05+A_B-\mu_B)}
\label{eq:32}
\end{equation}
for $\mu_B$ in units of mag arcsec$^{-2}$. (Extinctions of $A_B=0.09$ mag, 0,
and 0 are adopted for M87, M49, and NGC 1399; see \markcite{bur84}Burstein \&
Heiles 1984.) The stellar mass-to-light ratios $\Upsilon_B$ for all three
galaxies come from \markcite{vdm91}van der Marel's (1991) kinematic analysis
of their cores; his values have been corrected for the distances $D$ assumed
here, and are taken to be independent of $R_{\rm gc}$. 

The comparisons of $\Sigma_{\rm cl}$ and $\Sigma_{\rm stars}$ in
Fig.~\ref{fig3} illustrate the $S_N$ ``problems'' in M87 and show that,
contrary to previous claims, {\it they do not exist} in M49 and NGC 1399.

As was indicated above, the global specific frequency of M49 is quite normal;
the most recent estimate places it at $S_N=4.7\pm0.6$ (\markcite{lee98}Lee et
al.~1998). Given a stellar mass-to-light ratio of $\Upsilon_V=7.5$ [which
follows from $\Upsilon_B=10.4$ and $(B-V)=1.0$], equation (\ref{eq:23}) then
implies that $M_{\rm gcs}/M_{\rm stars}=(1.7\pm0.2)\times10^{-3}$ globally
in this galaxy. In keeping with this, the middle panel of Fig.~\ref{fig3}
shows directly that $\Sigma_{\rm cl}/\Sigma_{\rm stars}\simeq2.3\times10^{-3}$
at essentially any $R_{\rm gc}\ga 15$ kpc. This is slightly higher than the
global GCS mass fraction because the globular cluster surface density profile
is shallower than the stellar distribution at {\it small} galactocentric radii.
As was discussed in \S2.2, such a feature might result, at least in part, from
the dynamical destruction of globulars in the densest regions of the galaxy.
[Again, the theoretical work of \markcite{mwb97}Murali \& Weinberg 1997b, or
of \markcite{agu88}Aguilar et al.~1988, suggests that the important spatial
scale is generally of order a stellar effective radius, $R_{\rm eff}$. These
are marked by vertical lines for the galaxies in Fig.~\ref{fig3}; note that
$R_{\rm eff}({\rm M49})\simeq16$ kpc.] This aside, M49 has been put forward as
an example of the second specific frequency problem (beginning with
\markcite{har86}Harris 1986, and continuing on to \markcite{lee98}Lee et
al.~1998) on the basis of comparisons at {\it large} $R_{\rm gc}\ga R_{\rm
eff}$ between $\Sigma_{\rm cl}$ and the $\Sigma_{\rm stars}$ {\it derived
from the surface photometry of \markcite{kin78}King (1978)}. King's data are
shown as the dashed line in the middle of Fig.~\ref{fig3}, and they clearly do
suggest that the projected stellar density falls off more steeply with radius
than the GCS density in M49. However, the solid line in this panel represents
the more recent $\mu_B$ data of \markcite{cao94}Caon et al.~(1994), which
include many more measurements at $R_{\rm gc}\ga30$ kpc and do {\it not} show
any GCS-halo discrepancy. The local specific frequency, which is proportional
to $\Sigma_{\rm cl}/\Sigma_{\rm stars}$, then takes on a roughly constant value
throughout the outer halo of M49, and the second $S_N$ problem does not appear
there.\footnotemark
\footnotetext{\markcite{lee98}Lee et al.~(1998) have recently claimed evidence
for a difference in the radial distributions $\Sigma_{\rm cl}$ of red
(metal-rich) and blue (or metal-poor) globular clusters in M49. They argue
that the former are more centrally concentrated, i.e., exhibit a steeper
fall-off in density with $R_{\rm gc}$, than the latter. They also suggest that
the ``red'' component of the M49 GCS  follows the galaxy's stellar profile
$\Sigma_{\rm stars}$, while the ``blue'' component is spatially more extended
(see also \markcite{for97}Forbes et al.~1997). However, since
\markcite{lee98}Lee et al.~were unable to determine the background densities
$N_b$ separately in the blue and red subsets that they defined for the GCS,
further observations are required to verify their claim (and comparisons then
should perhaps be made with the surface brightness profile of
\markcite{cao94}Caon et al.~1994). Moreover, any real differences in the
{\it volume} density profiles of red and blue globulars will be artificially
amplified in projection. If \markcite{lee98}Lee et al.'s result is confirmed,
it will be important to understand in detail; but {\it all} of the GCSs
discussed here are considered only in their entireties, without any
attempt to distinguish between subpopulations.}
The differences between the light profiles of \markcite{kin78}King (1978) and
\markcite{cao94}Caon et al.~(1994) illustrate the difficulty of accurate sky
subtraction at faint surface brightness levels, and make the point that
background corrections are an important consideration in any discussion of
galaxy (or GCS) structure on large spatial scales. The surface photometry of
\markcite{cao94}Caon et al.~(1994) is adopted here for M49, and thus
$\Sigma_{\rm cl}\propto\Sigma_{\rm stars}$ at large $R_{\rm gc}$; but this
important caveat must be kept in mind.

In \S3.2, it is shown that the gas in M49 adds up to a small fraction of its
stellar mass, in which case the local cluster formation efficiency can be
estimated by $\wepsilon_{\rm cl}=\rho_{\rm cl}/(\rho_{\rm gas}+\rho_{\rm
stars})\simeq\rho_{\rm cl}/\rho_{\rm stars}$. The fact that $\Sigma_{\rm cl}
\propto\Sigma_{\rm stars}$ therefore implies that $\epsilon_{\rm cl}$ was
essentially independent of galactocentric radius beyond about $R_{\rm eff}
\simeq16$ kpc. An extrapolation of this result to smaller distances requires
that the systematic decrease in $\Sigma_{\rm cl}/\Sigma_{\rm stars}$ there has
resulted entirely from the gradual destruction of (initially bound) globulars
over a Hubble time. As {\it possible} anecdotal support for this
interpretation, recall that globally $M_{\rm gcs}/M_{\rm stars}\simeq0.0017$
for M49, while locally $\Sigma_{\rm cl}/\Sigma_{\rm stars}\simeq0.0023$ at
large radii. If the latter number is assumed to have held all the way to the
center of the galaxy initially, then a global $M_{\rm gcs}^{\rm init}/M_{\rm
stars}^{\rm init}\simeq0.0023$ is implied. In such a scenario, if $M_{\rm
stars}$ were roughly conserved over a Hubble time while globulars were
dynamically disrupted (their remains adding nothing significant to the total
stellar mass), the global ratio of current to initial GCS masses would be
$M_{\rm gcs}/M_{\rm gcs}^{\rm init}\simeq0.0017/0.0023\sim0.75$. This is
surprisingly close to the result of the general GCS mass-function arguments in
\S2.1 (see eq.~[\ref{eq:28}]). However, this could be just a coincidence; it
remains possible that $\epsilon_{\rm cl}$ really was lower at small
galactocentric radii, i.e., that the GCS {\it formed} with a ``core'' which
was somewhat larger than that of the unclustered stellar distribution, and
which was only enhanced over time. The current data cannot settle the question
either way.

The bottom panel of Fig.~\ref{fig3} shows that the second $S_N$ problem does
not exist beyond a stellar $R_{\rm eff}$ in NGC 1399, either. Every study of
this GCS has appreciated this point, and it is not a new result
(see the references cited above). (Note also that the surface brightness
profile used here for NGC 1399, from \markcite{cao94}Caon et al.~1994, is
consistent with that of \markcite{kil88}Killeen \& Bicknell 1988 at large
$R_{\rm gc}$.) On the other hand, the global $S_N$ in this galaxy---and
therefore the local $\Sigma_{\rm cl}/\Sigma_{\rm stars}$ at any single
$R_{\rm gc}$---has always been thought to be significantly higher than the
normal, M49-like value of $\sim$5: the latest calculation claims $S_N=12\pm3$
(\markcite{kis97}Kissler-Patig et al.~1997).\footnotemark
\footnotetext{After this paper was submitted, a new study of NGC 1399 by
\markcite{ost98}Ostrov, Geisler, \& Forte (1998) found that the specific
frequency of its GCS is a much more modest $S_N=5.6\pm1.0$. This result is
consistent with that obtained here, and in fact is even closer to the value
for M49; it supports the conclusion that NGC 1399 does not exhibit the first
$S_N$ problem.}
However, it is immediately apparent from Fig.~\ref{fig3} that this is {\it not}
the case: the bulk scaling of $1/0.0029$ required to match the GCS radial
distribution to the stellar mass profile in NGC 1399 is identical, within the
observational uncertainties, to the $1/0.0023$ needed in M49. Since the gas
mass is negligible compared to the stellar mass on $<70$-kpc scales in NGC
1399, as it is in M49, $\wepsilon_{\rm cl}\simeq\Sigma_{\rm cl}/\Sigma_{\rm
stars}$ holds for both systems, and their cluster formation efficiencies are
inferred to have been essentially identical. The spuriously high global $S_N$
values typically quoted for NGC 1399 can be traced back to the practice of
assigning an absolute magnitude of $M_V\simeq-21.6$ to the galaxy (from the
Third Reference Catalogue: \markcite{dev91}de Vaucouleurs et al.~1991); but,
in fact, $M_V\simeq-22.1$ is indicated by direct integration (to $R_{\rm gc}=
70$ kpc, assuming $D=16.5$ Mpc) of the surface brightness profiles of either
\markcite{cao94}Caon et al.~(1994) or \markcite{kil88}Killeen \& Bicknell
(1988). Given this brighter magnitude, and an estimated GCS population of
${\cal N}_{\rm tot}=4\,700\pm400$ projected to within 70 kpc of NGC 1399 (see
\S5), equation (\ref{eq:21}) yields $S_N({\rm NGC\ 1399})=7.0\pm0.6$, much
closer to the value for M49 (see also \markcite{ost98}Ostrov et al.~1998).
This is consistent with the alternate representation of the data in
Fig.~\ref{fig3}, and confirms that the {\it first} specific frequency problem
is not an issue in NGC 1399.

It appears, therefore, that there are no $S_N$ problems in either M49 or
NGC 1399. There is no evidence for any fundamental variation in $\epsilon_{\rm
cl}$ from the total populations of these two GCSs, and the local
$\wepsilon_{\rm cl}$ in each is independent of galactocentric radius beyond
$\simeq1\,R_{\rm eff}$. These results lend some weight to the claim (\S2.2)
that the first and second specific frequency problems seem likely to appear,
and should be solved, only together.

Both of these problems are genuine in M87, as the top panel of
Fig.~\ref{fig3} attests. (The surface photometry of \markcite{dvn78}de
Vaucouleurs \& Nieto 1978, which is the standard, is well sampled and is
supported by the independent studies of \markcite{oem76}Oemler 1976 and
\markcite{kin78}King 1978; and the total $M_V$ used to calculate the
galaxy's $S_N$ is taken from integration of its $\mu_B$ profile.) The
second specific frequency problem is evidenced by the shallower decline of
$\Sigma_{\rm cl}$ compared to $\Sigma_{\rm stars}$ at $R_{\rm gc}\ga R_{\rm
eff}\simeq7$ kpc. This also causes the first problem, since scaling the total
GCS density upwards by the factors appropriate for M49 or NGC 1399 obviously
puts $\Sigma_{\rm cl}$ well in excess of $\Sigma_{\rm stars}$ throughout most
of the galaxy. Indeed, the most recent estimate for the global $S_N$ of M87 is
$14.1\pm1.6$ (\markcite{har98}Harris et al.~1998), which is fully a factor of
three higher than that of M49. Here it is necessary to consider the
distribution of hot gas---which is known to exist in great quantities around
M87---in evaluating $\wepsilon_{\rm cl}$ from equation (\ref{eq:210}).

\subsection{Volume Densities}

The volume density profiles of circumgalactic and intracluster gas are
routinely derived in the course of X-ray studies of elliptical galaxies. To
include such a halo component in this discussion, it is first necessary either
to deproject observed GCS and stellar surface densities, or to compute the gas
surface density via a standard projection integral. In general, the first of
these options potentially has the most insight to offer, since projected
quantities always mix information from a wide range of physical radii.

An Appendix therefore develops a nonparametric
deprojection algorithm which is explicitly designed to handle GCS data that
have been binned in projected radius, $R_{\rm gc}$ (i.e., counted in circular
annuli). This procedure returns a coarse volume density profile---$n_{\rm cl}$
vs.~three-dimensional $r_{\rm gc}$---expressed as a series of average
densities in a number of concentric spherical shells. The algorithm is
preferably applied to surface density profiles that extend to fairly large
radii (so that corrections for cluster populations beyond the observed field
of view can be estimated with confidence, and are as small as possible), and
it makes most sense for $N_{\rm cl}$ distributions that are monotonically
decreasing functions of $R_{\rm gc}$ (otherwise, the derived $n_{\rm cl}$ can
be zero or negative in some shells). After re-binning to smooth some of the
original data, as described above, the composite GCS radial distributions in
Tables \ref{tab2} through \ref{tab4} satisfy each of these criteria. Equation
(\ref{eq:37}) of the Appendix has therefore been applied to each of these
$N_{\rm cl}$ profiles, with $f(R_{i-1},R_i)$ (a sort of ``residual background''
correction) calculated by assuming that $n_{\rm cl}\propto r_{\rm gc}^{-3}$
beyond the last observed radius of each GCS. (This reflects the fact that the
projected $N_{\rm cl}$ shows roughly an $R_{\rm gc}^{-2}$ behavior at large
$R_{\rm gc}$ in each case. In any event, only the outermost one or two points
in each deprojected profile are affected---and then not greatly---by this
choice.) The resulting $n_{\rm cl}$ are listed in the fifth columns of Tables
\ref{tab2}, \ref{tab3}, and \ref{tab4}. The radii in the first columns, which
delimit projected annuli in reference to the surface density data, are
identified with the inner and outer three-dimensional radii of spherical
shells in reference to the volume densities.

\begin{figure*}[tb]
\centering \leavevmode
\epsfysize=4.0truein
\epsfbox{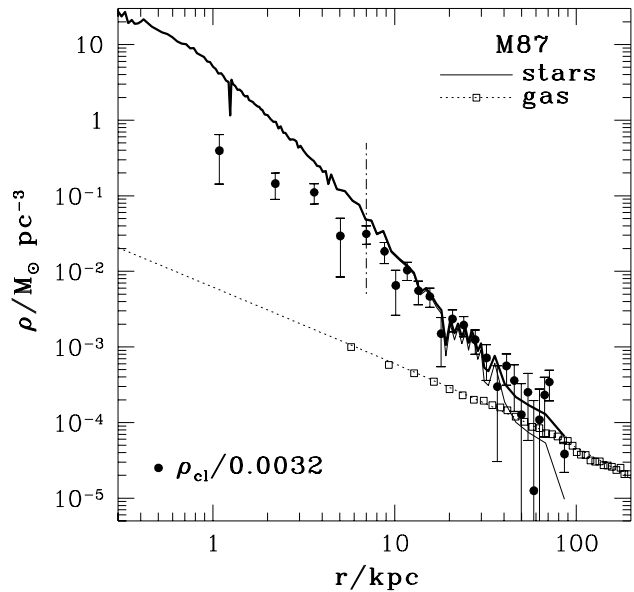}
\caption{Comparison of the (mass) volume density profiles of the
GCS (filled circles), halo stars (thin solid line), and X-ray gas (open squares
and broken line) in M87. The volume densities of globular clusters and field
stars follow from deprojections of the surface density profiles in
Fig.~\ref{fig3}. To deproject the galaxy surface photometry, it is assumed
that $\rho_{\rm stars}\propto r_{\rm gc}^{-4}$ at large radii $r_{\rm gc}\ga
100$ kpc (see eqs.~[\ref{eq:36}], [\ref{eq:37}], and [\ref{eq:a3}] of the
Appendix). The vertical line segment
is placed at the projected effective radius of the stellar light, as in
Fig.~\ref{fig3}. Thick black line is the sum of the stellar and gas densities.
Note that $\rho_{\rm cl}$ is directly proportional to $(\rho_{\rm stars}+
\rho_{\rm gas})$ beyond $r_{\rm gc}\simeq R_{\rm eff}$, giving a constant
estimated cluster formation efficiency of $\wepsilon_{\rm cl}=0.0032\pm0.0005$
there. The sharp feature at $r_{\rm gc}\simeq20$ kpc marks the onset of the cD
envelope in the stars and GCS of M87; it is visible as a smoother ``hump'' in
projection (de Vaucouleurs \& Nieto 1978; McLaughlin et al.~1993).
\label{fig4}}
\end{figure*}

The deprojection algorithm is an iterative one: the volume density in a
spherical shell with some mean radius is computed by first subtracting, from
the projected density at that radius, the contribution of any globulars that
actually lie at larger physical $r_{\rm gc}$. Thus, the uncertainties of the
derived $n_{\rm cl}$ at any two radii in one system are not independent. The
(1-$\sigma$) errorbars quoted in Tables \ref{tab2}, \ref{tab3}, and \ref{tab4}
are the standard deviations of results from 1000 different deprojections in
each case. Each trial deprojection in each GCS begins with a different $N_{\rm
cl}$ profile, in which the surface density for any given annulus is drawn at
random from a Gaussian distribution that is centered on the value given in
Column 3 of the appropriate Table, and that has a dispersion equal to the
observational uncertainty there.

\begin{figure*}[tb]
\centering \leavevmode
\epsfysize=4.0truein
\epsfbox{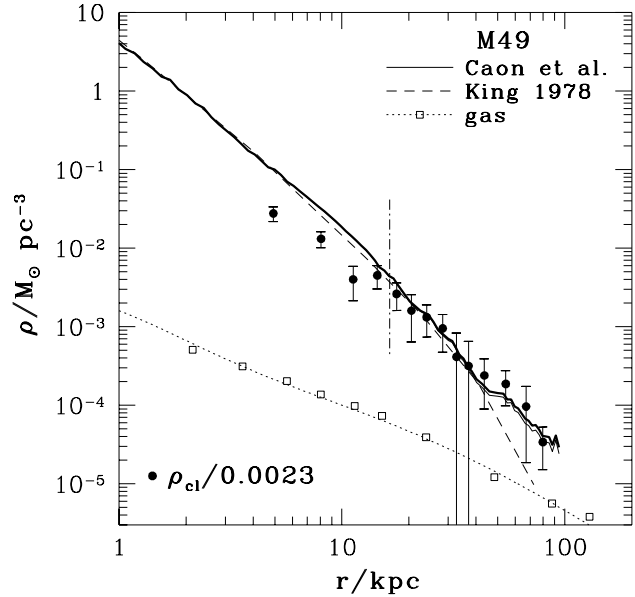}
\caption{Same as Fig.~\ref{fig4}, but for M49. Two separate
stellar density profiles are shown in this case, derived by deprojecting the
surface brightness profiles of King (1978; dashed line) and of Caon et
al.~(1994; thin solid line). The projected half-light radius of the latter
is marked with a vertical line. For both deprojections, it is assumed that
$\rho_{\rm stars}\propto r_{\rm gc}^{-2}$ at large radii. The bold line is
the sum of $\rho_{\rm stars}$ (from the profile of Caon et al.) and $\rho_{\rm
gas}$ (open squares and dotted line). At radii $r\ga R_{\rm eff}$, the ratio
$\rho_{\rm cl}/(\rho_{\rm stars}+\rho_{\rm gas})$ is again constant with
radius, giving a best estimate of $\wepsilon_{\rm cl}=0.0023\pm 0.0005$. This
is within 30\% of the value found in M87, and the two are identical within
their errors.
\label{fig5}}
\end{figure*}

\begin{figure*}[tb]
\centering \leavevmode
\epsfysize=4.0truein
\epsfbox{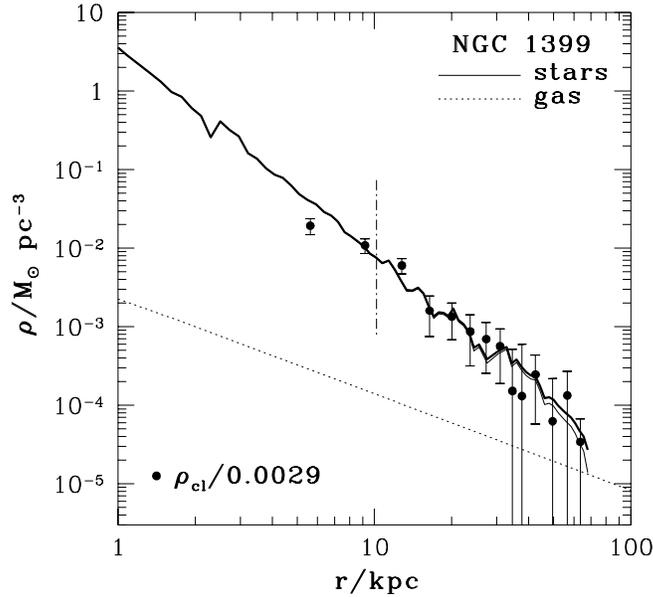}
\caption{Same as Figs.~\ref{fig4} and \ref{fig5}, but for NGC
1399. Filled circles are again the GCS densities; thin solid line, the
stellar densities (obtained by deprojection of $B$-band surface brightnesses,
correcting for $\rho_{\rm stars}\propto r_{\rm gc}^{-2}$ beyond about 70 kpc);
dotted line, the gas density; and bold solid line, the sum $(\rho_{\rm gas}+
\rho_{\rm stars})$. As before, the vertical line is placed at the stellar
$R_{\rm eff}$. The estimated cluster formation efficiency is again
independent of galactocentric radius: $\wepsilon_{\rm cl}=0.0029\pm0.0008$.
\label{fig6}}
\end{figure*}

The tabulated number densities $n_{\rm cl}$ have the same limiting magnitudes
as the corresponding surface densities, and are therefore multiplied by the
same two factors (so effectively scaling to $V_{\rm lim}=\infty$, and applying
the mean globular cluster mass $\langle m\rangle_{\rm cl}=2.4\times10^5\,
M_\odot$) to find the total GCS mass densities $\rho_{\rm cl}(r_{\rm gc})$.
These distributions are plotted as the filled circles in Figs.~\ref{fig4},
\ref{fig5}, and \ref{fig6}, where the density in each spherical shell has been
put at a radius $\overline{r}_{\rm gc}$, given in Columns 4 of Tables
\ref{tab2}--\ref{tab4},
calculated according to equation (\ref{eq:39}). As in Fig.~\ref{fig3}, the
$\rho_{\rm cl}$ profiles have been scaled up by factors of a few hundred in
order to compare them directly with the stellar mass densities $\rho_{\rm
stars}(r_{\rm gc})$ (thin solid lines in Figs.~\ref{fig4} to \ref{fig6}).
These have been obtained by applying the algorithm of the Appendix to the
projected profiles $\Sigma_{\rm stars}(R_{\rm gc})$ derived above from
$B$-band surface photometry (see eq.~[\ref{eq:32}]).\footnotemark
\footnotetext{All of the surface photometry used here has been averaged in
circular annuli, as is required by the deprojection method in the Appendix.
[Effective circular radii $R_{\rm gc}=(ab)^{1/2}$ are assigned to elliptical
isophotes with semi-major and -minor axes $a$ and $b$.] Note that these data
are generally published just as tables of $\mu_B$ vs.~$R_{\rm gc}$, with no
information given on the inner and outer radii of any annuli used in the
measurements. Thus, to apply equation (\ref{eq:37}), annuli have been defined
in an ad hoc manner, by setting boundaries at the geometric means $(R_1 R_2)
^{1/2}$ of successive pairs of projected radii in the published data. Data
points are generally spaced so closely as to make the difference between this
and any other choice unimportant.}
It is immediately apparent that the good agreement seen in Fig.~\ref{fig3}
between the GCS and stellar densities in NGC 1399 and M49 (for the data of
\markcite{cao94}Caon et al.~1994) persists after deprojection, as should be
the case. More noteworthy is the fact that the discrepancy between $\rho_{\rm
cl}$ and $\rho_{\rm stars}$ in M87, while certainly still present,
is somewhat less pronounced than that between $\Sigma_{\rm cl}$ and $\Sigma_
{\rm stars}$. For example, the truly local picture given by Fig.~\ref{fig4}
shows that the second $S_N$ problem here is confined to galactocentric radii
$r_{\rm gc}\ga30$--40 kpc, and does not extend inward to the $\sim15$ kpc
scales suggested by Fig.~\ref{fig3}. This just reflects the fact that the
projection of any two volume densities inevitably amplifies any small
differences that may actually exist between them, and artificially associates
disparities at large $r_{\rm gc}$ with smaller projected radii. A similar
comment applies to the comparison of $\rho_{\rm cl}$ with $\rho_{\rm stars}$
in M49: it is now clear, from Fig.~\ref{fig5}, that any argument for the
second $S_N$ problem in this galaxy could only be made for three-dimensional
radii greater than about 50 kpc, and then (as in \S3.1) only if
\markcite{kin78}King's (1978) $\mu_B$ profile---which includes just {\it a
single data point} at such large $r_{\rm gc}$---were preferred over that of
\markcite{cao94}Caon et al.~(1994).

The gas densities in these systems have been obtained from published {\it
electron} number densities by way of the relation
\begin{equation}
{{\rho_{\rm gas}}\over{M_\odot\,{\rm pc}^{-3}}}=
2.865\times10^{-5}\,\left({\mu\over{0.60}}\right)\left({{n_e}\over{10^{-3}\,
{\rm cm}^{-3}}}\right)\ ,
\label{eq:310}
\end{equation}
where a mean particle mass $\mu\simeq0.6$ (in units of $m_H$) is equally
appropriate for hot ($T\sim10^7$--$10^8$ K) and highly ionized plasmas of any
composition from primordial to solar.

The open squares for the M87 gas in Fig.~\ref{fig4} come from scaling the
electron densities of \markcite{nul95}Nulsen \& B\"ohringer (1995) for the
different Virgo distance assumed here (15 Mpc vs.~their 20), according to
$n_e\propto D^{-1/2}$. (These data were published as volume densities after
Nulsen \& B\"ohringer applied a deprojection algorithm, similar to that
derived in the Appendix, to their {\it ROSAT} spectra.) The dotted line in
Fig.~\ref{fig4} is a fit taken from \markcite{mcl99}McLaughlin (1999):
\begin{equation}
\rho_{\rm gas}({\rm M87})=5.60\times10^{-6}\,M_\odot\,{\rm pc}^{-3}\,
\left({r_{\rm gc}\over{1070\,{\rm kpc}}}\right)^{-1}
\left(1+{r_{\rm gc}\over{1070\,{\rm kpc}}}\right)^{-3}\ ,
\label{eq:311}
\end{equation}
so that, essentially, $\rho_{\rm gas}\propto r_{\rm gc}^{-1}$ over the area of
interest here. This fit has been used to compute the total $\rho_{\rm gas}+
\rho_{\rm stars}$ at every radius with a measured stellar density; that sum is
drawn as the bold solid line in the Figure.

Similarly, the open squares in Fig.~\ref{fig5} here use the $n_e$ read from
\markcite{irw96}Irwin \& Sarazin's (1996) Figure 6, after scaling to $D=15$
Mpc from their assumed M49 distance of 25.8 Mpc. The dotted line in this
Figure is a scaled fit (from \markcite{bri98}Brighenti \& Mathews 1998) to
these data plus those of \markcite{tri86}Trinchieri et al.~(1986):
\begin{equation}
\rho_{\rm gas}({\rm M49})=\left[
{{3.57\times10^{-3}}\over{{1+(r_{\rm gc}/0.807\,{\rm kpc})^2}}}\,+\,
{{2.24\times10^{-4}}\over{{1+(r_{\rm gc}/7.18\,{\rm kpc})^{1.14}}}}\,-\,
{{1.50\times10^{-5}}\over{{1+(r_{\rm gc}/75.6\,{\rm kpc})^{1.19}}}}\right]
\ \ M_\odot\,{\rm pc}^{-3}\ .
\label{eq:312}
\end{equation}
The heavy black line again represents the sum of $\rho_{\rm gas}+\rho_{\rm
stars}$.

The broken line for $\rho_{\rm gas}$ in NGC 1399 (Fig.~\ref{fig6}) comes from
\markcite{tsi93}Tsai's (1993) fit to the X-ray observations of
\markcite{kil88}Killeen \& Bicknell (1988), corrected for $D({\rm Fornax})=
16.5$ Mpc:
\begin{equation}
\rho_{\rm gas}({\rm NGC\ 1399})=1.03\times10^{-2}\,M_\odot\,{\rm pc}^{-3}\,
\left[1+\left({r_{\rm gc}\over{0.304\,{\rm kpc}}}\right)^2\right]^{-0.615}\ .
\label{eq:313}
\end{equation}

It is clear that stars outweigh the gas everywhere in M49 and NGC 1399:
locally, $\rho_{\rm gas}/\rho_{\rm stars}\la0.3$ even at the largest radii in
Figs.~\ref{fig5} and \ref{fig6}; globally, $M_{\rm gas}/M_{\rm stars}\sim0.05$
within $r_{\rm gc}=100$ kpc and 70 kpc respectively. In both
systems, then, $M_{\rm gcs}/(M_{\rm gas}+M_{\rm stars})\simeq M_{\rm gcs}/
M_{\rm stars}$ and (albeit to slightly lower accuracy at very large $r_{\rm
gc}$) $\rho_{\rm cl}/(\rho_{\rm gas}+\rho_{\rm stars})\simeq\rho_{\rm cl}/
\rho_{\rm stars}$. It is indeed valid to use standard specific frequencies as
direct measures of globular cluster formation efficiencies in these two cases,
as was done at the end of \S3.1 above. To be complete, however, all of the
stars, gas, and globular clusters can now be used to estimate $\epsilon_{\rm
cl}$ more precisely, according to the basic equation (\ref{eq:210}). By
summing the data at large $r_{\rm gc}$ in each of M49 and NGC 1399, then,
average $\wepsilon_{\rm cl}$ values are obtained:
\begin{equation}
{{M_{\rm gcs}(r_{\rm gc}>12.5\,{\rm kpc})}\over{M_{\rm gas}(r_{\rm gc}>12.5\,
{\rm kpc})+M_{\rm stars}(r_{\rm gc}>12.5\,{\rm kpc})}}=
0.0023\pm0.0005\ \ \ \ \ \ {\rm (M49)}
\label{eq:314}
\end{equation}
and
\begin{equation}
{{M_{\rm gcs}(r_{\rm gc}>11\,{\rm kpc})}\over{M_{\rm gas}(r_{\rm gc}>11\,{\rm
kpc})+M_{\rm stars}(r_{\rm gc}>11\,{\rm kpc})}}=
0.0029\pm0.0008\ \ \ \ \ \ {\rm (NGC\,1399)}\ ,
\label{eq:315}
\end{equation}
which account for the scalings applied to $\Sigma_{\rm cl}$ and $\rho_{\rm cl}$
in Figs.~\ref{fig3}, \ref{fig5}, and \ref{fig6}. As has already been discussed,
these are equally well interpreted as either local or global formation
efficiencies, since the ratio $\rho_{\rm cl}/(\rho_{\rm gas}+\rho_{\rm stars})$
shows no variations with $r_{\rm gc}$ beyond a stellar effective radius
(or even inside this, in NGC 1399), i.e., outside those regions of the GCSs
which could be significantly dynamically evolved. (As in Fig.~\ref{fig3},
the stellar $R_{\rm eff}$ are marked with vertical broken lines in
Figs.~\ref{fig4}, \ref{fig5}, and \ref{fig6}.)

The situation is somewhat different for M87: referring to Fig.~\ref{fig4},
$\rho_{\rm gas}/\rho_{\rm stars}>1$ at all radii $r_{\rm gc}\ga40$ kpc (which
is where the second $S_N$ problem begins in the deprojected data), and $M_{\rm
gas}/M_{\rm stars}\simeq0.35$--$0.40$ globally (i.e., integrated over scales
$r_{\rm gc}\leq100$ kpc). The hot gas around M87 is therefore an important
factor in calculating its cluster formation efficiency. As before, outside of
the central regions of the galaxy it is found that
\begin{equation}
{{M_{\rm gcs}(r_{\rm gc}>11\,{\rm kpc})}\over{M_{\rm gas}(r_{\rm gc}>11\,
{\rm kpc})+M_{\rm stars}(r_{\rm gc}>11\,{\rm kpc})}}=
0.0032\pm0.0005\ \ \ \ \ \ {\rm (M87)}\ ,
\label{eq:316}
\end{equation}
explaining the scaling factor applied to the M87 GCS densities in
Figs.~\ref{fig3} and \ref{fig4}. Figure \ref{fig4} shows that this average
value is valid also as the local $\wepsilon_{\rm cl}=\rho_{\rm cl}/(\rho_{\rm
gas}+\rho_{\rm stars})$ at any galactocentric position beyond about $R_{\rm
eff}=7$ kpc; as in M49 and NGC 1399, the inferred globular cluster formation
efficiency in M87 was independent of $r_{\rm gc}$ at large radii. In this
case, however, if the gas were not included in the calculation, much
larger---and radially varying---formation efficiencies would have been
indicated: for example, $M_{\rm gcs}(r_{\rm gc}>11\,{\rm kpc})/M_{\rm stars}
(r_{\rm gc}>11\,{\rm kpc})=0.0056\pm0.0008$, and $M_{\rm gcs}(r_{\rm gc}>40\,
{\rm kpc})/M_{\rm stars}(r_{\rm gc}>40\,{\rm kpc})=0.0095\pm0.0023$. Note that
the first of these ratios, $M_{\rm gcs}/M_{\rm stars}=0.0056$ for $r_{\rm gc}>
11$ kpc, is a factor of 2.4 higher than the corresponding quantity in M49. In
combination with the different stellar mass-to-light ratios in these two
galaxies ($\Upsilon_B=14.6$ vs.~10.4; see \markcite{vdm91}van der Marel 1991,
or Fig.~\ref{fig3} above), this just corresponds to the factor of 3 difference
observed in their global $S_N$ values. The high local and global specific
frequencies in M87 are thus seen to result from the presence of a population
of globular clusters that are intimately associated with the hot gas there.

\subsection{Gas Surface Densities}

As was mentioned earlier, an alternate comparison of a galaxy's gas
distribution with its stars and globular clusters could make use of only
surface densities---$\Sigma_{\rm stars}$, $\Sigma_{\rm cl}$, and a derived
$\Sigma_{\rm gas}$---as functions of projected $R_{\rm gc}$. The advantage of
such an approach is that $\Sigma_{\rm gas}$ is on an equal physical footing
with $\Sigma_{\rm stars}$ and $\Sigma_{\rm cl}$, and the latter are basically
unmanipulated data with smaller relative uncertainties than the derived
$\rho_{\rm stars}$ and $\rho_{\rm cl}$ profiles. The cost, aside from
possible uncertainties in any extrapolations required to obtain
$\Sigma_{\rm gas}$ from $\rho_{\rm gas}$, is that differences between
projected distributions are not easily related to local effects at some true,
three-dimensional radius (e.g., recall the comparison between projected
$\Sigma_{\rm cl}$ vs.~$\Sigma_{\rm stars}$ and deprojected $\rho_{\rm cl}$
vs.~$\rho_{\rm stars}$ in M87). More to the point, if the ratio $\Sigma_{\rm
cl}/(\Sigma_{\rm gas}+\Sigma_{\rm stars})$ varies as a function of $R_{\rm gc}$
in a galaxy, then variations in the physical quantity of real interest, the
local $\wepsilon_{\rm cl}=\rho_{\rm cl}/(\rho_{\rm gas}+\rho_{\rm stars})$,
are clearly indicated, but {\it cannot be properly quantified} without
resorting to a deprojection of some sort. For this reason, the procedure of
\S3.2 must be favored as the best way, generally speaking, to measure local
$\wepsilon_{\rm cl}$.

That said, however, in the galaxies studied here, $\wepsilon_{\rm cl}$ is found
to be independent of radius at large $r_{\rm gc}$, i.e., $\rho_{\rm cl}\propto
(\rho_{\rm gas}+\rho_{\rm stars})$. In this specific instance, it should then
also happen that $\Sigma_{\rm cl}\propto(\Sigma_{\rm gas}+\Sigma_{\rm stars})$
at large $R_{\rm gc}$, with a constant of proportionality that is exactly
$\wepsilon_{\rm cl}$. It is therefore useful to compute $\Sigma_{\rm gas}$ and
the projected analogue of $\wepsilon_{\rm cl}$ in M87, M49, and NGC 1399, in
order both to verify the results of \S3.2 (which depend in good part on the
behavior of $\rho_{\rm cl}$ at large $r_{\rm gc}$, where the deprojected data
are noisiest) and to obtain potentially higher-precision estimates of the
cluster formation efficiencies in these systems.

First, then, the projected mass densities of gas are obtained from the
volume density profiles in equations (\ref{eq:311}), (\ref{eq:312}), and
(\ref{eq:313}), by way of the standard integral
\begin{equation}
\Sigma_{\rm gas}(R_{\rm gc})=2\int_0^{\sqrt{r_{\rm max}^2-R_{\rm gc}^2}}
\rho_{\rm gas}(r_{\rm gc})\,dz\ ,
\label{eq:317}
\end{equation}
where the coordinate $z=(r_{\rm gc}^2-R_{\rm gc}^2)^{1/2}$ measures distance
along the line of sight, and  $r_{\rm max}$ represents the total physical
extent of the X-ray corona. For M87, this is taken to be the virial radius of
the Virgo Cluster: $r_{\rm max}\simeq1.5$ Mpc (see \markcite{mcl99}McLaughlin
1999). For M49, which is not at the center of Virgo, $r_{\rm max}=500$ kpc
is adopted. NGC 1399 lies at the center of the Fornax Cluster, which is
somewhat less massive than Virgo (e.g., \markcite{jon97}Jones et al.~1997),
and $r_{\rm max}$ is set at 1 Mpc there. There is a certain arbitrariness to
each of these choices for $r_{\rm max}$, but any of them can be changed by a
factor of a few without affecting the conclusions which follow.

\begin{figure*}[tb]
\centering \leavevmode
\epsfysize=4.0truein
\epsfbox{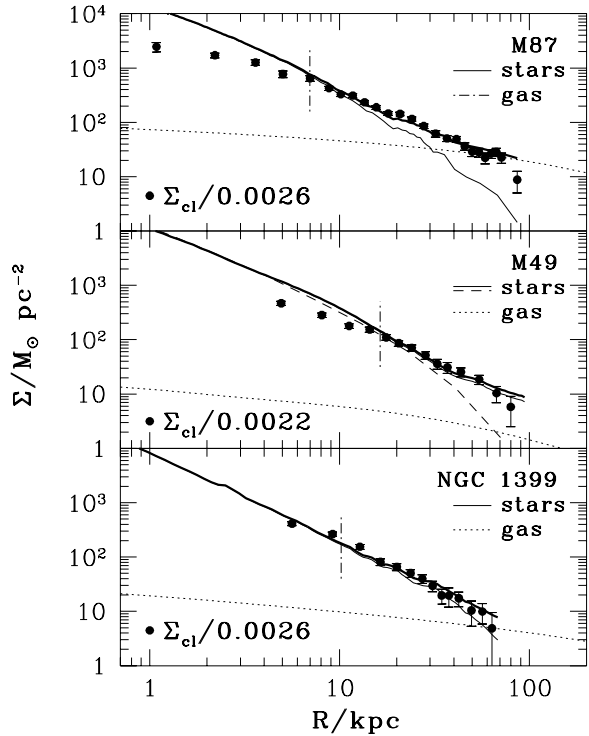}
\caption{Comparison of GCS, stellar, and gas {\it surface} densities in
M87, M49, and NGC 1399. The bold solid line in each panel traces the sum
$\Sigma_{\rm gas}+\Sigma_{\rm stars}$, and the galaxies' effective radii are
marked as in Fig.~\ref{fig3}. Also as in Figs.~\ref{fig3} and \ref{fig5},
the solid and dashed lines for the stellar densities in M49 refer to the
surface photometry of Caon et al.~(1994) and King (1978), respectively. The
implied values of $\wepsilon_{\rm cl}$ are essentially the same as those
derived from the volume density profiles in all three galaxies, although the
$\simeq20\%$ lower value in M87 may be significant. The cluster formation
efficiency is again inferred to have been locally invariant in each of these
systems (at least beyond $R_{\rm gc}\simeq R_{\rm eff}$), and not significantly
different from one to the other.
\label{fig7}}
\end{figure*}

The resulting $\Sigma_{\rm gas}$ profiles are shown in Fig.~\ref{fig7}, along
with the same $\Sigma_{\rm stars}$ and $\Sigma_{\rm cl}$ profiles plotted in
Fig.~\ref{fig3} above (although the $\Sigma_{\rm cl}$ have been scaled up by
slightly different factors in the current Figure). It is easily seen that
$\Sigma_{\rm cl}\propto(\Sigma_{\rm gas}+\Sigma_{\rm stars})$ at large
$R_{\rm gc}$ in each galaxy, just as expected following the analysis
in \S3.2. Moreover, the constants of proportionality are in reasonable
agreement with the $\wepsilon_{\rm cl}$ derived from the deprojected data: by
taking the direct mean of the surface-density ratios beyond the galaxies'
effective radii (so as once again to avoid a zone of potential GCS erosion), it
is found that
\begin{equation}
\left\langle{{\Sigma_{\rm cl}}\over{\Sigma_{\rm gas}+\Sigma_{\rm stars}}}
\right\rangle_{R_{\rm gc}\ga R_{\rm eff}}\,=\,\left\{
\begin{array}{ll}
0.0026\pm0.0003\ , & \ \ {\rm M87} \\
0.0022\pm0.0004\ , & \ \ {\rm M49} \\
0.0026\pm0.0006\ , & \ \ {\rm NGC\,1399}\ .
\end{array}
\right.
\label{eq:318}
\end{equation}
The result for M49 is indistinguishable from that obtained in either \S3.1 or
\S3.2, owing to the truly low gas densities in this galaxy. Similarly, the
projected $\wepsilon_{\rm cl}$ in NGC 1399 is identical, within the
uncertainties, to the deprojected quantity $\rho_{\rm cl}/(\rho_{\rm gas}+
\rho_{\rm stars})=0.0029\pm0.0008$.
The number in equation (\ref{eq:318}) is formally 10\% smaller because of the
apparently large projected gas densities at radii $R_{\rm gc}\ga40$--50 kpc in
NGC 1399. However, these values are almost certainly overestimates: in deriving
$\Sigma_{\rm gas}$, the shallow volume density profile of equation
(\ref{eq:312}), which roughly has $\rho_{\rm gas}\propto r_{\rm gc}^{-1.2}$,
has been assumed to hold even on Mpc spatial scales in Fornax---well beyond
the range of actual X-ray observations---and this is likely unrealistic. On
the other hand, the projected $\wepsilon_{\rm cl}=0.0026\pm0.0003$ indicated
for M87 in Fig.~\ref{fig7} might genuinely be slightly smaller than the
$0.0032\pm0.0005$ inferred from deprojected quantities. The 20\% difference
between these two numbers is significant at just the 1-$\sigma$ level, however,
and---although it is interesting that the former is actually closer to the
values for M49 and NGC 1399---it is not obvious which is more reliable. The
X-ray gas density around M87 (eq.~[\ref{eq:311}]) has again been extrapolated
to unobserved radii to compute $\Sigma_{\rm gas}$, but in this case the basic
profile actually follows the dark matter distribution in Virgo
(\markcite{mcl99}McLaughlin 1999), and exhibits a realistic, steepening
slope at large $r_{\rm gc}$. Thus, although it is possible that $\Sigma_{\rm
gas}$ is overestimated at large $R_{\rm gc}$ in M87, any argument for this is
not as compelling as in NGC 1399.

The best measure of $\wepsilon_{\rm cl}$ in M87 will therefore be taken as
the error-weighted mean of the values (eqs.~[\ref{eq:316}] and [\ref{eq:318}])
obtained by analyzing the deprojected and projected data there; in M49, the
choice is of no consequence, but for definiteness the deprojected
$\wepsilon_{\rm cl}$ will be adopted; and the deprojected estimate will also
be adopted for NGC 1399, to avoid any concerns that its gas surface density
might have been overestimated at large $R_{\rm gc}$. Thus,
\begin{equation}
\wepsilon_{\rm cl}=\left\{
\begin{array}{ll}
0.0028\pm0.0004\ , & \ \ {\rm M87} \\
0.0023\pm0.0005\ , & \ \ {\rm M49} \\
0.0029\pm0.0008\ , & \ \ {\rm NGC\ 1399}\ .
\end{array}
\right.
\label{eq:319}
\end{equation}
Once again, these can be seen as estimates of either local or global cluster
formation efficiencies if the inner several kpc of M87 and M49 are excluded
from the discussion.

\section{A Universal Efficiency for Cluster Formation}

The results of the previous Section clearly show that {\it working in terms of
stellar masses rather than luminosities and including the gas in M87 removes
both the first and second specific frequency problems there}: in all three of
M87, M49, and NGC 1399, it is found that $\rho_{\rm cl}\propto
(\rho_{\rm gas}+\rho_{\rm stars})$, in identical proportions within the
uncertainties, outside of the central, possibly dynamically depleted regions
of the GCSs. (At the most, differences between the constants of proportionality
in M87 and M49 might remain at the $\sim$20\% level; but again, this is easily
taken up in the observational uncertainties, and in any case is a significant
improvement over the 300\% difference in their specific frequencies.) The
interpretation put to this is that globular clusters formed with an
essentially constant efficiency throughout all of these galaxies, at the level
of
\begin{equation}
\epsilon_{\rm cl}({\rm globular})\equiv
{{M_{\rm gcs}^{\rm init}}\over{M_{\rm gas}^{\rm init}}}\simeq
{{M_{\rm gcs}}\over{M_{\rm gas}+M_{\rm stars}}}=
0.0026\pm0.0005\ ,
\label{eq:41}
\end{equation}
which is the error-weighted mean of equation (\ref{eq:319}).

It should be kept in mind that these are all large galaxies with complex
evolutionary histories. Thus, as was discussed in \S2.1, what has been
measured here is necessarily a {\it mass-weighted average} of $\epsilon_{\rm
cl}$, possibly over multiple distinct episodes of in situ cluster formation
and/or of the accretion of globulars formed elsewhere.
Nevertheless, the uniformity of the observational estimates $\wepsilon_{\rm
cl}$ in these systems---both as a function of $r_{\rm gc}$ within each and
from one entire system to another---suggests that these concerns may not be
too severe, and that equation (\ref{eq:41}) may still serve as a reliable
guide to a single, possibly universal efficiency for cluster formation.
This impression can now be checked quantitatively, albeit to rather lower
precision than has proved possible with the excellent datasets for M87, M49,
and NGC 1399, by appealing to observations of many other GCSs. All indications
are that $\epsilon_{\rm cl}$ is indeed similar, at least to first order, from
galaxy to galaxy and from the protogalactic epoch to the present.

\subsection{Other Ellipticals and BCGs}

Given the definition for $\wepsilon_{\rm cl}$ (eq.~[\ref{eq:210}]), and
denoting by $G$ the global ratio $M_{\rm gas}/M_{\rm stars}$ in a galaxy, the
total population of any GCS may be written as
\begin{equation}
{\cal N}_{\rm tot}=2.92\times10^6\ \wepsilon_{\rm cl}\,(1+G)\,
\left({{L_{V,{\rm gal}}}\over{10^{11}\,L_\odot}}\right)\,
\left({{\Upsilon_V}\over{7\,M_\odot\,L_\odot^{-1}}}\right)\,
\left({{{\langle m\rangle}_{\rm cl}}\over{2.4\times10^5\,M_\odot}}\right)^{-1}
\ .
\label{eq:42}
\end{equation}
This can, of course, also be derived from equation (\ref{eq:22}) above,
which explicitly relates $S_N$ to $\wepsilon_{\rm cl}$ when $G=0$.

Now, in the general discussion of \S2, the stellar mass-to-light ratio was
taken to be roughly similar from galaxy to galaxy. This is not exactly true,
however: as was seen in \S3, for example, the recovery of similar
$\wepsilon_{\rm cl}$ in M87 and M49 from their significantly different global
and local $S_N$ depends on the $\simeq$40\% larger $\Upsilon_B$ of M87 as well
as on its greater gas mass. Thus, if other ellipticals are to be brought
properly into this discussion, it is necessary first to account for the fact
that $\Upsilon$ depends systematically on galaxy luminosity. This is a well
known result of many fundamental-plane analyses of bright ellipticals, and it
can be incorporated quantitatively in this analysis by referring to the study
of \markcite{vdm91}van der Marel (1991; and references therein). Specifically,
for a sample of 37 early-type systems, \markcite{vdm91}van der Marel obtains a
mean $R$-band mass-to-light ratio of $\Upsilon_R=3.32h_{50}\,M_\odot\,L_
\odot^{-1}$, corresponding to $\Upsilon_B=5.93h_{50}$, at a mean magnitude of
$M_B\simeq-22.22+5\,\log\,h_{50}$. For a galaxy color of $(B-V)=1.0$ (typical
of large ellipticals) and $H_0=70$ km s$^{-1}$ Mpc$^{-1}$, this result
corresponds to $\Upsilon_V=6.0$ at an absolute magnitude of $M_V\simeq-22.5$,
or $L_{V,{\rm gal}}=8.5\times10^{10}\,L_\odot$. In addition to this,
\markcite{vdm91}van der Marel (1991) finds that $\Upsilon$ increases with
$L_{\rm gal}$ as a power law with exponent $0.35\pm0.05$. Since other
studies of the fundamental plane have claimed a slightly shallower exponent of
0.25--0.30 (e.g., \markcite{fab87} Faber et al.~1987; \markcite{pah95}Pahre,
Djorgovski, \& de Carvalho 1995, and references therein), $\Upsilon_V\propto
L_{V,{\rm gal}}^{0.3}$ is adopted here. Thus,
\begin{equation}
{{\Upsilon_V}\over{M_\odot\,L_\odot^{-1}}}=6.3\left({{L_{V,{\rm gal}}}
\over{10^{11}\,L_\odot}}\right)^{0.3}\ .
\label{eq:43}
\end{equation}
It should be noted again that these mass-to-light ratios refer to the stellar
populations in the cores of the galaxies, and are not unduly influenced by any
dark matter on larger scales. If equation (\ref{eq:43}) is used in
(\ref{eq:42}), the dependence ${\cal N}_{\rm tot}\propto L_{V,{\rm gal}}^
{1.3}$, or $S_N\propto L_{V,{\rm gal}}^{0.3}$, emerges naturally for constant
$\wepsilon_{\rm cl}$ and $G$. This is essentially the scaling found
empirically by \markcite{kis97}Kissler-Patig (1997).

The other point of concern is, of course, the hot gas content of ellipticals
in general. To get a rough handle on this, notice that the (soft) X-ray
luminosities of early-type systems scale with their blue luminosities as
$L_X\propto L_B^x$, with $x\simeq2$--3 (and with significant scatter about the
mean trend: e.g., \markcite{for85}Forman et al.~1985; \markcite{brw98}Brown \&
Bregman 1998). This scaling can be related to the ratio $G=M_{\rm gas}/M_{\rm
stars}$ by noting that $L_X\propto \int \rho_{\rm gas}^2T_{\rm gas}^{1/2}\,dV
\propto M_{\rm gas}^2T_{\rm gas}^{1/2}R^{-3}$ for the bremsstrahlung emission
that dominates $L_X$ in this context. Also, the slope $x$ of the $L_X$--$L_B$
relation is the same if $L_V$ is used instead, so the implication is that $G^2
\propto R^3\, T_{\rm gas}^{-1/2} \Upsilon_V^{-2} L_V^{x-2}$. To go further,
$T_{\rm gas}$ bears some relation to the stellar velocity dispersion: $T_{\rm
gas}\propto \sigma^{2y}$, with $y\simeq0.7$--1.5 (\markcite{dav96}Davis \&
White 1996; \markcite{brw98}Brown \& Bregman 1998); the $V$-band fundamental
plane gives for the stellar quantities, $\sigma\propto L_V^{0.64}R^{-0.48}$ and
$R\propto L_V^{1.1}$ (e.g., de Carvalho \& Djorgovski \markcite{dec89}1989,
\markcite{dec92}1992; \markcite{pah95}Pahre et al.~1995); and, from equation
(\ref{eq:43}) just above, $\Upsilon_V\propto L_V^{0.3}$. Putting all of this
together, $G^2\propto L_V^{x+0.7-0.11y}$. Evidently, the uncertainties in $x$
and $y$ are not crippling to this analysis, as $G\propto L_V^{1.5\pm0.25}$ is
fairly representative of the range of possibilities. The scaling can be
normalized to apply specifically to brightest cluster galaxies (BCGs) by using
the observations of M87, which lies at the spatial and dynamical center of the
Virgo Cluster.\footnotemark
\footnotetext{The term BCG is used loosely here, to refer to the centrally
dominant galaxies in groups as well as in rich clusters, even when these are
not strictly ranked first by luminosity.}
As was mentioned in \S3.2, the global ($r_{\rm gc}\leq100$ kpc) ratio of gas
to stars in M87 is $\simeq0.4$ by mass. Since the $V$-band luminosity of the
galaxy is about $8\times10^{10}L_\odot$ (from \markcite{dvn78}de Vaucouleurs
\& Nieto 1978, for $D=15$ Mpc), it is then reasonable to set
\begin{equation}
G\approx0.55\left({{L_{V,{\rm gal}}}\over{10^{11}\,L_\odot}}\right)^{1.5}\ .
\label{eq:44}
\end{equation}
Putting this and equation (\ref{eq:43}) both into equation (\ref{eq:42}), and
holding $\wepsilon_{\rm cl}$ fixed, it can be seen that the mean trend
expected for ${\cal N}_{\rm tot}$ between $L_{V,{\rm gal}}=3\times10^{10}
L_\odot$ and $2\times10^{11}L_\odot$ (so between an implied $G$ of $\simeq$0.1
and 1.6) is roughly ${\cal N}_{\rm tot}\propto L_{V,{\rm gal}}^{1.75}$.
Despite the crudeness of the estimate of $G$ used here, the power 1.75 is just
that inferred from observational scalings presented by \markcite{har98}Harris
et al.~(1998) for a large sample of BCGs.

\begin{figure*}[tb]
\centering \leavevmode
\epsfysize=4.0truein
\epsfbox{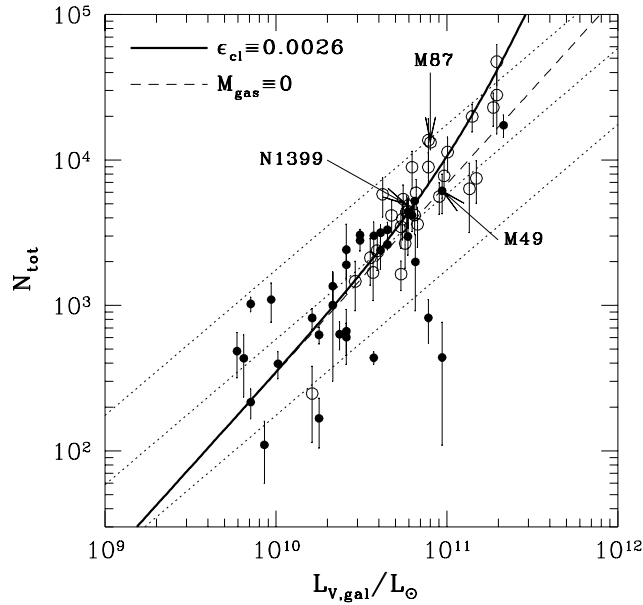}
\caption{Observed total GCS populations, ${\cal N}_{\rm tot}$,
vs.~total galaxy luminosity, $L_{V,{\rm gal}}$, for a sample of 65 giant
ellipticals and BCGs (filled and open circles; from
Harris et al.~1998, and references therein). Dotted lines have ${\cal N}_{\rm
tot}\propto L_{V,{\rm gal}}$, or constant $S_N=15$, 5, and 1.5. The systematic
departure of BCGs from $S_N\approx5$, at $L_{V,{\rm gal}}\ga3-5\times10^{10}
L_\odot$, is an illustration of the first specific frequency problem. The
bold, solid curve is the prediction of an assumed universal cluster formation
efficiency $\epsilon_{\rm cl}=0.0026$; the thin, dashed line is what that
prediction would be if all ellipticals were completely devoid of gas. The two
most extreme outliers are NGC 3557 and NGC 5018, each of which is relatively
isolated and has $L_V\sim10^{11}L_\odot$ ($M_V\simeq -22.5$) but $S_N<1$.
\label{fig8}}
\end{figure*}

The full relation between ${\cal N}_{\rm tot}$ and $L_{V,{\rm gal}}$ is drawn
as the heavy black line in Fig.~\ref{fig8}, where the total GCS populations of
65 early-type galaxies are plotted against their $V$ luminosities. These data
are the same as those plotted in Figure 11 of \markcite{har98}Harris et
al.~(1998), and the original references are given in that paper. The open
circles in Fig.~\ref{fig8} here denote BCGs in a large number of rich clusters
and a few poor groups; the filled symbols refer to other ellipticals, some in
clusters and some more isolated. The bold curve has $\wepsilon_{\rm cl}=M_{\rm
gcs}/(M_{\rm gas}+M_{\rm stars})=0.0026$, constant in all galaxies. Also
shown, as the dashed line, is the result of keeping this uniform
$\wepsilon_{\rm cl}$, but setting $G\equiv0$ in equation (\ref{eq:42}) while
still having $\Upsilon_V$ increase with luminosity as in equation
(\ref{eq:43}). The three dotted straight lines are loci of constant specific
frequency, ${\cal N}_{\rm tot}\propto L_{V,{\rm gal}}$, as given by equation
(\ref{eq:21}) for (from top to bottom in the Figure) $S_N=15$, 5, and 1.5.

It is immediately obvious why a roughly constant $S_N\simeq5$ is often assumed
for large ellipticals such as these. It is equally clear, however, that such a
simple characterization tends to overpredict ${\cal N}_{\rm tot}$ at low
$L_{V,{\rm gal}}$, and systematically underestimates it at high luminosities.
Accounting for $\Upsilon_V\propto L_{V,{\rm gal}}^{0.3}$ goes some of the way
towards explaining this while allowing for a universal formation efficiency
$\epsilon_{\rm cl}$; and referring the GCS populations in BCGs to the galaxies'
total stellar {\it plus gas} masses seems indeed to remove the rest of the
first specific frequency problem in general.

It is encouraging that these rather broad considerations lead to a
significantly improved description of the observed ${\cal N}_{\rm tot}$--$L_{V,
{\rm gal}}$ dependence, in which there are no obvious {\it systematic}
deviations from a uniform $\wepsilon_{\rm cl}\simeq0.0026$ (see also \S4.4).
There remains a good deal of scatter about the mean line in Fig.~\ref{fig8},
but this could well reflect departures of individual galaxies from the
fundamental plane, the mean $L_X$--$L_B$ (or $G$--$L_{V,{\rm gal}}$) relation,
or the assumed $(B-V)=1.0$, rather than any large variations in the fundamental
efficiency of cluster formation. Note, for example, that the observed
mass-to-light ratio of M87 is $\Upsilon_V\simeq10.5$, some 80\% higher than
that expected from equation (\ref{eq:43}). Correcting for this would bring the
$\wepsilon_{\rm cl}=0.0026$ curve in Fig.~\ref{fig8} just up to the M87 data
point there. To rather a lesser degree, some of the scatter could also arise
from GCS-to-GCS differences in the relative magnitude of any dynamical
reduction of $M_{\rm gcs}^{\rm init}$ over a Hubble time.

It should be noted that this explanation of the first $S_N$ problem differs
somewhat from those previously proposed (see \S2.2). In particular,
\markcite{bla97}Blakeslee (1997) and \markcite{btm97}Blakeslee et al.~(1997)
showed a correlation between the specific frequencies on 40-kpc scales in BCGs,
and the X-ray luminosities over much larger volumes ($R_{\rm gc}\le500$ kpc)
in their host clusters. These authors, along with \markcite{wes95}West et
al.~(1995) and \markcite{har98}Harris et al.~(1998), therefore interpret the
phenomenon of high specific frequency as one associated in some way with entire
clusters of galaxies. The analysis presented here, however, has attempted to
address the problem more locally, by referring to the gas and stellar masses
on similar spatial scales (typically $r_{\rm gc}\la100$ kpc) to estimate $G$
(eq.~[\ref{eq:44}]) in individual systems; nowhere has any appeal been made to
more global properties of galaxy clusters. This general approach is motivated,
of course, by the observed situation in M87 specifically (\S3). The correlation
that does exist between BCG $S_N$ and cluster-wide $L_X$ (and other such
relations; see the references cited) may simply stem from the fact that the
gas around BCGs in rich clusters blends smoothly into the intracluster medium
as a whole. Note that a relation of some kind between ``local'' $L_X$ and
${\cal N}_{\rm tot}$ is also implied by \markcite{san93}Santiago \& Djorgovski
(1993), although they attach no specific interpretation to it.

Interestingly, the bold line in Fig.~\ref{fig8} departs significantly from
linearity (i.e., from a simple power-law scaling between ${\cal N}_{\rm tot}$
and $L_{V,{\rm gal}}$) for $M_{\rm gas}/M_{\rm stars}\ga 0.1$--0.2,
corresponding to $L_V\ga3$--$5\times10^{10}L_\odot$ or $M_V\la-21.5$. This
magnitude is similar to that which appears to divide ellipticals into two broad
classes in terms of their kinematics, isophote shapes, and light distributions.
(Lower-luminosity ellipticals generally show rapid rotation, flattened and
disky isophotes, and steep cusps in their central densities, while brighter
systems are slowly rotating, with round and boxy isophotes and shallower
central density profiles; see, e.g., \markcite{dav83}Davies et al.~1983;
\markcite{ben89}Bender et al.~1989; \markcite{nie91}Nieto, Bender, \& Surma
1991; \markcite{tre96}Tremblay \& Merritt 1996; \markcite{geb96}Gebhardt et
al.~1996; \markcite{fab97}Faber et al.~1997.) This bears directly on the
recent discussion of \markcite{kis97}Kissler-Patig (1997; see also
\markcite{vdb98}van den Bergh 1998), who used observed trends in GCS specific
frequencies and radial distributions as further evidence for the empirical
``dichotomy'' of ellipticals. In particular, \markcite{kis97}Kissler-Patig
noted that the cluster systems in early-type galaxies with $M_V\ga-21.5$ tend
to have lower specific frequencies and steeper $N_{\rm cl}$ profiles than
those in brighter galaxies. Although he went on to argue for a discontinuous
change in the cluster formation efficiency for galaxies with $L_V$ below and
above $\approx5\times10^{10} L_\odot$, Fig.~\ref{fig8} now shows that all of
the data are well represented by a {\it continuously} increasing $S_N$ that
derives from a {\it constant} $\epsilon_{\rm cl}$ and a {\it smoothly
increasing} gas mass fraction, $G$; the nonlinearity intrinsic to the
resulting ${\cal N}_{\rm tot}$--$L_{V,{\rm gal}}$ relation just happens to
roughly mimic a break of sorts around $M_V\simeq-21.5$.

This interpretation can also account for another correlation suggested by
\markcite{kis97}Kissler-Patig (1997), namely, that $S_N$ appears to be lower
in (faint) ellipticals whose isophotes show disk-like deviations from pure
ellipses, and higher in those (brighter) galaxies with more box-like shapes.
These isophotal perturbations are themselves correlated with the
X-ray luminosities of ellipticals (\markcite{ben89}Bender et al.~1989), in the
sense required qualitatively to explain this observation of GCSs again in terms
of a constant $\epsilon_{\rm cl}$ and a systematically increasing
gas-to-stellar mass ratio. This raises the interesting question of which, of
gas content or isophote shape, is the more fundamental property of elliptical
galaxies; the present discussion, at least, would appear to favor the first
alternative.

Finally, although the slopes of GCS radial distributions cannot be addressed
directly by Fig.~\ref{fig8}, it seems natural that, given a fixed
$\epsilon_{\rm cl}$, ellipticals with large gas contents (and thus high global
$S_N$) should generally have shallower GCS density distributions, relative to
their stellar profiles, at large galactocentric radii (simply because
$\rho_{\rm gas}/\rho_{\rm stars}$ usually increases with $r_{\rm gc}$ in
early-type galaxies). As was suggested in \S2.2, then, the first and second
$S_N$ problems should be closely related, with $G=M_{\rm gas}/M_{\rm stars}$
the controlling factor in both. Again, this is consistent with the observed
trend (noted by \markcite{for97}Forbes et al.~1997;
\markcite{kis97}Kissler-Patig 1997; and \markcite{vdb98}van den Bergh 1998)
towards weaker dependences of $N_{\rm cl}$ on $R_{\rm gc}$ in brighter and
higher-$S_N$ galaxies. This proposal should, of course, be checked on a
case-by-case basis, as it was for M87 in \S3.2 above, in as many systems as
possible.

\subsection{Dwarf Ellipticals}

If the bold line in Fig.~\ref{fig8} is extrapolated into the regime of dwarf
ellipticals, $L_{V,{\rm gal}}\la2\times10^9L_\odot$, it clearly implies that
global $S_N<2$ should be observed in these faint galaxies. This is
completely at odds with the data of \markcite{dur96}Durrell et al.~(1996) and
\markcite{mil98}Miller et al.~(1998), who find an average $S_N\approx5$ for a
large sample of dE's in the Local Group, in other groups nearby, and in the
Virgo and Fornax Clusters. In addition, \markcite{mil98}Miller et al.~claim
evidence for an {\it increasing $S_N$ towards fainter $L_{V,{\rm gal}}$}.

This can be explained by first recalling that dwarf ellipticals do not fall on
the fundamental plane of bright galaxies; thus, the scalings (\ref{eq:43}) and
(\ref{eq:44}) do not apply to them. Also, these galaxies are bluer than their
giant counterparts, so that the conversion to $V$-band mass-to-light ratios
from measured $\Upsilon_R$, as described above, is again inapplicable; and in
any case, even the cores of dwarfs appear to be dominated by dark matter, so
that their dynamical $\Upsilon_V$ are not necessarily representative of the
stellar values. Most importantly---and likely at the heart of these other
points---dE's present sufficiently shallow potential wells that they should
have suffered significant amounts of mass loss during supernova-driven winds in
early bouts of star formation. Thus, even though these galaxies are gas-poor
at the present, the appropriate observable $\wepsilon_{\rm cl}=M_{\rm gcs}/
M_{\rm stars}$ {\it is not an accurate estimator of the true cluster formation
efficiency} $\epsilon_{\rm cl}=M_{\rm gcs}^{\rm init}/M_{\rm gas}^{\rm init}$
(see \markcite{dur96}Durrell et al.~1996 for the first suggestion that this
is an important consideration in understanding dwarf GCSs). Even still
assuming that the GCSs now have $M_{\rm gcs}\simeq M_{\rm gcs}^{\rm init}$,
allowance has to be made for an additional factor $M_{\rm stars}/M_{\rm gas}^
{\rm init}<1$. The relation analogous to equation (\ref{eq:42}) in this case
is therefore
\begin{equation}
{\cal N}_{\rm tot}\simeq1.7\times10^4\,
{{M_{\rm gcs}}\over{M_{\rm gas}^{\rm init}}}\,
\left({{M_{\rm gas}^{\rm init}}\over{M_{\rm stars}}}\right)\,
\left({{L_{V,{\rm gal}}}\over{2\times10^9\,L_\odot}}\right)\,
\left({{\Upsilon_V}\over{2\,M_\odot\,L_\odot^{-1}}}\right)\,
\left({{\langle m\rangle_{\rm cl}}\over{2.4\times10^5\,M_\odot}}\right)^{-1}\ ,
\label{eq:45}
\end{equation}
where $\Upsilon_V\simeq2$ should be a good representation of the stellar
populations in systems fainter than about $2\times10^9L_\odot$.

\begin{figure*}[tb]
\centering \leavevmode
\epsfysize=4.0truein
\epsfbox{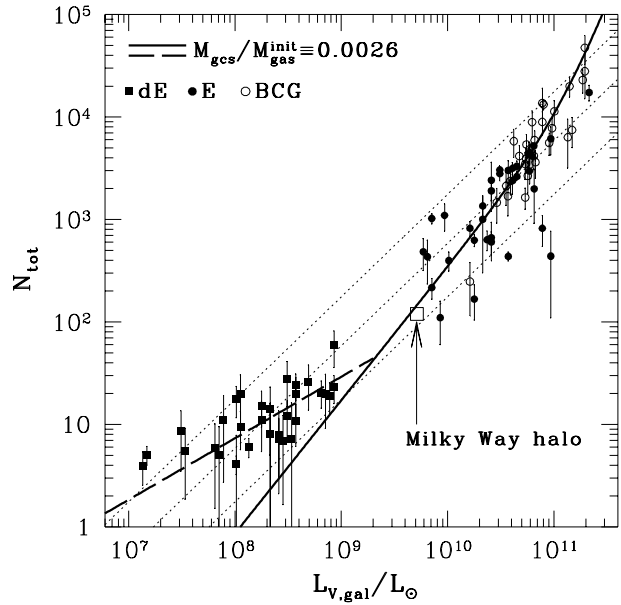}
\caption{${\cal N}_{\rm tot}$ vs.~$L_{V,{\rm gal}}$ for
early-type systems, now including 32 dwarf ellipticals with $L_{V,{\rm gal}}\la
2\times10^9L_\odot$ (solid squares; Durrell et al.~1996 and Miller et
al.~1998). Filled and open circles represent the same giant ellipticals and
BCGs plotted in Fig.~\ref{fig8}. The dotted lines of constant $S_N=15$, 5, and
1.5, and the heavy solid curve with $\epsilon_{\rm cl}\equiv0.0026$ are also
the same as those shown in Fig.~\ref{fig8}. The bold, dashed line is the
prediction for the same constant $\epsilon_{\rm cl}=M_{\rm gcs}^{\rm init}/
M_{\rm gas}^{\rm init}=0.0026$, after correcting for significant gas loss in
supernova-driven winds from dwarfs fainter than $M_V\simeq-18.5$. The large
open square is plotted at the observed $L_V$ and ${\cal N}_{\rm tot}$ of the
Milky Way {\it spheroid}; see \S4.3 of the text.
\label{fig9}}
\end{figure*}

The specific model of \markcite{dek86}Dekel \& Silk (1986), which considers
supernova-driven winds in dwarf galaxies that are dynamically dominated by
dark matter halos, can be used to evaluate the ratio of initial gas to present
stellar mass. That study finds that the winds from systems with velocity
dispersions less than about 100 km s$^{-1}$, corresponding to $M_V\ga-18.5$ at
the current epoch, should have expelled whatever gas remained after a single
burst of star formation. Moreover, as one moves towards shallower potential
wells and fainter $L_{V,{\rm gal}}$, the fractional mass lost, relative to the
initial total, is naturally expected to increase. The predicted scaling, in
the limit of very large mass loss and for a power spectrum similar to that in
a standard cold dark matter cosmology, is roughly $M_{\rm gas}^{\rm init}/
M_{\rm stars}\propto L_{\rm gal}^{-0.4}$. Taking the crude step of
extrapolating this all the way to the critical velocity dispersion (i.e., to
$L_{V,{\rm gal}}\simeq2\times10^9L_\odot$), and assuming that any gas driven
from brighter and more massive galaxies was a negligible fraction of the total,
the implication is that
\begin{equation}
{{M_{\rm gas}^{\rm init}}\over{M_{\rm stars}}}\simeq \left({{L_{V,{\rm gal}}
\over{2\times10^{9}\,L_\odot}}}\right)^{-0.4}\ ,\ \ \ \ \ 
L_{V,{\rm gal}}\le2\times10^9\,L_\odot\ .
\label{eq:46}
\end{equation}
Substituting this in equation (\ref{eq:45}), and assuming a constant stellar
$\Upsilon_V=2$, yields ${\cal N}_{\rm tot}\propto L_{V,{\rm gal}}^{0.6}$ and
$S_N\propto L_{V,{\rm gal}}^{-0.4}$ for dwarf ellipticals. Although the picture
of dE's as single-burst populations is now known to be incomplete (see, e.g.,
\markcite{mat98}Mateo 1998), this result is nevertheless consistent with the
observations of \markcite{mil98}Miller et al.~(1998).

Figure \ref{fig9} uses the data from \markcite{dur96}Durrell et al.~(1996)
and \markcite{mil98}Miller et al.~(1998) to place 32 nucleated and
non-nucleated dwarf ellipticals into the $L_{V,{\rm gal}}-{\cal N}_{\rm tot}$
plane with the brighter galaxies from Fig.~\ref{fig8}. (Of these, 16 are taken
from \markcite{dur96}Durrell et al., and 16 from \markcite{mil98}Miller et al.
Only GCSs detected at the $\ge1$-$\sigma$ level are used.)
The bold dashed line in the lower left of this plot comes from
equations (\ref{eq:45}) and (\ref{eq:46}), assuming a constant $\epsilon_{\rm
cl}=M_{\rm gcs}/M_{\rm gas}^{\rm init}=0.0026$. The heavy solid line is the
same as that in Fig.~\ref{fig8}---i.e., it also has $\epsilon_{\rm cl}=0.0026$,
but applies to fundamental-plane ellipticals---and the light, dotted lines
correspond, as before, to constant $S_N=15$, 5, and 1.5. Once again, all the
data appear consistent with a universal globular cluster formation efficiency.
It will be important to properly confirm this with detailed models for the
dE's in particular, and to flesh it out with observations of systems at
$L_{V,{\rm gal}}\sim(1-5)\times10^9L_\odot$ in general.

\subsection{Globular Clusters in the Milky Way}

One object that falls just inside this observational ``gap'' in luminosity is
the stellar {\it spheroid}---disk excluded---of the Milky Way. The total
$V$-band luminosity here has been estimated by \markcite{dvp78}de Vaucouleurs
\& Pence (1978) to be $L_V\simeq 5.1\times10^9 L_\odot$ [after applying a
color of $(B-V)=0.65$ to their $M_B=-18.8$]. At the same time, the total
number of globular clusters that are genuine halo objects, rather than part of
the  (thick) disk, is ${\cal N}_{\rm tot}\simeq110$--120. (Halo globulars are
identified on the basis of their metallicity, ${\rm [Fe/H]}\leq-0.80$; refer
back to Fig.~\ref{fig2} above.) This is the point plotted as the large open
square in Fig.~\ref{fig9}.

In fact, \markcite{dvp78}de Vaucouleurs \& Pence (1978) model the Galactic
spheroid with an $R^{1/4}$ density law with effective radius $R_{\rm eff}=
2.67$ kpc. Thus, in addition to predicting the absolute magnitude that would
be seen by an external observer, it is also possible to parametrize the density
profile of the entire stellar halo. To do this in three dimensions,
\markcite{you76}Young's (1976) deprojection of the $R^{1/4}$ law may be used:
\begin{equation}
{{\rho_{\rm stars}(r_{\rm gc})}\over{M_\odot\,{\rm pc}^{-3}}}=52.195\,
\left({{\Upsilon_B}\over{M_\odot\,L_\odot^{-1}}}\right)\,
\left({{L_B^{\rm tot}}\over{L_\odot}}\right)\,
\left({{R_{\rm eff}}\over{{\rm pc}}}\right)^{-3}\,
\exp\left[-7.669\left({{r_{\rm gc}}\over{R_{\rm eff}}}\right)^{1/4}\right]\,
\left({{r_{\rm gc}}\over{R_{\rm eff}}}\right)^{-7/8}\ ,
\label{eq:47}
\end{equation}
which is valid for $r_{\rm gc}\ga0.2 R_{\rm eff}$. Substitutions can be made
for $L_B^{\rm tot}$ and $R_{\rm eff}$ directly from the work of
\markcite{dvp78}de Vaucouleurs \& Pence. One way to estimate the $B$-band
mass-to-light ratio of the Galactic Population II is to adopt the same
$\Upsilon$--$L_B$ scaling used in \S4.1, but to normalize the relation by 
adopting solar colors rather than the redder ones that apply in ellipticals.
\markcite{vdm91}van der Marel's (1991) mean $\Upsilon_R=3.32h_{50}$ then
corresponds to $\Upsilon_B=4.65 M_\odot L_\odot^{-1}$ at $M_B=-21.49$ (for
$H_0=70$ km s$^{-1}$ Mpc$^{-1}$ assumed), and $\Upsilon_B\propto L_B^{0.3}$.
The implied $\Upsilon_B({\rm halo})\simeq2.2 M_\odot L_\odot^{-1}$ for
$M_B=-18.8$ is consistent with the mean $\Upsilon_V=2$ for globular clusters,
and with the $\Upsilon_V=1.75$ adopted for the spheroid by
\markcite{bah80}Bahcall \& Soneira (1980) in their model of the Galaxy.
Equation (\ref{eq:47}) is therefore evaluated here with $\Upsilon_B=2 M_\odot
L_\odot^{-1}$.

The dashed line in the bottom panel of Fig.~\ref{fig2} in \S3.1 above shows
this density profile, scaled down by a factor of
\begin{equation}
\langle \wepsilon_{\rm cl}\rangle_{\rm MW}=0.0027\pm0.0004\ .
\label{eq:48}
\end{equation}
This value is obtained as the mean of the ratio $\rho_{\rm cl}/\rho_{\rm
stars}$ over all galactocentric radii $2\le r_{\rm gc}\le 40$ kpc, and it
accounts for the position of the Milky Way halo in Fig.~\ref{fig9}. Such a
basic correspondence between our Galaxy and early Hubble types was previously
suggested, although developed from a somewhat different argument, by
\markcite{dev93}de Vaucouleurs (1993).

It is also evident from Fig.~\ref{fig2} that there is no second $S_N$ problem
in the Milky Way: $\rho_{\rm cl}\propto \rho_{\rm stars}$ obtains throughout
the entire halo at $r_{\rm gc}\ga3$ kpc---so again, beyond roughly an effective
radius, within which the original GCS might have suffered significant
dynamical depletion. This point has long been appreciated, although it is
usually discussed in the context of the fact that the same rough rule $\rho
\propto r_{\rm gc}^{-3.5}$ is inferred both for the spheroid, from counts of
RR Lyrae and horizontal-branch stars (e.g., \markcite{pre91}Preston, Schectman,
\& Beers 1991; \markcite{kin94}Kinman, Suntzeff, \& Kraft 1994), and for the
GCS, from simple power-law fits to its density profile (\markcite{har76}Harris
1976; \markcite{zin85}Zinn 1985; \markcite{djo94}Djorgovski \& Meylan 1994).
This power law is shown as the dotted line in the bottom panel of
Fig.~\ref{fig2}.

\subsection{Scatter in the $L_V$--${\cal N}_{\rm tot}$ Plane}

\begin{figure*}[tb]
\centering \leavevmode
\epsfysize=4.0truein
\epsfbox{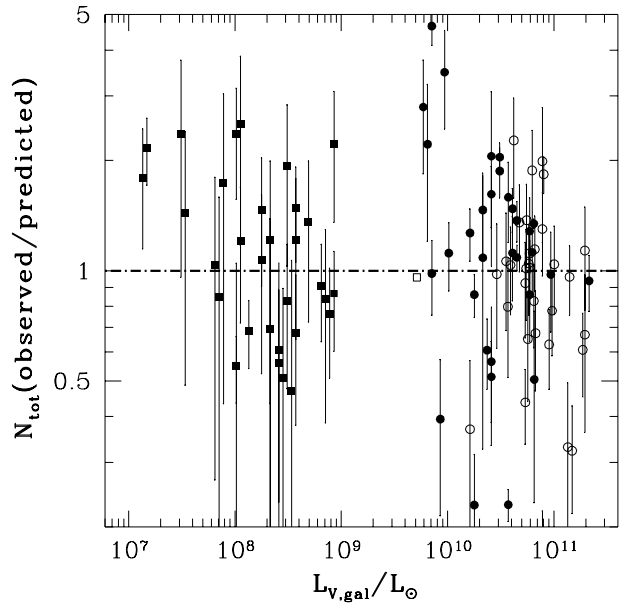}
\caption{Ratio of observed ${\cal N}_{\rm tot}$ to the predicted values
for all GCSs represented in Fig.~\ref{fig9}, using $\wepsilon_{\rm cl}=0.0026$
in equation (\ref{eq:42}) or $M_{\rm gcs}/M_{\rm gas}^{\rm init}=0.0026$ in
equation (\ref{eq:45}). Point types have the same meaning as in
Fig.~\ref{fig9}. The ratios for NGC 3557 and NGC 5018 are 0.07 and 0.17, and
do not appear on the scale of this plot.
\label{fig10}}
\end{figure*}

As a final test of the general viability of a single globular cluster
formation efficiency, Fig.~\ref{fig10} plots the ratio of observed to
predicted GCS ${\cal N}_{\rm tot}$ as a function of host galaxy luminosity for
all of the 98 systems discussed thus far. The predicted ${\cal N}_{\rm tot}$
for fundamental-plane ellipticals and the Milky Way spheroid (filled circles
and open square in the Figure) are given by equation (\ref{eq:42}), with the
gas mass fraction $G$ set identically to zero; for BCGs (open circles), by the
same relation but with $G$ as in equation (\ref{eq:44}); and for dE's (filled
squares), by equation (\ref{eq:45}). In all cases, it has been assumed that
$\epsilon_{\rm cl}=0.0026$. The median of the resulting ratios ${\cal R}\equiv
{\cal N}_{\rm tot}^{\rm obs}/{\cal N}_{\rm tot}^{\rm pred}$ is 1.04, and the
geometric mean is 0.98. Close to 80\% (77/98) of these galaxies have
$\frac{1}{2}\le{\cal R}\le2$.

As was also mentioned in \S4.1, but can perhaps be seen more clearly here than
in Fig.~\ref{fig8} or Fig.~\ref{fig9}, the deviations of observed ${\cal N}_
{\rm tot}$ from the uniform-$\epsilon_{\rm cl}$ prediction show no significant
dependence on either galaxy luminosity or Hubble type. There are
possibly two exceptions to this claim: (1) the GCS populations in the
brightest BCGs, $L_{V,{\rm gal}}\ga 10^{11}\,L_\odot$, appear to fall below
the values predicted with $\epsilon_{\rm cl}=0.0026$, by perhaps $\sim50\%$
on average; and (2) the two faintest dwarfs, which happen to be the Local
Group spheroidals Fornax and Sagittarius, both have ${\cal N}_{\rm tot}$
greater than predicted. These rough impressions may well not remain, in the
mean, after future additions of GCS data from other very high- and
low-luminosity galaxies; but even if they do, they need not imply the
existence of fundamental variations in $\epsilon_{\rm cl}$. Rather, there is
likely enough uncertainty in the adopted relation between gas mass fraction
and $L_{V,{\rm gal}}$ (eq.~[\ref{eq:44}]) to account for the first item; and
the second could reflect either a deficiency in the very simple correction
(\S4.2) for gas blow-out from very faint dwarfs, or the neglect of possibly
extensive post-formation dynamical evolution in such low-mass galaxies (such
as tidal stripping of their field stars; recall, for example, that Sagittarius
is in the process of being accreted by the Milky Way).

Even if these issues are ignored completely, a least-squares fit to all of the
data shown in Fig.~\ref{fig10} reveals an exceedingly weak dependence of
${\cal N}_{\rm tot}^{\rm obs}/{\cal N}_{\rm tot}^{\rm pred}$ on galaxy
luminosity: ${\cal R}\propto L_{V,{\rm gal}}^{-0.04\pm0.03}$. Fitting to
only the fundamental-plane ellipticals and BCGs yields a steeper
${\cal R}\propto L_{V,{\rm gal}}^{-0.2}$ or so, but this result appears to
be driven by just three gE's with anomalously low or high GCS populations
(cf.~Fig.~\ref{fig8}): NGC 3557, with $L_V=9.4\times10^{10}\,L_\odot$ and
$S_N=0.4\pm0.3$; NGC 5018, which has $L_V=7.8\times10^{10}\,L_\odot$ and
$S_N=0.9\pm0.3$; and NGC 4278, with $L_V=7.1\times10^9\,L_\odot$ but $S_N=
12.3\pm1.4$. If these systems are excluded, then ${\cal R}({\rm gE + BCG})
\propto L_{V,{\rm gal}}^{0.02\pm0.04}$ is indicated instead. Similarly,
the slopes of ${\cal R}$ vs.~$L_{V,{\rm gal}}$ do not differ significantly
from 0 for any of the individual Hubble-type samples represented in
Fig.~\ref{fig10}.

Thus, current observations of total GCS populations do show some scatter,
and it will be important to understand---likely through detailed analyses,
along the lines of \S3, of more systems {\it individually}---whether or not
this reflects an intrinsic scatter in the underlying globular cluster formation
efficiency (see also \S4.1). However, there is no clear evidence at present
for any systematic departures from a mean global efficiency of $\epsilon_{\rm
cl}\approx0.0026$.

\subsection{Open Clusters}

As was discussed in \S2.1, the formation of stellar clusters continues to be
an important element of star formation in general, and the existence of an
apparently universal formation efficiency for globular clusters must reflect a
generic piece of this process. However, the picture may still be incomplete:
although the discussion to this point has included clusters that formed in a
wide variety of environments---at galactocentric distances $r_{\rm gc}\sim
10$--100 kpc, in galaxies spanning four orders of magnitude in total
luminosity and inhabiting both poor groups and rich clusters---all of these
formed at essentially a single epoch, and under physical conditions that were
possibly much more extreme than those in, say, the disk of the Milky Way today.
It is important, therefore, to compare the protogalactic $\epsilon_{\rm cl}
\simeq0.0026\pm0.0005$ to the frequency with which bound star clusters form in
a typical quiescent galaxy at the current epoch. Disk clusters in the Milky Way
present one opportunity to do this.

The rate of open cluster formation in the solar neighborhood has been
estimated by \markcite{elm85}Elmegreen \& Clemens (1985) as ${\cal S}_{\rm cl}
\simeq(2.5\pm1)\times10^{-7}$ clusters kpc$^{-2}$ yr$^{-1}$. If the masses of
the clusters range between $100\,M_\odot$ and $5\,000\,M_\odot$ according to
$d{\cal N}/dm\propto m^{-1.5}$ (from the luminosity spectrum $d{\cal N}/dL$ of
\markcite{vdl84}van den Bergh \& Lafontaine 1984), then a mean mass of
$\simeq700\,M_\odot$ is indicated, and the formation rate becomes ${\cal S}_
{\rm cl}\simeq(1.8\pm0.7)\times10^{-4}M_\odot$ kpc$^{-2}$ yr$^{-1}$. This is
to be compared with the {\it total} star formation rate in the solar
neighborhood, which includes those stars appearing in loose groups and unbound
associations (or even in relative isolation) as well as in bound clusters.
This rate is ${\cal S}_{\rm tot}\simeq4.5\times10^{-3}M_\odot$ kpc$^{-2}$
yr$^{-1}$ (e.g., \markcite{mck89}McKee 1989).

Since the mean stellar mass is no different in clusters than in the field,
the ratio of these two rates can be taken directly, to assess the local {\it
number} fraction of stars born in clusters: ${\cal S}_{\rm cl}/{\cal S}_{\rm
tot}\sim0.04$. This is consistent with the classic estimate (e.g.,
\markcite{rob57}Roberts 1957) of about 10\%. However, the globular cluster
$\wepsilon_{\rm cl}$ has been obtained here as a {\it mass} ratio, since this
is the observable quantity that can be related directly to its formation value
(see eq.~[\ref{eq:28}]). To make a proper comparison with open cluster
formation, it is therefore necessary to refer the mass going into newborn 
clusters (per unit area, per unit time) to the total mass of gas available for
any kind of star formation. This latter figure can be derived from ${\cal S}_
{\rm tot}$ using the mean star formation efficiency, $\langle {\rm SFE}\rangle=
\langle M_{\rm stars}/M_{\rm gas}\rangle$, averaged over many of the dense,
star-forming (not necessarily cluster-forming) cores in giant molecular
clouds. This number is only poorly known, but statistical arguments (e.g.,
\markcite{elm83}Elmegreen 1983) suggest that it lies at the level of
$\langle {\rm SFE}\rangle\sim5$\%--10\%. Thus,
\begin{equation}
\epsilon_{\rm cl}({\rm open})\simeq
{{{\cal S}_{\rm cl}}\over{{\cal S}_{\rm tot}/\langle {\rm SFE}\rangle}}
\sim0.002-0.004\ ,
\label{eq:49}
\end{equation}
which is in reasonably good agreement with the result for globular cluster
systems. The numerical value in equation (\ref{eq:49}) clearly should
not be taken as definitive, given the very rough, order-of-magnitude nature
of the argument that has produced it. That said, however, it is just the
order-of-magnitude agreement between equations (\ref{eq:49}) and (\ref{eq:41})
which suggests that the basic mechanism of cluster formation may not have
changed appreciably from protogalactic times to the present.

Another obvious place to check the current level of $\epsilon_{\rm cl}$ is
in starbursts and merging galaxies, which are forming massive young clusters
in substantial numbers (e.g., \markcite{whs95}Whitmore \& Schweizer 1995;
\markcite{meu95}Meurer et al.~1995). Complete inventories of the star-forming
gas mass in individual systems, and of the total mass in those of their young
clusters that can be shown to be gravitationally bound (which to date has been
done rigorously for only two objects: Ho \& Filippenko 1996\markcite{hoa96}a,
\markcite{hob96}1996b), would be of considerable interest. The results should
be of particular relevance to the oft-made claim, which is now at the center
of some debate (\markcite{ste98}Sternberg 1998; \markcite{bro98}Brodie et
al.~1998), that these super star clusters are the modern-day equivalents of
young globulars. If it turns out, for example, that strongly bound clusters in
starbursts are forming in numbers very much in excess of those expected on
the basis of $\epsilon_{\rm cl}\simeq0.25\%$ (as, e.g., the UV observations of
\markcite{meu95}Meurer et al.~1995 appear to suggest), then these spectacular
events might {\it not} be representative of a ubiquitous phase in the
formation of average elliptical galaxies.

\section{Discussion}

The fact that the $\epsilon_{\rm cl}$ values inferred here are everywhere
so similar---over large ranges in radius inside M87, M49, and NGC 1399;
from dwarf ellipticals to brightest cluster galaxies; and from massive, old
globular clusters to smaller, much younger open clusters---should serve as a
strong constraint on theories of star formation. The simplest interpretation
is that the probability of attaining a cumulative star formation efficiency of
${\rm SFE}\ga20\%$--50\% (see \S2.1) in any dense clump of gas more massive
than $\sim10^2$--$10^3\,M_\odot$ (so from open clusters on up) can depend only
weakly, on average, on local environment (i.e., the density and pressure of
ambient gas) or on factors such as global background density (meaning, e.g.,
in the protogalactic context, a 1-$\sigma$ vs.~a 3-$\sigma$ density
fluctuation; or a poor group of galaxies vs.~a rich cluster). Moreover,
implicit in plots like Figs.~\ref{fig8} and \ref{fig9} is the fact that
cluster formation
efficiencies are also rather insensitive to metallicity: the mean abundances
of the globulars in the galaxies represented there range over two orders of
magnitude, from $[{\rm Fe/H}]\simeq-2$ to $[{\rm Fe/H}]\simeq0$
(see \markcite{dur96}Durrell et al.~1996; \markcite{for97}Forbes et al.~1997).
Over a remarkably broad spectrum of physical conditions, therefore, it appears
that similar fractions of star-forming gas---always about 0.25\% by mass---are
able to produce bound stellar clusters. The implicitly mass-averaged nature of
the observational estimates of $\epsilon_{\rm cl}$ has to be kept in mind, but
would seem to be of little concern in most cases. It is therefore reasonable to
place the efficiency of cluster formation alongside the globular cluster mass
function $d{\cal N}/dm$ (\markcite{mcl96}McLaughlin \& Pudritz 1996) and the
mean globular cluster mass $\langle m\rangle_{\rm cl}$ (e.g., Fig.~\ref{fig2}
above) as robust and nearly universal physical quantities that must be derived
from the most general aspects of the star formation process.

As a corollary to this, one implication of Figs.~\ref{fig8} and \ref{fig9}
above is that the efficiency of {\it unclustered} star formation could {\it
not} have been universal. That is, in both dwarf elliptical and brightest
cluster galaxies, where the data are consistent with $\epsilon_{\rm cl}=M_{\rm
gcs}^{\rm init}/M_{\rm gas}^{\rm init}$ having been constant in the mean,
globular clusters apparently did form in precisely the numbers expected for
the total amount of gas initially available to these systems. But then, the
higher than average specific frequencies there are inferred to reflect larger
ratios of $M_{\rm gas}^{\rm init}/M_{\rm stars}$, so that lower than average
amounts of this gas were converted into field stars. The data for dwarfs agree
with a simple scenario in which most of the unused gas supply is driven
entirely out of the galaxies by a strong wind following one major burst of
star and cluster formation. If this is basically correct, then the globulars
in dE's quite possibly were able to form completely before the onset of such a
wind, and they could conceivably have been instrumental in driving it; at the
very least, if the protoclusters were still largely gaseous, they must have
been already sufficiently dense and well defined that they could survive such a
catastrophic event essentially intact. At the same time, field star formation
must have been truncated by the wind, before reaching what would have been a
``normal'' efficiency of $M_{\rm stars}/M_{\rm gas}^{\rm init}=1-\epsilon_{\rm
cl}$. Thus, {\it the massive and dense clumps of gas which ultimately formed
bound star clusters had to have collapsed more rapidly than those which
produced unbound groups and associations}. This conclusion may hold quite
generally---one possible interpretation being that only those gas clouds
in the highest-density tail of some distribution, in any setting, are able to
achieve a high cumulative SFE---and it may be related to the fact that the
globular clusters in dwarfs are generally somehat bluer, and presumably more
metal-poor, than the bulk of the field stars there (\markcite{dur96}Durrell et
al.~1996; \markcite{mil98}Miller et al.~1998)

A similar feedback argument could also apply to BCGs, where it may be that
especially strong early bursts of star and cluster formation led to the
premature virialization of large amounts of protogalactic gas, and perhaps
drove slow, {\it partial} galactic winds, most effectively in the low-density
environs at large galactocentric radii (cf.~\markcite{har98}Harris et
al.~1998). The unused gas would have to remain hot to the present day, and
more or less in the vicinity of the parent galaxy, in order to enter the
observed $L_X$--$L_B$ relation as inferred in \S4.1. This requirement is
consistent with the fact that BCGs are situated at or near the centers of
their galaxy clusters, and thus at the bottoms of very deep potential wells.
In this scenario, again, globulars would have had to form, in just the numbers
expected of them, somewhat before field stars were able to do the same; and
this truncation of unclustered star formation would have had to be more severe
at large galactocentric radii (where the feedback is strongest). This could
then account explicitly for the connection between the first and second
specific frequency problems, i.e., between high global $S_N$ and local
$\Sigma_{\rm cl}/\Sigma_{\rm stars}$ ratios that increase with $R_{\rm gc}$.
It also gives a more specific context to the general claim
(\markcite{har86}Harris 1986) that the shallowness of GCS radial distributions
relative to the stellar light profiles in some ellipticals resulted from the
globulars having formed slightly in advance of the stars there. The main
demand on this picture for BCG formation is that it must be able to explain
why $M_{\rm gas}/M_{\rm stars}$ increases with galaxy luminosity, i.e., why
more massive BCGs apparently had larger fractions of their initial gas mass
virialized early on. As \markcite{har98}Harris et al.~(1998)
discuss in some detail, this issue is likely related to the fact that brighter
BCGs are generally found in more massive galaxy clusters, which may have
presented higher-density and more turbulent environments that led to
systematically higher star formation rates, and more violent feedback.
Although speculative, these ideas are suggested directly by the GCS data, and
are consistent with all those available. Since they are also obviously related
to questions on the star-formation histories of dE's, on the origin of the
$L_X$--$L_B$ correlation in the brightest ellipticals, and on the nature of
the intracluster medium, they should be addressed with considerably more rigor
than has been applied here.

It is particularly striking that the cluster formation efficiency derived in
\S4.3 for the Milky Way halo is essentially indistinguishable from that which
applied in M87, M49, and NGC 1399, and which appears to have held quite
generally in ellipticals of most any description. This calls into serious
question one of the primary motivations for the model of
\markcite{ash92}Ashman \& Zepf (1992) and \markcite{zep93}Zepf \& Ashman
(1993), who posit that ellipticals are formed primarily by the mergers of
gas-rich spirals, and invoke preferential cluster formation during these events
to account for what they claim is a typical factor of two difference in the
observed ratios $M_{\rm gcs}/M_{\rm stars}$ of late- and early-type systems.
Although it is true that spirals have specific frequencies that are lower than
those in bright E galaxies, by factors of perhaps 2--3 on average (as is
exemplified by the Milky Way in Fig.~\ref{fig9} above; see also
\markcite{har91}Harris 1991), it is now apparent that this is not necessarily a
reflection of similar discrepancies in their more fundamental $\wepsilon_{\rm
cl}$. If the Milky Way is typical, it suggests that---even though stellar
clusters certainly can form in merging galaxies---{\it mergers are not
required to explain the relative GCS populations of spirals and ellipticals in
general}. Conversely, whenever an elliptical does form by the major merger of
two spirals, this is allowed to include cluster formation only at the standard
efficiency of $\epsilon_{\rm cl}\simeq0.0025$.

\markcite{ash92}Ashman \& Zepf (1992) and \markcite{zep93}Zepf \& Ashman
(1993) also attribute the second specific frequency problem to the formation
of ellipticals by mergers of spirals with pre-existing GCSs. However, it is
not clear that this is necessary either, if the extended spatial distribution
of globulars in the brightest ellipticals is generally related to their
association with hot gas at rather large galactocentric radii. In fact, a
basic expectation of the original Ashman-Zepf scenario is that the second
$S_N$ problem should be {\it less} pronounced in galaxies with higher global
$S_N$ (which they hypothesize to have suffered greater numbers
of major mergers). The sense of this predicted trend is {\it opposite} to the
one implied here---with empirical support directly from the evidence in M87,
and indirectly from the correlations of \markcite{for97}Forbes et al.~(1997)
and \markcite{kis97}Kissler-Patig (1997)---in which a constant $\epsilon_{\rm
cl}$ combines with the larger $M_{\rm gas}/M_{\rm stars}$ in higher-$S_N$
ellipticals, to produce greater contrasts in their projected $\Sigma_{\rm cl}$
and $\Sigma_{\rm stars}$ profiles. Although radial-profile and $S_N$ data by
themselves may not rule out the merger model for GCSs altogether, a revision
to take account of the hot gas in large ellipticals is essential if it is to
be at all viable. (See also \markcite{fir97}Forbes et al.~1997 and
\markcite{kis98}Kissler-Patig et al.~1998 for critical discussions of this and
other aspects of the basic idea.)

{\small
\begin{deluxetable}{lccclcclcc}
\tablecaption{Parametric Fits to GCS Volume Density Profiles\tablenotemark{1}
\label{tab5}}
\tablewidth{0pt}
\tablehead{ & & & & & & & \nl
 & \multispan3{\hfill $n_{\rm cl}=n_0[1+(r/a)^2]^{-\gamma/2}$ \hfill} & &
\multispan2{\hfill $n_{\rm cl}=n_0(r/a)^{-1}(1+r/a)^{-2}$ \hfill} & &
\multispan2{\hfill $n_{\rm cl}=n_0(r/a)^{-1}(1+r/a)^{-3}$ \hfill} \nl
 & & & & & & & & & \nl
 & \colhead{$\gamma$} & \colhead{$a$} & \colhead{log $n_0$} & &
\colhead{$a$} & \colhead{log $n_0$} & &
\colhead{$a$} & \colhead{${\cal N}_{\infty}=2\pi a^3 n_0$}}
\startdata
M87\tablenotemark{2} &
2.53$\pm$0.17 & 0.73$\pm$0.27 & 2.177$^{+0.248}_{-0.193}$ & &
2.10$\pm$0.40 & 1.522$^{+ 0.220}_{- 0.180}$ & &
$\ldots$ & $\ldots$ \nl
 & & & & & & & & & \nl
M49\tablenotemark{3} &
2.73$\pm$0.27 & 1.41$\pm$0.56 & 1.170$^{+0.356}_{-0.187}$ & &
2.25$\pm$0.45 & 0.985$^{+0.285}_{-0.219}$ & &
6.18$\pm$1.32 & 3260$\pm$475 \nl
 & & & & & & & & & \nl
N1399\tablenotemark{4} &
3.15$\pm$0.35 & 1.54$\pm$0.46 & 1.466$^{+0.233}_{-0.124}$ & &
1.15$\pm$0.45 & 1.859$^{+0.543}_{-0.351}$ & &
3.78$\pm$0.92 & 3630$\pm$460 \nl
% & & & & & & & & & \nl
\enddata
\tablenotetext{1}{For radii in arcminutes and densities in number per
arcmin$^{3}$.}
\tablenotetext{2}{Normalizations $n_0$ apply to $V\leq23.9$; divide by 0.5572
to obtain population over all magnitudes.}
\tablenotetext{3}{Normalizations $n_0$ apply to $V\leq23.3$; divide by 0.3646
to obtain population over all magnitudes.}
\tablenotetext{4}{Normalizations $n_0$ apply to $V\leq24.0$; divide by 0.5332
to obtain population over all magnitudes.}
\end{deluxetable}
}

Finally, the GCS density profiles derived specifically for M87, M49, and NGC
1399 in \S3 also are relevant to issues of the generic spatial structure of
GCSs, and to the specific matter of the possible existence of intergalactic
globulars in clusters of galaxies. To
facilitate a discussion of these points, and for reference, Table \ref{tab5}
presents the results of least-squares fits of a few different parametric
functions to the spatial distributions of the M87, M49, and NGC 1399 GCSs, as
they are given in Tables \ref{tab2}, \ref{tab3}, and \ref{tab4} above. The
functions include (from left to right in Table \ref{tab5}) a power law with a
constant-density core region; the cusped density profile suggested by Navarro,
Frenk, \& White (\markcite{nav96}1996, \markcite{nav97}1997) as appropriate for
cold dark matter halos; and the law popularized by \markcite{her90}Hernquist
(1990) and applied widely to the light profiles of elliptical galaxies. In all
cases, fits were performed by minimizing the residuals of the functions' {\it
projected} density profiles relative to the $N_{\rm cl}$ data, which have
smaller observational uncertainties than the volume densities do. The best-fit
parameters thus obtained were then used to overlay the original,
three-dimensional functions on the deprojected $n_{\rm cl}$ and verify that
they provided good descriptions of those as well. One consequence of this
procedure is that no fit is given for the \markcite{her90}Hernquist (1990)
profile, $n_{\rm cl}\propto (r/a)^{-1}(1+r/a)^{-3}$, to the M87 GCS. Although
it is possible to fit this function directly to the $n_{\rm cl}$ data, its
projection affords a poor description of the full $N_{\rm cl}$ distribution.
This almost certainly reflects the fact that the GCS profile in M87 appears
about to become significantly shallower, at least temporarily, beyond the
last radius ($r_{\rm gc}\simeq100$ kpc) for which cluster counts exist (because
Fig.~\ref{fig4} suggests essentially that $\rho_{\rm cl}\propto\rho_{\rm gas}$
there). Such a change in slope is obviously not allowed by Hernquist's density
profile. Thus, the failure of this model in this case makes the point that
{\it any} of the density fits in Table \ref{tab5} should be taken at face
value only over the ranges of $r_{\rm gc}$ that have been directly observed;
extrapolations to much larger radii could potentially introduce appreciable
systematic errors.

Table \ref{tab5} has three interesting implications: (1) None of the
available data {\it require} the existence of truly constant-density ``cores''
in GCSs. This was also mentioned by \markcite{mcl95}McLaughlin (1995)
and it is, of course, plainly seen in Figs.~\ref{fig4} through \ref{fig6}.
Evidence is emerging for a similar situation in elliptical galaxies generally
(\markcite{cra93}Crane et al.~1993; \markcite{geb96}Gebhardt et al.~1996), and
apparently also in galaxy clusters (\markcite{mer94}Merritt \& Tremblay
1994; \markcite{car97}Carlberg et al.~1997; \markcite{mcl99}McLaughlin 1999).
(2) It has been suggested, particularly in the contexts of some dynamical
analyses (e.g., \markcite{huc87}Huchra \& Brodie 1987; \markcite{wei97}Weil,
Bland-Hawthorn, \& Malin 1997) that the M87 GCS directly reflects the
distribution of dark matter in and around that galaxy. However, this is not
true in any simple sense; neither of the $n_{\rm cl}$ fits given in Table
\ref{tab5} provides an adequate description of $\rho_{\rm dark}(r_{\rm gc})$ in
M87. Instead, it is found there that $\rho_{\rm dark}\propto\rho_{\rm gas}$
(see \markcite{mcl99}McLaughlin 1999), in which case Fig.~\ref{fig4} shows
that the GCS becomes a reliable tracer of the dark matter halo {\it alone}
only on fairly large spatial scales (where the number densities become quite
low). (3) The analytical fits to $N_{\rm cl}$ and $n_{\rm cl}$ offer a
convenient way to extrapolate the observed densities to (slightly) smaller and
larger radii than those observed, so as to compute the total GCS populations
of these three galaxies. Thus, in M87, it is found that ${\cal N}_{\rm tot}=
13\,600\pm500$ clusters (over all magnitudes) are {\it projected} to distances
$R_{\rm gc}\leq25\arcmin\simeq110$ kpc from the galaxy's center; in M49, there
are ${\cal N}_{\rm tot}=6\,850\pm550$ globulars within the same projected
$R_{\rm gc}$; and in NGC 1399, ${\cal N}_{\rm tot}=4\,700\pm400$ clusters have
$R_{\rm gc}\leq14\farcm5\simeq70$ kpc. These total ``projected populations''
are in good agreement with those estimated in many other studies (see the
references cited in \S3). However, it is now possible to determine the
three-dimensional radius of the sphere which actually contains all of these
clusters in each system. From either of the M87 fits in Table \ref{tab5}, it
is found that 13\,600 clusters are contained within $r_{\rm gc}\simeq150$--160
kpc; in M49, 6\,850 globulars come from the sphere $r_{\rm gc}\leq140$--150
kpc; and the 4\,700 globulars in NGC 1399 all have true galactocentric
positions $r_{\rm gc}\leq90$--95 kpc. To put this another way, all three GCSs
are sufficiently centrally concentrated that, in each case, some 85\%--90\% of
the globular clusters projected onto the galaxy---so again, to within
$R_{\rm gc}\leq110$ kpc for M87 and M49, and $R_{\rm gc}\leq70$ kpc for NGC
1399---really are located at those galactocentric distances in three
dimensions.

This last point specifically can clarify and constrain the suggestion
(\markcite{whi87}White 1987; \markcite{wes95}West et al.~1995;
\markcite{cot98}C\^ot\'e et al.~1998) that the
high specific frequencies in BCGs like M87 are due to the ``contamination'' of
their GCSs, in projection, by a significant number of globular clusters that
are associated with a galaxy cluster as a whole, rather than bound to the
central galaxy itself. \markcite{har98}Harris et al.~(1998) tried to test this
idea in some detail for M87 in particular, working from the assumption
(following \markcite{whi87}White 1987) that any population of intergalactic
globular clusters (IGCs) should generally be associated with the diffuse
stellar light of the cD envelopes that surround many BCGs (i.e., these
structures were considered to comprise primarily intracluster stars).
This hypothesis leads to inconsistencies that prompted \markcite{har98}Harris
et al.~to reject IGCs as the main cause of the first specific frequency
problem. While their arguments still stand within the framework that they
constructed, it is now seen that the high $S_N$ of M87 certainly, and
likely those of other BCGs as well, can be traced to subsets of their GCSs that
are associated with hot gas, and not necessarily with stars at all. This does
appear to favor the basic notion of IGCs, since the gas around M87 extends to
very large spatial scales, and traces out the potential well of the Virgo
Cluster as a whole (\markcite{mcl99}McLaughlin 1999). If a constant
$\wepsilon_{\rm cl}$ holds all the way out,
then, a cluster-wide population of globulars could be expected. However, since
{\it all} of the 13\,600 globular clusters which give M87 its high specific
frequency in projection are physically associated with a volume of radius
$\sim$150 kpc around the galaxy, it is not obvious that any of them are truly
``intergalactic'' objects at the current epoch. Projection effects are {\it
not} solely responsible for the high $S_N$ of M87, and although intergalactic
globular clusters may well exist, it is not clear that any have yet been
observed.

Of course, since M87 is at the dynamical center of Virgo, it can be difficult
to distinguish meaningfully between the galaxy and the cluster on 100-kpc
scales. In the context of IGCs, the question evidently comes down to one of
origin: did the globulars now seen right around M87 actually form there, or
were many of them somehow brought in from much further than 150 kpc away? Any
self-consistent answer to this must be able to account for the observed fact
that $\rho_{\rm cl}=0.003\,(\rho_{\rm gas}+\rho_{\rm stars})$ at galactocentric
radii $r_{\rm gc}\sim10$--100 kpc in M87; that is, both the very existence
and the normalization of this proportionality have to be explained. Although
essentially any scenario which is able to do this would be allowed by the
current data, none could be {\it preferred} over what remains the simplest
option: that the bulk of the M87 GCS formed in situ, with a thoroughly
standard and spatially constant $\epsilon_{\rm cl}$. Ultimately, only detailed
evolutionary models, tailored specifically to the properties of M87 and Virgo
and considering both stars {\it and gas}, can really decide what fraction of
the galaxy's GCS might have come originally from the cluster at large.

\section{Summary}

Stars form mainly in groups, some of which emerge from their natal clouds
of gas as gravitationally bound clusters. An understanding of cluster formation
will therefore be integral to any full theory of star formation, and an
important element of this is simply the frequency with which it occurs. That
is, what is the likelihood that a star-forming cloud of gas---whether this be
a dense clump in a Galactic giant molecular cloud, or a larger one in a
protogalactic fragment---will achieve the high cumulative star formation
efficiency (viz.~${\rm SFE}\ga20\%$--50\%) that allows its stars to remain
bound as a cluster after they have cleared away the gas? This paper has used
observations of the globular cluster systems in galaxy halos to address this
question, and to empirically evaluate the efficiency of cluster formation: by
mass, $\epsilon_{\rm cl}\simeq0.25$\%. This is, by all appearances, nearly a
universal number, and it should therefore serve as a strong constraint on
theories of star and cluster formation in any context.

To arrive at this result, it was first shown (\S2.1) that the total (global)
{\it masses} of GCSs are quite robust, even over a Hubble time, against the
dynamical processes that work to destroy low-mass globular clusters in the
central regions of galaxies. With an observed $M_{\rm gcs}$ thus useful as
a measure of the initial quantity, the specific frequencies of GCSs, $S_N
\propto {\cal N}_{\rm tot}/L_{V,{\rm gal}}\propto M_{\rm gcs}/M_{\rm stars}$,
serve as {\it crude} first estimates of the basic formation efficiency
$\epsilon_{\rm cl}\equiv M_{\rm gcs}^{\rm init}/M_{\rm gas}^{\rm init}$. To
the {\it limited} extent that $S_N\approx5$ is an adequate description of
early-type galaxies, the implication is that $\epsilon_{\rm cl}\sim2\times
10^{-3}$. The limiting factor in this simplest treatment of GCS data is the
assumption that a galaxy's current stellar mass always suffices as an estimate
of its total initial gas mass; if taken too far, this quickly leads to some
perplexing conclusions.

As was reviewed in \S2.2, systematic variations in $S_N$ exist along the
entire sequence of early-type galaxies, from dwarf ellipticals (where ${\cal
N}_{\rm tot}\propto L_{V,{\rm gal}}^{0.6}$ or so) through normal giants
(${\cal N}_{\rm tot}\propto L_{V,{\rm gal}}^{1.3}$) and on to the brightest
ellipticals, including many of the central galaxies in groups and rich
clusters (${\cal N}_{\rm tot}\propto L_{V,{\rm gal}}^{1.8}$). If specific
frequency were indeed a good estimator of $\epsilon_{\rm cl}$ in general, the
implication would seem to be that this also varied, for unknown reasons and in
a non-monotonic fashion, as a function of galaxy luminosity. Closely related to
this is the fact that the GCSs of some, but not all, large ellipticals are
spatially more extended than their stellar halos. That is, local $S_N$ values
increase with projected $R_{\rm gc}$ in some systems, which naively would
suggest that---again, for unknown reasons, and counter to any intuition---the
efficiency of cluster formation {\it sometimes} increased with galactocentric
radius.

An important clue to the true nature of both these ``specific frequency
problems''---which are most likely two aspects of a single phenomenon---lies
in the observation that the global $S_N$ in central cluster galaxies increases
with the X-ray luminosity of their parent clusters. Some of the globulars in
these systems ought then to be associated with the hot gas there; and $M_{\rm
gas}^{\rm init}$ is approximated better by the present-day $(M_{\rm gas}+M_{\rm
stars})$ than by $M_{\rm stars}$ alone. $S_N$ then systematically overestimates
the true cluster formation efficiency $\epsilon_{\rm cl}$, which instead is
given empirically by $\wepsilon_{\rm cl}=M_{\rm gcs}/(M_{\rm gas}+M_{\rm
stars})$. In addition, since a galaxy's hot gas generally follows a shallower
density profile than its stellar halo does, the ``extra'' clusters in very
gas-rich (and high-$S_N$) ellipticals might well be expected to also show a
radial distribution, $\rho_{\rm cl}$ or $\Sigma_{\rm cl}$, that is distended
relative to $\rho_{\rm stars}$ or $\Sigma_{\rm stars}$.

These ideas, which extend similar suggestions in the recent literature, have
been tested in detail for each of M87, M49, and NGC 1399. Data were collected
from the literature on the GCSs, surface brightness profiles, and X-ray gas
densities of these galaxies, covering spatial scales $r_{\rm gc}\sim1$--100
kpc. In \S3.1, new comparisons of the projected densities $\Sigma_{\rm cl}$ and
$\Sigma_{\rm stars}$ in M49 and NGC 1399 (which are gas-poor) showed that,
contrary to previous claims, neither galaxy suffers from either of the two
specific frequency problems: The global $S_N$ of M49 has a ``standard'' value
of $\simeq$5, and the local ratio $\Sigma_{\rm cl}/\Sigma_{\rm stars}$ is
constant with radius (outside of a ``core'' region, $R_{\rm gc}\la R_{\rm
eff}$, that possibly has been dynamically depleted) if the recent
surface-brightness profile of \markcite{cao94}Caon et al.~(1994) is used in
the comparison rather than the older $\mu_B$ measurements of
\markcite{kin78}King (1978). The global specific frequency of NGC 1399 is
similarly ``average'', $S_N=7.0\pm0.6$, if its GCS population is normalized to
an absolute magnitude for the galaxy ($M_V\simeq-22.1$) that is significantly
brighter than the value more commonly adopted. The ratio $\Sigma_{\rm cl}/
\Sigma_{\rm stars}$ is also constant with radius in this galaxy, and has a
value equal to that found in M49.

M87, at the center of the Virgo Cluster, is much richer in gas than either of
M49 or NGC 1399, and both $S_N$ problems are in evidence there. A discrete,
geometrical deprojection algorithm was developed to enable a direct comparison
between the volume densities $\rho_{\rm cl}$, $\rho_{\rm stars}$, and
$\rho_{\rm gas}$ at all three-dimensional galactocentric radii in M87. It was
found in \S3.2 that $\rho_{\rm cl}\propto(\rho_{\rm gas}+\rho_{\rm stars})$,
beyond $r_{\rm gc}\sim8$ kpc, and that the constant of proportionality is the
same as that in M49 and NGC 1399. {\it The cluster formation efficiency was
constant with galactocentric radius inside each of these systems, and the same
from one galaxy to another}. Specifically,
$$\epsilon_{\rm cl}\simeq
{{\rho_{\rm cl}}\over{\rho_{\rm gas}+\rho_{\rm stars}}}=
{{M_{\rm gcs}}\over{M_{\rm gas}+M_{\rm stars}}}=
0.0026\pm0.0005\ .$$
(Observed departures from this ratio in the central regions $r_{\rm gc}\la
R_{\rm eff}$ of M87 and M49 may have resulted, at least in part, from the
dynamical destruction of globular clusters there.) Many globulars in M87 are
indeed associated with its X-ray gas, and this is in fact responsible for the
appearance of both specific frequency problems. This single case leads to the
basic expectation---which should be tested in as many individual systems as
possible---that real discrepancies between $\Sigma_{\rm cl}$ and $\Sigma_{\rm
stars}$ should appear only in gas-rich ellipticals that also have high global
specific fequencies.

The obvious generalization of these results is that most galaxies might have
been subject to a single, common $\epsilon_{\rm cl}$. This possibility was
explored, and borne out, in \S4. Observations of global $S_N$ (i.e., total GCS
populations and galaxy luminosities) in 97 giant ellipticals, brightest
cluster galaxies, and dwarf ellipticals {\it all} were shown to be consistent
with the predictions of a universal $\epsilon_{\rm cl}$ at the same level as
that observed directly in M87, M49, and NGC 1399. The different scalings of
$S_N$ with $L_{V,{\rm gal}}$ among normal giant ellipticals, central cluster
galaxies, and dwarfs stem entirely from fundamentally different relations
between $L_{V,{\rm gal}}$ and $M_{\rm gas}^{\rm init}$ in these different
physical regimes. These relations have been obtained from scaling arguments
that employ the optical fundamental plane of bright ellipticals; the observed
correlation between X-ray and optical luminosities in early-type systems; and
the fractional gas loss expected in a specific model (\markcite{dek86}Dekel
\& Silk 1986) for winds from a single burst of star formation in dark-matter
dominated dwarf ellipticals. Observations of ${\cal N}_{\rm tot}$ vs.~$L_{V,
{\rm gal}}$ still show some {\it scatter} about the mean prediction of constant
$\epsilon_{\rm cl}$, and it will be important to clarify the significance and
the origins of this. However, there are no {\it systematic} deviations from a
universal cluster formation efficiency, in galaxies that span four orders of
magnitude in total luminosity and GCS population.

Exactly the same $\epsilon_{\rm cl}$ was shown, in \S4, to have obtained for
globular clusters in the spheroid, or stellar halo, of the Milky Way; and it
was argued that it appears also to apply now, to the formation of {\it open}
clusters locally in the Galactic disk. This remarkable robustness, in which a
single cluster formation efficiency was and is realized in a great variety of
local and global environments, demands a very general theoretical explanation.
From a more empirical standpoint, it bears directly on current questions
involving GCS and galaxy formation and evolution on large scales. Some of these
were discussed in \S5. Of considerable interest is the fact that, with
$\epsilon_{\rm cl}$ constant but $M_{\rm gcs}/M_{\rm stars}$ not, the
efficiency of {\it unclustered} star formation could {\it not} have been
universal; field star formation must have been suppressed, in a sense, in
dwarf ellipticals and at large galactocentric radii in the brightest, most
gas-rich ellipticals. The possibility that these points could be understood in
terms of feedback processes seems promising (\S5), and should be explored in
quantitative models.

\acknowledgments

I am grateful to Bill Harris for providing the data plotted in Figures
\ref{fig8} and \ref{fig9}; to Ivan King and Chris McKee for helpful comments
on an earlier draft of the paper; and to an anonymous referee for several
valuable suggestions, including the addition of Figures \ref{fig7} and
\ref{fig10}. This work was supported by NASA through grant number
HF-1097.01-97A awarded by the Space Telescope Science Institute, which is
operated by the Association of Universities for Research in Astronomy, Inc.,
for NASA under contract NAS5-26555.

\appendix
\def\thefigure{\arabic{figure}}

\section{Deprojection Algorithm}

The surface density data in Tables \ref{tab2}, \ref{tab3}, and \ref{tab4}
lend themselves well to a geometrical deprojection of the sort that is often
applied to the X-ray emission from ellipticals and galaxy clusters
(e.g., \markcite{fab81}Fabian et al.~1981; \markcite{kri83}Kriss, Cioffi, \&
Canizares 1983; \markcite{nul95}Nulsen \& B\"ohringer 1995). This technique
makes only one assumption---that of circular symmetry---and is fully
nonparametric. It also avoids the explicit differentiation of $N_{\rm cl}$
that is required in the standard Abel-transform approach to deprojection.
Such a differentiation can be performed if an ad hoc parametrization is adopted
for the global form of the density profile from the outset, but it is clearly
desireable to minimize any such a priori restrictions. Although sophisticated,
nonparametric algorithms that resort finally to an Abel integral have also
been developed (e.g., \markcite{mer94}Merritt \& Tremblay 1994), they present
a greater degree of complexity than is necessary here; and, in any case, they
seem poorly suited to density profiles that are constructed from multiple
observations with different fields of view and statistical corrections for
different levels of background contamination and photometric incompleteness.

The assumption of spherical symmetry made here is not perfect (the isopleths
of the M87 GCS, for one, are clearly non-circular in projection; see
\markcite{mcl94}McLaughlin et al.~1994), but neither is it nonsensical: it
takes azimuthally averaged surface densities and returns volume densities that
are properly interpreted as angular averages at various galactocentric radii.

\begin{figure*}[tb]
\centering \leavevmode
\epsfysize=4.0truein
\epsfbox{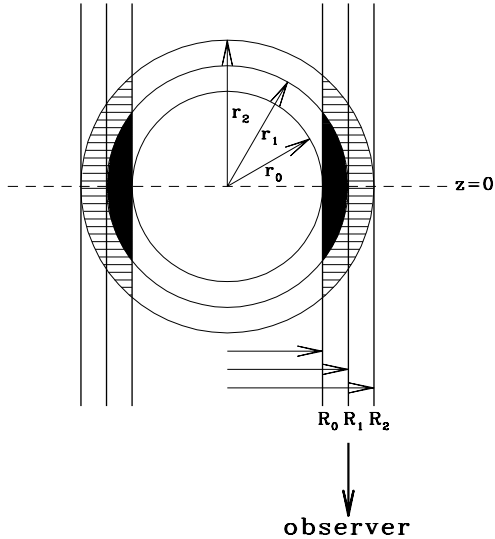}
\caption{Illustration of a geometrical deprojection algorithm.
The plane of the sky is perpendicular to the page. Radii $r_0$, $r_1$, and
$r_2$ are three-dimensional quantities, and define spherical shells. Radii
$R_0$, $R_1$, and $R_2$ are projected quantities referring to cylindrical
shells aligned along the line of sight; these correspond to circular annuli
on the plane of the sky.
\label{fig11}}
\end{figure*}

A simple example of the basic method, which should serve to define its
geometrical set-up and mathematical notation, is illustrated in
Fig.~\ref{fig11}. The plane of the sky there is perpendicular to the plane of
the page, and the two intersect in the line $z=0$. Thus, the lines on the
right, labelled by projected radius $R_0$, $R_1$, and $R_2$, are lines of
sight for an observer situated off the bottom of the page. The bands $R_0\le
R\le R_1$ and $R_1\le R\le R_2$, along with their mirror images on the left
side of the Figure, are axial cross sections of cylinders that extend to
$z=\pm\infty$; in the plane of the sky, of course, the corresponding cross
sections are circular annuli. The circular annuli shown here are not those on
the plane of the sky, however; rather, they are equatorial cross sections of
spherical shells with three-dimensional radii $r_0\le r\le r_1$ and $r_1\le
r\le r_2$. Evidently, these shells have been chosen such that $r_0=R_0$, and
so on. The origin $r=0$ (through which passes the axis $R=0$) represents the
center of a spherical distribution of globular clusters, which are supposed to
have been counted up in discrete cylinders $R_0\le R\le R_1$ and $R_1\le R\le
R_2$ on the sky, to form the surface densities $N_{\rm cl}(R_0,R_1)$ and
$N_{\rm cl}(R_1,R_2)$. From such data, it is possible to compute the fraction
of clusters in each cylinder that actually reside in the spherical shells
$r_0\le r\le r_1$ and $r_1\le r\le r_2$, and thus to form the average volume
densities $n_{\rm cl}(r_0,r_1)$ and $n_{\rm cl}(r_1,r_2)$.

Suppose, for the moment (this will be relaxed just below), that $r_2=R_2$
marks the outer edge of the GCS in question, i.e., $n_{\rm cl}(r>r_2)\equiv0$
and $N_{\rm cl}(R>R_2)\equiv0$. (It is assumed that the density $N_b$ of any
uniform distribution of background objects has already been subtracted to form
$N_{\rm cl}$ from the raw data.) In this case, all of the objects seen in the
outermost cylinder actually come from the outer spherical shell, i.e., from
the hatched regions between $R_1$ and $R_2$ in Fig.~\ref{fig11}. Since the
number of globulars in this cylinder is just
$${\cal N}_{\rm cl}(R_1,R_2)=N_{\rm cl}(R_1,R_2)\times\pi(R_2^2-R_1^2)\ ,$$
all that is required is a calculation of the volume of the shell $r_1\le r\le
r_2$ which is intersected by the cylinder $R_1\le R\le R_2$. Denoting this by
$V_{\rm int}(r_1,r_2;R_1,R_2)$, the volume density is obviously
$$n_{\rm cl}(r_1,r_2)={{{\cal N}_{\rm cl}(R_1,R_2)}\over{V_{\rm int}(r_1,r_2;
R_1,R_2)}}\ .$$
Moving inwards to the cylinder $R_0\le R\le R_1$, the total number of clusters
observed there includes a contribution from the spherical shell $r_1 \le r\le
r_2$ (hatched regions between $R_0$ and $R_1$), as well as one from the shell
of interest, $r_0\le r\le r_1$ (solid region). The volume density in the outer
shell has already been determined, and the volume of intersection between that
shell and the inner cylinder can be calculated [call this $V_{\rm int}
(r_1,r_2;R_0,R_1)$], so the volume density in the inner shell is given by
$$n_{\rm cl}(r_0,r_1)={{{\cal N}_{\rm cl}(R_0,R_1)\ -\ 
n_{\rm cl}(r_1,r_2)\,V_{\rm int}(r_1,r_2;R_0,R_1)}\over
{V_{\rm int}(r_0,r_1;R_0,R_1)}}\ ,$$
where ${\cal N}_{\rm cl}(R_0,R_1)$ and $V_{\rm int}(r_0,r_1;R_0,R_1)$ have the
same meanings as the corresponding quantities in the first cylinder, $R_1\le
R\le R_2$.

Clearly, this procedure can be extended and applied iteratively, from the
outside in, to the projected number counts in any series of concentric,
non-overlapping, circular annuli; it is only required that there be no gaps
between any two consecutive annuli. Thus, let the annuli be defined on the
plane of the sky by projected radii $R_0<R_1<R_2<\ldots<R_m$, i.e., set up $m$
rings with $R_0\le R\le R_1$; $R_1\le R\le R_2$; and so on. Given average
surface densities $N_{\rm cl}$ in each of these rings, the volume densities
$n_{\rm cl}$ in the corresponding spherical shells $r_0<r_1<r_2<\ldots<r_m$
can be obtained once the volume of intersection between some generic shell
$r_{j-1}\le r\le r_j$ and a cylinder $R_{i-1}\le R\le R_i$ is known. A sketch
similar to Fig.~\ref{fig11} readily shows that this volume is, quite generally,
\begin{equation}
V_{\rm int}(r_{j-1},r_j;R_{i-1},R_i)={{4\pi}\over{3}}\,
\left[(r_j^2-R_{i-1}^2)^{3/2}-(r_j^2-R_i^2)^{3/2}
+(r_{j-1}^2-R_i^2)^{3/2}-(r_{j-1}^2-R_{i-1}^2)^{3/2}\right],
\label{eq:33}
\end{equation}
for $i,j=1,\ldots,m$. This holds for any set of radii $\{r_{j-1},r_j,R_{i-1},
R_i\}$, with the understanding that any term in parentheses which evaluates as
negative is to be set to 0. An important application is to the case shown in
Fig.~\ref{fig11}, where, e.g., $r_1=R_1$ and $r_2=R_2$ and the volume common to
the shell $R_1\le r\le R_2$ and the cylinder $R_1\le R\le R_2$ is $V_{\rm int}
(R_1,R_2;R_1,R_2)=(4\pi/3)(R_2^2-R_1^2)^{3/2}$. It is most appropriate to
define shells such that $r_i=R_i$ for all $i$, in which case the results given
just above for $m=2$ generalize to
\begin{eqnarray}
n_{\rm cl}^{\prime}(R_{i-1},R_i) & = &
{{{\cal N}_{\rm cl}(R_{i-1},R_i)-\sum_{j=i+1}^{m} \left[n_{\rm cl}(R_{j-1},R_j)
\,V_{\rm int}(R_{j-1},R_j;R_{i-1},R_i)\right]}
\over{V_{\rm int}(R_{i-1},R_i;R_{i-1},R_i)}}\nonumber \\
 & = & {3\over{4(R_i^2-R_{i-1}^2)^{1/2}}}\left[
N_{\rm cl}(R_{i-1},R_i)-\sum_{j=i+1}^{m} {{n_{\rm cl}(R_{j-1},R_j)\,V_{\rm int}
(R_{j-1},R_j;R_{i-1},R_i)}\over{\pi(R_i^2-R_{i-1}^2)}}\right]. \nonumber \\
%{\hbox{\qquad}}
 & &
\label{eq:34}
\end{eqnarray}
Again, this assumes that any uniform background has already been subtracted to
obtain the surface densities $N_{\rm cl}$ in every annulus $i=1,\ldots,m$.

This equation is not quite final, however, since its derivation has been
facilitated by treating the outermost annulus seen in projection, $R_{m-1}\le
R\le R_m$ (or the outermost spherical shell, $R_{m-1}\le r\le R_m$), as if it
ended just at the ``true edge'' of some GCS. Of course, this will never
actually be the case---if not because it makes little sense to associate
sharp boundaries with such distributions in the first place, then because real
observations and number counts can cover small fields of view within GCSs
that are known to extend considerably further. This means that there
is a second type of ``background'' contamination to be dealt with in any
dataset, quite apart from that (already accounted for) due to a uniform
distribution of unassociated galaxies and stars; corrections have to be made,
in every annulus or shell, for the possible presence of bona fide globular
clusters at radii $r>R_m$, beyond the last annulus for which a surface density
can be directly measured. If the density distribution of such objects is
denoted by $n_{\rm cl}(r)$, their contribution to the projected number density
in a cylinder $R_{i-1}\le R\le R_i$ is given by
\begin{equation}
N_{\rm cl}^{\rm back}(R_{\rm i-1},R_i) = {4\over{R_i^2-R_{i-1}^2}}\,
\int_{R_{i-1}}^{R_i} R\,dR\,\int_{z_0(R)}^{\infty} n_{cl}(r)\,dz
\ ;\ \ \ \ \ \ z_0(R)\equiv(R_m^2-R^2)^{1/2}\ ,
\label{eq:35}
\end{equation}
which is just an average of the projected densities along every line
of sight contained in the cylinder. [As usual, the coordinate $z=(r^2-R^2)^
{1/2}$ here measures distance along these lines of sight; $z=0$ defines the
plane of the sky, as in Fig.~\ref{fig11}.] It is most convenient to express
this ``background'' surface density as a fraction of the observed $N_{\rm cl}$
in the outermost annulus $R_{m-1}\le R\le R_m$:
\begin{equation}
f(R_{i-1},R_i)\equiv{{N_{\rm cl}^{\rm back}(R_{i-1},R_i)}\over{N_{\rm cl}
(R_{m-1},R_m)}}={{R_m^2-R_{m-1}^2}\over{R_i^2-R_{i-1}^2}}\,\left[
{{\int_{R_{i-1}}^{R_i} R\,dR\,\int_{z_0(R)}^{\infty} n_{\rm cl}(r)\,
dz}\over{\int_{R_{m-1}}^{R_m} R\,dR\,\int_{0}^{\infty} n_{\rm cl}(r)\,dz}}
\right],
\label{eq:36}
\end{equation}
where $z_0(R)$ is defined as in equation (\ref{eq:35}), and the second equality
relies on the basic definition of average surface density in a circular annulus
of finite width. The advantage of this formulation is that the form of $n_{\rm
cl}(r)$ cannot be directly measured for $r>R_m$, so the calculation of
$N_{\rm cl}^{\rm back}$ necessarily depends on an extrapolation of some sort.
Without any constraints on this extrapolation, it is conceivable that an
artificial and unphysical situation such as $N_{\rm cl}^{\rm back}(R_{m-1},R_m)
>N_{\rm cl}(R_{m-1},R_m)$ could arise. This and similar difficulties are
avoided, however, by only ever evaluating $N_{\rm cl}^{\rm back}$ in terms of
$f$ (which is guaranteed always to be $<$1 when $i=m$) and the known
$N_{\rm cl}$ at the largest projected radii observed.

Equation (\ref{eq:36}) should therefore be used to subtract $N_{\rm cl}^{\rm
back}$ from the total $N_{\rm cl}$ in every annulus before deprojecting the
distribution according to equation (\ref{eq:34}). The corrected volume
densities in the spherical shells $i=1,\ldots,m$ may then be written as
\begin{equation}
n_{\rm cl}(R_{i-1},R_i)=n_{\rm cl}^{\prime}(R_{i-1},R_i)\ -\ 
{3\over{4(R_i^2-R_{i-1}^2)^{1/2}}}\,f(R_{i-1},R_i)N_{\rm cl}(R_{m-1},R_m)\ .
\label{eq:37}
\end{equation}
Typically, when measured GCS surface densities extend to sufficiently large
radius, an adequate estimate of $f$ can be made by assuming a power-law
distribution $n_{\rm cl}\propto r^{-\alpha}$ in the outer reaches of the
system, $r>R_m$; and in the case that $\alpha$ is an integer, equation
(\ref{eq:36}) can be evaluated analytically. With $\alpha=2$, for example,
\begin{equation}
f(R_{i-1},R_i)={{R_{m-1}+R_m}\over{R_{i-1}+R_i}}
\left[1-{2\over{\pi}}\,
{{\left(\cos^{-1}{{R_i}\over{R_m}}-{{R_{i-1}}\over{R_i}}\cos^{-1}{{R_{i-1}}
\over{R_m}}\right)-\left(\sqrt{{{R_m^2}\over{R_i^2}}-1}-\sqrt{{{R_m^2}\over
{R_i^2}}-{{R_{i-1}^2}\over{R_i^2}}}\right)}\over{1-R_{i-1}/R_i}}
\right] .
\label{eq:a1}
\end{equation}
For $\alpha=3$,
\begin{equation}
f(R_{i-1},R_i)={{R_m^2-R_{m-1}^2}\over{R_i^2-R_{i-1}^2}}\,
%{1\over{\ln(R_m/R_{m-1})}}\,
\left[{{\sqrt{1-{{R_{i-1}^2}\over{R_m^2}}}-\sqrt{1-{{R_i^2}\over{R_m^2}}}
-\ln\left({{1+\sqrt{1-R_{i-1}^2/R_m^2}}\over{1+\sqrt{1-R_i^2/R_m^2}}}\right)}
\over{\ln(R_m/R_{m-1})}}
\right] .
\label{eq:a2}
\end{equation}
And for $\alpha=4$,
$$f(R_{i-1},R_i)=
\left[{{(R_{m-1}+R_m)R_mR_{m-1}}\over{(R_{i-1}+R_i)R_iR_{i-1}}}\right]\ 
\ \ \ \times
\qquad\qquad\qquad\qquad\qquad\qquad\qquad\qquad\qquad\qquad\qquad$$
\begin{equation}
\qquad\qquad\qquad\qquad
\left[1-{2\over{\pi}}\,
{{\left({{R_i}\over{R_{i-1}}}\cos^{-1}{{R_{i-1}}\over{R_m}}-\cos^{-1}{{R_i}
\over{R_m}}\right)-{{R_i}\over{R_m}}\left(\sqrt{1-{{R_{i-1}^2}\over{R_m^2}}}-
\sqrt{1-{{R_i^2}\over{R_m^2}}}\right)}\over{R_i/R_{i-1}-1}}
\right] .
\label{eq:a3}
\end{equation}
These three power-law exponents are expected to bracket the conditions that
hold in real GCSs.

Finally, the volume densities derived in this way are a series of averages of
a continuous $n_{\rm cl}(r)$ in spherical shells of finite thickness. The
question therefore arises as to how a single radius $\overline{r}_i$ should be
assigned to each shell so that the discrete distribution, $n_{\rm cl}(R_{i-1},
R_i)$ vs.~$\overline{r}_i$, represents the continuous one as faithfully as
possible. (See \markcite{har86}Harris 1986 or \markcite{kin88}King 1988 for a
discussion of this issue in the two-dimensional context.) Clearly, the answer
is to choose each $\overline{r}_i\in [R_{i-1},R_i]$ such that
\begin{equation}
n_{\rm cl}(R_{i-1},R_i)={3\over{R_i^3-R_{i-1}^3}}
\int_{R_{i-1}}^{R_i} n_{\rm cl}(r)\,r^2\,dr\,=\,
n_{\rm cl}(\overline{r}_i)\ .
\label{eq:38}
\end{equation}
If it is assumed that $n_{\rm cl}\propto r^{-\alpha}$, the second equality is
easily solved for $\overline{r}_i$ in every shell. However, $\alpha$ will
generally not be a constant from shell to shell; that is, a real GCS need not
obey a single power-law density distribution over its entire extent (as is
evident, for example, in Fig.~\ref{fig3} above). Moreover, $\alpha$ cannot be
known a priori, and any attempts to derive it from the data must rely on fits
to the average deprojected densities---fits which themselves depend on the
values of $\overline{r}_i$. Fortunately, a complicated iterative procedure can
be avoided by noting that observed radial distributions suggest that $0\la
\alpha\la3$ anywhere in a typical GCS. This in turn suggests, and numerical
experiments confirm, that simply using $\alpha=3/2$ everywhere should lead to
$\overline{r}_i$ estimates that are acceptable substitutes for the exact
solutions of equation (\ref{eq:38}) in any realistic situation. Thus, the
formula adopted here is
\begin{equation}
\overline{r}_i=\left[(R_{i-1}^{3/2}+R_i^{3/2})/2\right]^{2/3}\ .
\label{eq:39}
\end{equation}

\end{document}